\pdfoutput=1 %this is for arXiv

\PassOptionsToPackage{dvipsnames}{xcolor}
\documentclass[10pt,showpacs,letterpaper,aps,nofootinbib,pra,onecolumn,superscriptaddress]{revtex4-2} 

\usepackage[margin=1.3in]{geometry}

\usepackage{mathtools,amssymb,amsthm,bm,centernot,dsfont}
\usepackage[title]{appendix}
\usepackage{xcolor}
\usepackage{enumitem}
	\setlist[itemize]{topsep=4pt,parsep=0pt,partopsep=0pt,itemsep=4pt}
	\setlist[enumerate]{topsep=0pt,parsep=0pt,partopsep=0pt,itemsep=4pt}
	\setlist[enumerate,1]{label=(\roman*)}
\usepackage[parfill]{parskip}		% Activate to begin paragraphs with an empty line rather than an indent
\usepackage{physics}
\usepackage[nottoc,numbib]{tocbibind}
\usepackage{chngcntr}
	\counterwithout{figure}{section}
\usepackage{soul}
\usepackage{todonotes}
	
\usepackage[colorlinks=false]{hyperref} 

\usepackage[capitalize]{cleveref}
\crefname{paragraph}{Sec.}{Sec.}
\usepackage{titlesec}
\titleformat{\paragraph}[hang]{\normalfont\normalsize\centering}{\theparagraph.}{1em}{}
\titlespacing*{\paragraph}{0pt}{3.25ex plus 1ex minus .2ex}{1.5ex plus .2ex}

\usepackage{csquotes}

\newtheorem{theorem}{Theorem}[section]
\newtheorem{corollary}[theorem]{Corollary}
\newtheorem{lemma}[theorem]{Lemma}

\theoremstyle{remark}

\newtheorem{notation}[theorem]{Notation}

\renewcommand{\emph}[1]{\textit{#1}}

\DeclareRobustCommand{\minwidthbox}[2]{%
  \ifmmode
    \expandafter\mathmakebox
  \else
    \expandafter\makebox
  \fi
  [\ifdim#2<\width\width\else#2\fi]{#1}%
}

\newcommand{\beq}{\begin{equation}}
\newcommand{\eeq}{\end{equation}}
\newcommand{\bea}{\begin{align}}
\newcommand{\eea}{\end{align}}

\newcommand{\dcs}{\mathord{\downarrow}}

\newcommand{\id}{\mathbb{I}}
\newcommand{\idmatrix}{\bm{I}}
\newcommand{\rmd}{\mathrm{d}}

\newcommand{\xz}{$\hat{x}$--$\hat{z}$\ }

\newcommand{\yz}{$\hat{y}$--$\hat{z}$\ }
\newcommand{\yzplane}{\yz plane}

\newcommand{\smallsqrt}[1]{\sqrt{\smash[b]{#1}}}

\newcommand{\afcov}{D_f}
\newcommand{\afcovext}{\mathrm{Ext(\afcov)}}

\let\oldvec\vec
\newlength{\vecwidth}
\renewcommand{\vec}[1]{%
	\minwidthbox{%
		\mkern-2mu\oldvec{\mkern2mu#1}%
	}{%
		\vecwidth}%
	}

\newcommand{\ph}{\mathord{\rule[-0.05em]{0.6em}{0.05em}}}		% Argument placeholder
\let\tr\relax
\DeclareMathOperator{\tr}{tr}

	\let\abs\relax
	\let\norm\relax
	\DeclarePairedDelimiter{\abs}{\lvert}{\rvert}
	\DeclarePairedDelimiter{\norm}{\lVert}{\rVert}
	\DeclarePairedDelimiterXPP{\pnorm}[2]{}{\lVert}{\rVert}{_{#1}}{#2}
	\makeatletter
		\let\oldabs\abs
		\def\abs{\@ifstar{\oldabs}{\oldabs*}}
		\let\oldnorm\norm
		\def\norm{\@ifstar{\oldnorm}{\oldnorm*}}
		\let\oldpnorm\pnorm
		\def\pnorm{\@ifstar{\oldpnorm}{\oldpnorm*}}
	\makeatother

	\providecommand{\given}{}			% Just to make sure the \given command exists.
	\newcommand{\SetSymbol}[1][]{%
		\nonscript\;\,#1\vert
		\allowbreak
		\nonscript\;\,
		\mathopen{}
	}
	\DeclarePairedDelimiterX{\Set}[1]{\{}{\}}{%
		\renewcommand{\given}{\SetSymbol[\delimsize]}
		#1
	}
	\makeatletter
		\let\oldSet\Set
		\def\Set{\@ifstar{\oldSet}{\oldSet*}}
	\makeatother

	\DeclarePairedDelimiterX{\Family}[1]{(}{)}{%
		\renewcommand{\given}{\SetSymbol[\delimsize]}
		#1
	}
	\makeatletter
		\let\oldFamily\Family
		\def\Family{\@ifstar{\oldFamily}{\oldFamily*}}

	\newsavebox{\numbox}%
	\newsavebox{\slashbox}%
	\newsavebox{\denbox}%
	\newlength{\slashlength}%
	\newlength{\faktorscale}%
	\DeclareDocumentCommand{\newfaktor}{m O{0.35} m O{-0.35}}{% \newfaktor{#1}[#2]{#3}[#4] -> #1/#3
		\savebox{\numbox}{\ensuremath{#1}}% Store numerator
		\savebox{\slashbox}{\ensuremath{\diagup}}% Store slash /
		\savebox{\denbox}{\ensuremath{#3}}% Store denominator
		\setlength{\faktorscale}{0.5\ht\numbox+0.5\ht\denbox}%
		\setlength{\slashlength}{2pt+0.8\faktorscale+#2\faktorscale-#4\faktorscale}%
		\raisebox{#2\ht\slashbox}{\usebox{\numbox}}% Numerator
		\mkern-2mu%
		\rotatebox{-30}{\rule[#4\ht\denbox]{0.4pt}{\slashlength}}% tilted rule as a slash
		\mkern9mu%
		\hspace{-0.44\slashlength}%
		\raisebox{#4\ht\denbox}{\usebox{\denbox}}% Denominator
	}
	\DeclareDocumentCommand{\linefaktor}{m O{0.08} m O{-0.08}}{% \newfaktor{#1}[#2]{#3}[#4] -> #1/#3
		\savebox{\numbox}{\ensuremath{#1}}% Store numerator
		\savebox{\slashbox}{\ensuremath{\diagup}}% Store slash /
		\savebox{\denbox}{\ensuremath{#3}}% Store denominator
		\setlength{\faktorscale}{0.5\ht\numbox+0.5\ht\denbox}%
		\setlength{\slashlength}{0.2\faktorscale+0.8\baselineskip}%
		\raisebox{#2\ht\slashbox}{\usebox{\numbox}}% Numerator
		\mkern-1mu%
		\raisebox{-0.8pt}{%
			\rotatebox{-30}{\rule[#4\ht\denbox]{0.4pt}{\slashlength}} % tilted rule as a slash
		}%
		\mkern-1mu%
		\hspace{-0.25\slashlength}%
		\raisebox{#4\ht\denbox}{\usebox{\denbox}}% Denominator
	}

\let\originalleft\left
\let\originalright\right
\renewcommand{\left}{\mathopen{}\mathclose\bgroup\originalleft}
\renewcommand{\right}{\aftergroup\egroup\originalright}

\begin{document}

%\title{Resource dependence relations: a simple example in the resource theory of asymmetry}
%\title{A prolegomenon to studying resource dependence relations}
\title{Conceptual and formal  groundwork for the study of resource dependence relations}
%\date{}                                           % Activate to display a given date or no date

\author{Y{\`i}l{\`e} Y{\=\i}ng}
\email{yying@pitp.ca}
\affiliation{Perimeter Institute for Theoretical Physics, Waterloo, Ontario, Canada, N2L 2Y5}
\affiliation{Department of Physics and Astronomy, University of Waterloo, Waterloo, Ontario, Canada, N2L 3G1}
\author{Tomáš Gonda}
\email{tomas.gonda@uibk.ac.at}
\affiliation{Institute for Theoretical Physics, University of Innsbruck, Austria}
\author{Robert Spekkens}
\email{rspekkens@pitp.ca}
\affiliation{Perimeter Institute for Theoretical Physics, Waterloo, Ontario, Canada, N2L 2Y5}
\affiliation{Department of Physics and Astronomy, University of Waterloo, Waterloo, Ontario, Canada, N2L 3G1}

\begin{abstract}

	A resource theory imposes a preorder over states, with one state being above another if the first can be converted to the second by a free operation, and where the set of free operations defines the notion of resourcefulness under study. 
	In general, the location of a state in the preorder of one resource theory can constrain its location in the preorder of a different resource theory. 
	It follows that there can be nontrivial dependence relations between different notions of resourcefulness. 
In this article, we lay out the conceptual and formal groundwork for the study of resource dependence relations.  In particular, we note that the relations holding among a set of monotones that includes a complete set for each resource theory provides a full characterization of resource dependence relations.
As an example, we consider three resource theories concerning the about-face asymmetry properties of a qubit along three mutually orthogonal axes on the Bloch ball, where about-face symmetry refers to a representation of $\mathbb{Z}_2$, consisting of the identity map and a $\pi$ rotation about the given axis. This example is sufficiently simple that we are able to derive a complete set of monotones for each resource theory and to determine all of the relations that hold among these monotones, thereby completely solving the problem of determining resource dependence relations.
Nonetheless, we show that even in this simplest of examples, these relations are already quite nuanced.
\end{abstract}

{\let\clearpage\relax \maketitle}
\tableofcontents

\section{Introduction}

A useful way to think about certain properties of quantum states is in terms of the paradigm of resource theories~\cite{COECKE201659,Chitambar2019}.  Examples of properties of quantum states that are studied as resources include entanglement~\cite{plenio2006introduction,Horodecki2009}, athermality~\cite{GOUR20151,Lostaglio_2019}, and asymmetry~\cite{GouR_2008,marvianExtending2014}. 

A quantum resource theory of states is defined by identifying a set of \emph{free operations}.
	These include quantum channels interpreted as manipulations implementable in a restricted scenario, e.g.\ with no communication (for entanglement), no source of work (for athermality), or no access to a reference frame (for asymmetry).
	A set of free operations then induces a hierarchy among quantum states{\,---\,}a state is higher up if it can be converted to ones below it via free operations.
	It is possible, for a given pair of states, that no free operation can convert between them in either direction.
	Therefore, the resulting resource hierarchy is generally a \emph{preorder relation} that need not be a \emph{total} preorder (such as the ordering of real numbers).

	We can characterize the hierarchy by a set of \emph{resource monotones} (a.k.a. resource measures), which assign a value to each state in an order-preserving way.
	Whenever we deal with a preorder that is not a total preorder, a complete characterization can be only achieved with more than one monotone.
	That is why, for instance, the property of quantum entanglement cannot be captured by a single entanglement measure.

In this article, we consider situations with multiple relevant restrictions, giving rise to multiple sets of free operations and multiple resource hierarchies.
	In particular, we are interested in questions such as: 
	If a given state is high in one of these resource hierarchies, what can be said about its location in the remaining ones?
	In other words, we study dependence relations\footnote{The word ``relation'' in the term ``dependence relation'' should not be understood as referring to the mathematical concept of a relation but rather in the same sense as ``relation'' in the term ``uncertainty relation'', which expresses constraints on the quantities appearing therein.} among different notions of resourcefulness.

	An example of such a dependence relation is the monogamy of entanglement~\cite{Coffman2000distributed}, which has practical implications for quantum cryptography~\cite{Pawlowski2010}.
	More precisely, for a joint quantum state of three systems ($A$, $B$, and $C$), monogamy says that the entanglement 
	between $A$ and $B$, between $A$ and $C$, and between $B$ and $C$ cannot be maximized simultaneously.
	Each of these pairs gives rise to a resource theory of bipartite entanglement defined on the three systems, in which the free operations are local operations and classical communication (LOCC) on the pair  and  the trivial set of operations (containing only identity) on the third system. 
	The monogamy of entanglement can thus  be cast as a resource dependence relation{\,---\,}a given tripartite quantum state cannot be at the top of the resource order 
	  in all three resource theories. Dependence relations between different entanglement properties have also been studied for the  case where the free operations are LOCC across different bipartitions of a tripartite system~\cite{GeShuming2024,liu2024polygon}. 
	
	There are also nontrivial dependence relations among asymmetry resources. 
For example, consider a Hermitian operator $L${\,---\,}an infinitesimal generator of an instance of the Lie group U(1).
	Free operations of the corresponding resource theory of asymmetry are those quantum channels which are covariant with respect to the canonical action of this group. 
	Defining the variance of the observable $L$ in the usual way, $V_L \coloneqq \langle L^2\rangle - \langle L\rangle^2$, one can show that its inverse $V_L^{-1}$ is an asymmetry monotone for pure states.
	The variances of two non-commuting generators $L_1$ and $L_2$ cannot be simultaneously minimized due to an uncertainty relation $ V_{L_1}V_{L_2}\geq \frac{1}{2} \abs*[\big]{ \big\langle [L_1, L_2] \big\rangle }$~\cite{Robertson1929}.
	Consequently, there can be no pure state that is maximally asymmetric relative to the actions of the two symmetry groups generated by exponentiating $L_1$ and $L_2$ respectively.
	The two resource theories of asymmetry carry a trade-off.

Variances are, however, not suitable for studying asymmetry properties in general, since they do not provide asymmetry monotones for impure states. 
	More general and physically meaningful dependence relations are those among true asymmetry monotones, applicable to all states~\cite{marvian2012symmetry}.
	A suitable resource monotone extending variance is provided by skew information \cite{marvianExtending2014}.
	In this sense, literature exploring uncertainty relations for skew information \cite{Luo2003,Furuichi2009TraceIO} indirectly studies dependence relations among asymmetry properties.
Other dependence relations for various particular asymmetry monotones have been also studied~\cite{Adesso2014,cheng_complementarity_2015,Singh_2016,shenQuantum2020,Rastegin_2021}. 

Asymmetry resource dependence relations are relevant for quantum metrology. In particular, the optimal degree of success achievable in a metrological task for a given state is necessarily an asymmetry monotone~\cite{marvian2012symmetry}. 	
	Consequently, asymmetry dependence relations imply constraints for the simultaneous achievability of multiple metrological tasks.

Dependence relations between resources of different kinds, e.g.\ between entanglement and asymmetry properties, have also been studied. For example, Ref.~\cite{Gisin_1999} indicates such a dependence in the sense that states of multiple spin-1/2 particles that maximize rotational asymmetry relative to a particular monotone are necessarily entangled.

	The above five paragraphs showcase three examples of resource dependence relations:  between entanglement properties for different bipartite subsystems of a tripartite system,  between asymmetry properties for different actions of a symmetry group, and between entanglement properties and asymmetry properties. 
	The first two are \emph{trade-offs} in the sense that maximizing one resource property entails that another resource property cannot be maximized. 
	The last one is instead an example of a \emph{positive dependence}, where maximizing one resource property entails that another resource property is bounded below.
	Some of these dependence relations also have practical applications.
	In this paper, we aim to set up a foundational framework for describing resource dependence relations and to inform the way to approach an investigation of these. 

	In order to illustrate our general scheme for deriving dependence relations, we apply it to a concrete and simple example. 
	Specifically, we restrict our attention to resource theories of asymmetry associated with actions of the discrete group $\mathbb{Z}_2$. 
	For any choice of a spatial axis $\hat{n}$ in $\mathbb{R}^3$, we consider the representation of $\mathbb{Z}_2$ given by identity and a $\pi$ rotation about $\hat{n}$.
	We refer to this representation as the \emph{about-face symmetry} relative to $\hat{n}$. 
	Interesting dependence relations then arise for a triple of resource theories of about-face asymmetry, for example, those corresponding to an orthogonal triad of axes, denoted by $\hat{x}$, $\hat{y}$, and $\hat{z}$ respectively. 

	The reason we use this example is that we wish to have a {\em complete} characterization of the resource dependence relations, which requires a complete set of monotones to be known.
	There are, however, very few resource theories for which this is the case.\footnote{One example is the resource theory of athermality for a single qubit, which also implies a characterization of the resource theory of $U(1)$-asymmetry of a single qubit~\cite{lostaglio2015quantum}.}
	As we show here, such a complete set \emph{can} be obtained for the resource theory of about-face asymmetry of qubit states for any axis (a novel contribution of this article).
	We then use these to fully characterize the dependence relations among the triple of about-face asymmetry properties. 

	Moreover, thanks to the relative simplicity of the resource theory of about-face asymmetry and their dependence relations, we obtain intuitive geometrical accounts thereof. 
	This gives us a clear conceptual understanding and operational implications of these relations. 	Therefore, although the practical significance of dependence relations among resource theories of $\mathbb{Z}_2$-asymmetry is not immediate, they allow us to illustrate 
	\begin{enumerate}
		\item what a complete characterization of resource dependence relations looks like,
		\item what kinds of results one can expect in general, and
		\item how to understand and interpret these results. 
	\end{enumerate}

	The article is organized as follows: In \cref{sec_prelim}, we establish the language and framework for discussing resource dependence relations and in \cref{sec_recipe}, we provide a detailed recipe for deriving and analyzing them. 
	In \cref{sec_aboutface0,sec_constraint1RT,sec_eqineq0,sec_interpret3way}, we summarize our characterization of dependence relations among about-face asymmetry properties for three mutually orthogonal axes, obtained following our recipe. 
	The derivations of these findings, together with some further explanations, are given in \cref{sec_aboutface,sec_relations}. 
	The proof for our novel complete characterization of the about-face resource ordering is deferred to \cref{app:qubitConversion}.
	Lastly, in \cref{sec_discussion}, we summarize the lessons learned from our investigation and suggest future research directions.

\section{The recipe for deriving and interpreting resource dependence relations}
\label{sec_summary}

\subsection{Preliminaries}\label{sec_prelim}

A resource theory is defined by a set $R$ of \emph{resource objects} (a.k.a.\ \emph{resources} for short) and a set $F$ of operations that are deemed to be realizable at no cost an arbitrary number of times, called the {\em free operations}.\footnote{In this paper, we follow the definition of resource theory given by Ref.~\cite{COECKE201659} in the framework of partitioned process theories.} In general, $R$ can include resource objects of various types, such as states, channels, instruments, combs, etcetera. The input and output spaces of the free operations can also be arbitrary, so the definition of $F$ may include a specification of the free subset of every different type of operation, including states, channels and so on. Any operation that is in the set $F$ and any resource that can be prepared by operations in $F$ is termed \emph{$F$-free}.   Otherwise, it is termed \emph{$F$-nonfree}.  

	A few examples demonstrate the spirit of this terminological convention. 
	If the set of operations that are covariant with respect to the action of the group $G$ is denoted by $G$-{cov}, then the nonfree resources and operations are termed $G$-{cov}-nonfree.  Often, these are referred to simply as  $G$-{asymmetric}~\cite{marvianExtending2014}. Similarly, if the operations that are thermal relative to a bath at temperature $T$ are denoted $T$-{thermal}, then the nonfree resources and operations are termed $T$-{thermal}-nonfree, or simply $T$-{athermal} \cite{GOUR20151}.  
	In a Bell scenario, one is primarily interested in the resource objects corresponding to processes with two classical inputs and two classical outputs that make use of a common (quantum) source~\cite{Wolfe_2020}.  In this case, the free resources are those that can be prepared by local operations and shared randomness (LOSR), and the remaining ones are termed LOSR-nonfree.\footnote{Under this terminological convention, notions of resourcefulness are described {\em negatively} in terms of the set of free operations.  This convention is not universally followed in the literature on resource theories.  For instance, states that are not realizable by local operations and classical communication (LOCC) are standardly referred to as {\em entangled}~\cite{Horodecki2009}.  A disadvantage of the standard terminological convention is that the set LOSR~\cite{schmid2023understanding} defines precisely the same set of nonfree resources as LOCC, so that one is equally justified in using the term `entangled' as a shorthand for LOSR-nonfree rather than LOCC-nonfree.  Meanwhile, LOCC and LOSR define different preorders, so that LOCC-nonfreeness and LOSR-nonfreeness are inequivalent resources.  To distinguish the two notions, therefore, it is necessary to introduce a distinction in types of entanglement, such as LOCC-entanglement and LOSR-entanglement~\cite{schmid2023understanding}, which amounts to the same as using LOCC-nonfree and LOSR-nonfree. }%
 
	It is useful to order resource objects in terms of their degree of resourcefulness. 
	This means determining the \emph{resource ordering} $\succeq$ induced by $F$-convertibility, which is a preorder relation. That is, a resource $\tt r$ is above a resource $\tt s$, denoted $\tt r \succeq \tt s$, if there is a process in $F$ that maps $\tt r$ to $\tt s$. 	
	Two resources are in the same equivalence class relative to $F$-convertibility if and only if each can be converted into the other using operations in $F$. 
	The corresponding resource ordering among the equivalence classes, also denoted by $\succeq$, is then a \emph{partial order}.
	If the set of free resources is nonempty, then it is the unique lowest element in the partial order of equivalence classes of resources.
	We use the term \emph{$F$-nonfreeness properties} of a resource to refer to those properties that determine its location in	the partial order induced by $F$-convertibility. For example, by completely specifying the $G$-asymmetry properties of a state, its location in the partial order of states induced by $G$-{\tt cov}-convertibility is fully determined. 
	
	Consider a fixed set $R$ of resource objects
	  and two distinct sets of free operations, $F_1$ and $F_2$. These define two resource theories of $R$ with associated resource orderings $\succeq_1$ and $\succeq_2$, respectively, and we can study the dependence relations between these two resource theories.
	Since the resource objects are the same, we abuse notation and denote the resource theory associated with resources in $R$ and free operations in $F_i$ simply by $F_i$. 

 	There are two extreme kinds of resource dependence relations between the resource orderings $\succeq_1$ and $\succeq_2$.
	To express them, let us denote by $\tilde{R}_i \coloneqq R / {\succeq_i}$ the partially ordered set of $\succeq_i$-equivalence classes and by $q_i$ the associated quotient map $R \to \tilde{R}_i$, which takes each resource ${\tt r} \in R$ to its equivalence class relative to the preorder relation $\succeq_i$.
	The dependence relations between the two \emph{partially ordered sets}, $(\tilde{R}_1, \succeq_1)$ and $(\tilde{R}_2, \succeq_2)$, are captured by the set of realizable \emph{pairs} of values within the Cartesian product $\tilde{R}_1 \times \tilde{R}_2$. 
	Specifically, this set can be defined as the image of the function $R \to \tilde{R}_1 \times \tilde{R}_2$ given by ${\tt r} \mapsto \bigl( q_1({\tt r}), q_2({\tt r}) \bigr)$, and called the \emph{joint-realizability relation} $\mathcal{J} \subseteq \tilde{R}_1 \times \tilde{R}_2$. 
	If $(\tilde{\tt r}_1,\tilde{\tt r}_2)$ is an element of $\mathcal{J}$, it means that there exists a resource $\tt r$ such that it belongs to both the equivalence class represented by $\tilde{\tt r}_1$ in $(\tilde{R}_1, \succeq_1)$ and the equivalence class represented by $\tilde{\tt r}_2$ in $(\tilde{R}_2, \succeq_2)$ and thus, $(\tilde{\tt r}_1,\tilde{\tt r}_2)$ can be jointly realized by $\tt r$.  
	
	Consider two resource theories whose respective orderings, $\succeq_1$ and $\succeq_2$, are nontrivial (i.e., $\tilde{R}_1$ and $\tilde{R}_2$ are not singletons):
	\begin{itemize}
		\item If the joint-realizability relation is the full relation, i.e., if ${\mathcal{J} = \tilde{R}_1 \times \tilde{R}_2}$ holds, then the two resource theories are \emph{independent}.
			In this case, fixing $F_1$-nonfreeness properties places no constraints on the compatible $F_2$-nonfreeness properties (and vice versa).
			
		\item If the joint-realizability relation defined a bijection between $\tilde{R}_1$ and $\tilde{R}_2$, i.e., if for each $\tilde{\tt r}_1 \in \tilde{R}_1$ there is precisely one $\tilde{\tt r}_2 \in \tilde{R}_2$ such that the pair $(\tilde{\tt r}_1,\tilde{\tt r}_2)$ is jointly realizable (and vice versa), then the two resource theories are \emph{fully dependent}.
			In this case, fixing $F_1$-nonfreeness properties completely specifies all $F_2$-nonfreeness properties (and vice versa).	
	\end{itemize}
	In general, the joint-realizability relation between two resource theories falls in between these two extreme cases, and a careful study of resource dependence relations is needed.

	As we can see, resource dependence relations essentially concern the joint-realizability of various nonfreeness properties \cite{fraser2023realizability}. Namely, the key question is: 
	Which nonfreeness properties can be achieved simultaneously by a resource?
	
	In the case where the two resource theories are not independent, one can further study how the \emph{resource ordering} in $\tilde{R}_1$ constraint the resource ordering in $\tilde{R}_2$ under the joint-realizability relation.
	To simplify the discussion, consider two fully dependent resource theories whose $\tilde{R}_1$ and $\tilde{R}_2$ are not singletons. For two fully dependent resource theories, their joint-realizability relation defines a bijection map $f_{\mathcal{J}} \colon \tilde{R}_1 \to \tilde{R}_2$. 
	\begin{itemize}
		\item If $f_{\mathcal{J}}$ is an order-isomorphism, 
		i.e.,\ if we have
		\begin{equation}
			\label{eq_jsy}
			\tilde{\tt r} \succeq_1 \tilde{\tt s}  \quad \iff \quad  f_{\mathcal{J}}(\tilde{\tt r}) \succeq_2 f_{\mathcal{J}}(\tilde{\tt s}),
		\end{equation}
		for all ${\tt r,s}\in \tilde{R}_1$, then the resource dependence relation is a \emph{synergy relation}.
		Thus, a resource is above another one in resource theory $F_1$ if and only if the same ordering relation holds in the other resource theory. 
		
		\item If $f_{\mathcal{J}}$ is an order-antiisomorphism 
		i.e.,\ if we have
		\begin{equation}
			\label{eq_jtr}
			\tilde{\tt r} \succeq_1 \tilde{\tt s}  \quad \iff \quad  f_{\mathcal{J}}(\tilde{\tt r}) \preceq_2 f_{\mathcal{J}}(\tilde{\tt s}),
		\end{equation}
		for all ${\tt r,s} \in \tilde{R}_1$, then the resource dependence relation is a \emph{trade-off relation}.
		Thus, a resource is above another one in resource theory $F_1$ if and only if the {\em opposite} ordering relation holds in the other resource theory.
	\end{itemize}

	However, even for two fully dependent resource theories, their dependence relations can be neither pure synergy nor pure trade-off.	As we will see later, for resource theories in general, the dependence relations among them can be nuanced and intricate, even in seemingly simple cases. 

	Fully characterizing the dependence relations among certain nonfreeness properties can, in general, be technically challenging and, at times, even infeasible. 
	Nevertheless, the general recipe that we outline in \cref{sec_recipe} describes the necessary steps to take, the types of questions to pose, the crucial distinctions to consider, and the potential answers to anticipate concerning resource dependence relations. 

	An important notion used in the recipe is a {\em resource monotone}~\cite{COECKE201659}. 
	Consider the preordered set $(R, \succeq)$ of resources with $\succeq$ induced by convertibility under the free operations defining the resource theory. 
	Any function $R \to \mathbb{R}$ from resources to reals, which is order-preserving, is termed a resource monotone. 
	That is, if a resource $\tt r$ can be converted to another resource $\tt s$ under $F$-free operations, then a resource monotone must assign a value to $\tt r$ that is not lower than the value assigned to $\tt s$.
	It follows that all resources in a fixed equivalence class are assigned the same values by a resource monotone.
	Therefore, a monotone is equivalently an order-preserving map $(\tilde{R}, \succeq) \to (\mathbb{R}, \geq)$.

	A set of resource monotones is called \emph{complete} if it completely captures the partially ordered set $(\tilde{R}, \succeq)$.
	In particular, the set of values of monotones in the complete set fully determines, for each resource, the equivalence class it belongs to. 
	The choice of a complete set of monotones is not unique. 
	For any such choice, however, an arbitrary resource monotone can be expressed as a function of the ones in the chosen complete set.
	Consequently, a complete set of monotones characterizes \emph{all the $F$-nonfreeness properties} of a resource while an individual monotone characterizes only one aspect of $F$-resourcefulness, or one $F$-nonfreeness property.

\subsection{The recipe}
\label{sec_recipe}

	Consider a fixed set of resource objects $R$ and a collection of free operations $\mathcal{F} = \{ F_1, F_2, \ldots, F_n\}$, each associated with a resource theory of $R$.
	To understand the resource dependence relations among them, we propose a recipe that involves the following steps. 
	
	\begin{enumerate}[label=(\arabic*)]
		\item Understand each resource theory individually, e.g.,\ in terms of a complete set of monotones and relations between them.
		\item Derive dependence relations among all monotones across the resource theories under consideration. 
		\item Establish conceptual understandings of these relations.
	\end{enumerate}
	
	This recipe can still be followed even if one cannot derive complete sets of monotones. In such cases, although it is impossible to fully characterize the dependence relations of the resource theories under interest, our recipe still provides guidance for finding partial characterizations of these relations. Note that when complete sets of monotones are unknown, extra care is needed to distinguish whether constraints on a set of monotones stem from existing constraints in each individual resource theory or if they indeed describe dependence relations among the resource theories.

	\subsubsection*{Step (1): Understanding each resource theory individually}
	\addcontentsline{toc}{subsubsection}{Step (1): Understanding each resource theory individually}
	
	Before studying dependence relations among the resource theories, we have to learn about each separately.
	For each resource theory $F_i \in \mathcal{F}$, the task is to
	\begin{enumerate}[label=(1\alph*)]
		\item Study the equivalence classes of resource objects in the resource theory $F_i$ and their ordering $\succeq_i$ in terms of a set of monotones $\mathcal{M}_i\coloneqq \{M_{i,\alpha}\}_{\alpha \in \mathcal{I}_i}$.
			This set of monotones defines an order-preserving map $\mathcal{M}_i \colon R \to \mathbb{R}^{\mathcal{I}_i}$  from the set of resources $R$ to the ordered real vector space $\mathbb{R}^{\mathcal{I}_i}$, a so-called generalized monotone. %\footnote{A generalized monotone is in general an order-preserving function $\mathcal{M}_i \colon (R, \succeq_i) \to (\mathbb{R}^k, \geq)$, where $\mathbb{R}^k$ is now an ordered vector space of dimension $k$. Two of its vectors are ordered, denoted by $(v_\alpha) \geq (w_\alpha)$, whenever $v_\alpha \geq w_\alpha$ holds for all $\alpha \in \{1,2,\ldots, k\}$. Since, for a fixed resource theory,  a set of monotones and a generalized monotone are equivalent concepts, we use them interchangeably.} %
			Abusing the notation, we denote both sets of monotones and the function it defines by $\mathcal{M}_i$. 
			Ideally, $\mathcal{M}_i$ is a complete set of monotones, in which case it induces an order-isomorphism between the preordered set $(R, \succeq)$ of resources and 
%			its image $\mathrm{im}(\mathcal{M}_i) \subseteq \mathbb{R}^{\mathcal{I}_i}$.
%			In particular, the elements of $\mathrm{im}(\mathcal{M}_i)$ are then in bijection with the equivalence classes of resources.
%			We call $\mathrm{im}(\mathcal{M}_i)$ the set of \emph{jointly realizable values} of the monotones in $\mathcal{M}_i$.
its image within $\mathbb{R}^{\mathcal{I}_i}$.
			In particular, the elements of this image are then in bijection with the equivalence classes of resources.			
			This image defines the set of \emph{jointly realizable values} of the monotones in $\mathcal{M}_i$, and consequently we will denote it by $\mathtt{JointRealize}(\mathcal{M}_i)$.

		\item Find the full set of constraints (equalities and inequalities) characterizing $\mathtt{JointRealize}(\mathcal{M}_i)$. 
			Whenever $\mathcal{M}_i$ is complete, these constraints capture all dependence relations that hold for nonfreeness properties \emph{within} the resource theory $F_i$. 
	 \end{enumerate}

	Equality constraints on $\mathtt{JointRealize}(\mathcal{M}_i)$ can sometimes signify that some monotones in the set $\mathcal{M}_i$ are redundant. For example, if an equality constraint is linear in the values of the monotones, then we can express one monotone $M_{i,\alpha}$ as a function of others.
	As a result, $M_{i,\alpha}$ can be removed from $\mathcal{M}_i$ without losing any information about the resource ordering $\succeq_i$.

	Monotones $M_{i,\alpha} \in \mathcal{M}_i$ and the constraints on them also reveal aspects of the partial order $\succeq_i$ of the resource theory $F_i$. 
	Let us discuss a few examples to illustrate this point. 
	If there is a unique element of $\mathbb{R}^{\mathcal{I}_i}$ for which all the monotones in $\mathcal{M}_i$ are maximized, then the partial order $(\tilde{R}, \succeq)$ has a unique maximal element{\,---\,}the so-called top element.
	If the complete set of monotones $\mathcal{M}_i$ cannot be reduced to a single element by equality constraints, then the partial order is not a total order (and vice versa). 
	Equivalently, one can then find two resources $\tt r$ and $\tt s$ such that neither ${\tt r} \succeq {\tt s}$ nor ${\tt s} \succeq {\tt r}$ holds{\,---\,}the two resources are \emph{incomparable}.
	In terms of monotones, this means that there must be $M_{i,\alpha}, M_{i,\beta} \in \mathcal{M}_i$ such that we have both
	\begin{equation}
		M_{i,\alpha}({\tt r}) > M_{i,\alpha}({\tt s})  \qquad \text{and} \qquad  M_{i,\beta}({\tt r}) < M_{i,\beta}({\tt s}).
	\end{equation}
	Additionally, inequality constraints tell us how the values of some monotones influence the range of possible values of other monotones. 
	If there exist two monotones $M_{i,\alpha}, M_{i,\beta} \in \mathcal{M}_i$ and two closed intervals $I_{\alpha}, I_\beta \subseteq \mathbb{R}$, such that all points in the rectangle $I_{\alpha} \times I_\beta$ are realizable, then the resource ordering has infinite width and is not weak (i.e., the incomparability relation is not transitive). The proof is provided in \cref{app_incompara}.

	Finding relations among monotones within each resource theory not only helps with our understanding of that specific theory, but also will subsequently enable us to distinguish between two types of constraints: those that pertain to nonfreeness properties within \emph{each individual theory} under consideration, and those that describe dependence relations \emph{among different resource theories}.

\subsubsection*{Step (2): Deriving dependence relations between resource theories}
\addcontentsline{toc}{subsubsection}{Step (2): Deriving dependence relations between resource theories}	

	To study relations between the resource theories in $\mathcal{F}$, we compile all the arenas for understanding the individual resource orderings into a single one.
	That is, we consider the union of all of the monotones (described in step (1)) from each of the resource theories in $\mathcal{F}$:
	\begin{align}
		\label{eq_defM}
		\mathcal{M} \coloneqq &\bigsqcup_{i=1}^n \mathcal{M}_i = \Set*[\big]{ M_{i,\alpha} \given  i \in \{1, \ldots, n\} , \; \alpha \in \mathcal{I}_i }
	\end{align}
	Let us denote the disjoint union of all their index sets $\mathcal{I}_i$ by $\mathcal{I}$.
	Then we get a function $\mathcal{M} \colon R \to \mathbb{R}^{\mathcal{I}}$ that maps a resource $\rho$ to the tuple composed of the values $M_{i,\alpha}(\rho)$ for each $i$ and each $\alpha \in \mathcal{I}_i$. 
	Notably, $\mathcal{M}$ is not a generalized monotone, since $\mathcal{M}_i$ corresponds to distinct resource theories.
	The task is to:
	\begin{enumerate}[label=(2\alph*)]
		\item Derive all the constraints (equalities and inequalities) characterizing the set $\mathrm{im}(\mathcal{M})$ as a subset of $\mathbb{R}^{\mathcal{I}}$.
			These generalize the joint-realizability relation from \cref{sec_prelim}, which only applies to the case of comparing two resource theories.
			Any such constraints that do not follow from those determined in step (1b) necessarily describe dependence relations among nonfreeness properties of different resource theories. 
	\end{enumerate}
	
	Similarly to step (1b), the constraints on the monotones in $\mathcal{M}$ can manifest either as equalities or as inequalities. 
	The equalities may inform us that some of the monotones can be expressed as functions of others while the inequalities describe the boundaries of $\mathrm{im}(\mathcal{M})$. 

	Given the full set of equality constraints, there is a (non-unique) choice of a set of inequality constraints to serve as a \emph{generating set}.
	These, together with the equalities, can be used to derive any other inequality constraint. 
	Consequently, the generating set of inequalities, in conjunction with the complete set of equalities, suffices to characterize \emph{all} dependence relations among the resource theories under consideration.

	Steps (1) and (2) are primarily about solving the technical problems related to resource dependence relations, such as finding complete sets of monotones and the mathematical constraints among them. 
	Some of the relevant techniques for the latter are those developed for quantifier elimination problems. 
	Since such problems are hard in general, it may be difficult to identify analytic expressions for complete sets of monotones and also difficult to derive the full set of constraints on these monotones.
	Nonetheless, even if one cannot derive complete sets of monotones or the full set of constraints, the recipe described here can be followed for whatever (incomplete) sets of monotones and for whatever constraints one {\em has} identified. 

\subsubsection*{Step (3): Establishing conceptual understandings of these dependence relations}
\addcontentsline{toc}{subsubsection}{Step (3): Establishing conceptual understandings of these dependence relations}	
\label{sec_step3}

Step (3), on the other hand, is about extracting useful conclusions from the technical developments.
	Here, we seek to analyze and gain a conceptual understanding of dependence relations among a collection of monotones. We mention three ways to extract useful information, labelled by (3a)--(3c). 

	\begin{enumerate}[label=(3\alph*)]
		\item Characterize the dependence relations among monotones across different resource theories.
	\end{enumerate}
	
	To classify dependence relations between monotones, we can once again distinguish between synergy and trade-off relations, just as in \cref{sec_prelim}. 
	This helps us uncover their conceptual significance, since each monotone represents an aspect of the nonfreeness properties in its respective resource theory.

	Since generic dependence relations are neither a strict synergy nor a strict trade-off, we define these notions relative to a subset $R_0$ of resources.
	The same set of monotones can thus manifest synergy in one region of the set $R$ of all resources and manifest trade-off in another region.
	This allows us to make more fine-grained statements about their dependence relation.

	Let us first consider dependence relations between merely two monotones $A, B \in \mathcal{M}$. 
	\begin{itemize}
		\item The pair of $A$ and $B$ is said to exhibit a \emph{synergy relation} within the subset $R_0 \subseteq R$ if for any two resources ${\tt r,s} \in R_0$, we have
		\begin{equation}
			\label{eq_mmsy}
			A({\tt r})\geq A({\tt s}) \quad \iff \quad B({\tt r})\geq B({\tt s}),
		\end{equation}
		and both $A$ and $B$ are nontrivial functions on $R_0$ (i.e., neither $A$ nor $B$ is a constant on $R_0$), and consequently, the pair of nonfreeness properties specified by $A$ and $B$ also exhibit a synergy relation within $R_0$.

		\item On the other hand, they exhibit a \emph{trade-off relation} within $R_0$ if for any two resources ${\tt r,s}\in R_0$, we have
		\begin{equation}
			\label{eq_mmtr}
			A({\tt r})\geq A({\tt s}) \quad \iff \quad B({\tt r})\leq B({\tt s}),
		\end{equation}
		and both $A$ and $B$ are nontrivial functions on $R_0$, and consequently, the pair of nonfreeness properties specified by $A$ and $B$ also exhibit a trade-off relation within $R_0$.
	\end{itemize}
	Whenever $A$ and $B$ belong to distinct resource theories, e.g., $A \in \mathcal{M}_1$ and $B\in \mathcal{M}_2$, such a synergy or trade-off relation is a dependence relation between an $F_1$-nonfreeness property and an $F_2$-nonfreeness property.

When considering the dependence relations among more than two monotones, the most general case in which one should say that a set of monotones exhibits a synergy or a trade-off relation is unclear to us. Nevertheless, in some special cases, determining whether there is a synergy or a trade-off relation is straightforward.

For example, consider a triple of nontrivial monotones $A, B, C \in \mathcal{M}$ such that for any ${\tt r}\in R_0$, we have 
$A({\tt r})=B({\tt r})= C({\tt r})$. 
This is a case where $A$, $B$, and $C$ clearly exhibit a synergy relation. 

Consider another example with a triple of nontrivial monotones $A, B, C \in \mathcal{M}$, and a constraint that implies that they cannot be made to simultaneously achieve the maximum values that they can each achieve individually. 
For instance, suppose that for any ${\tt r}\in R_0$, we have $A({\tt r})+B({\tt r})+C({\tt r})=c$, where
\begin{equation}
	c<\max_{{\tt r}\in R_0}A({\tt r})+\max_{{\tt r}\in R_0}B({\tt r})+\max_{{\tt r}\in R_0}C({\tt r}),
\end{equation}
i.e., $c$ is a constant \emph{strictly} smaller than the sum of the individual maximum values of $A$ $B$ and $C$.
This is a case where one may say that $A$, $B$, and $C$ exhibit a trade-off relation. 
The pair of $A$ and $B$ among these can still exhibit synergy for certain choices of the domain $R_0$.
However, if we restrict the domain $R_0$ to a subset 
\begin{equation}
	\Set*[\big]{\mathtt{r} \in R_0  \given  C(\mathtt{r}) = x_C  }
\end{equation}
on which $C$ monotone has a fixed value $x_C$ (and on which $A$ and $B$ are nontrivial), then $A$ and $B$ necessarily exhibit a trade-off relation.

Inspired by these two examples, we list the following \emph{sufficient} (not necessary) conditions for a set of monotones $\mathcal{S}\subseteq \mathcal{M}$ to exhibit a synergy or a trade-off relation.
\begin{itemize}
	\item A sufficient condition for $\mathcal{S}$ to exhibit \emph{synergy} within $R_0$: \\
	For any $ A \in \mathcal{S} $ and all resources ${\tt r}, {\tt s} \in R_0$, we have
	\begin{equation}
		A ({\tt r}) \geq A({\tt s})  \quad \iff \quad \forall \, B \in \mathcal{S} \setminus \{A\} \; : \; B ({\tt r}) \geq B({\tt s}) \\
	\end{equation}
	and that all monotones in $\mathcal{S}$ are nontrivial functions on $R_0$. 
	\item A sufficient condition for $\mathcal{S}$ to exhibit \emph{trade-off} within $R_0$: \\
	For any pair of monotones $A, B \in \mathcal{S}$, and for any set of (fixed) real numbers $\{x_C\}_{C\in \mathcal{S} \setminus \{A, B\}}$ that are in $\mathtt{JointRealize}(\mathcal{S} \setminus \{A, B\})$, i.e., are jointly realizable values of monotones in $\mathcal{S} \setminus \{A, B\}$,
				we have 
				that $A$ and $B$ exhibit trade-offs within the set of resources given by
				\begin{equation}\label{eq:monotone_fibre}
					\Set*[\big]{\mathtt{r} \in R_0  \given  \forall \, C \in \mathcal{S} \setminus \{A, B\} \; : \; C(\mathtt{r}) = x_C  }.
				\end{equation}
				That is, assuming there are $n$ monotones in $\mathcal{S}$, whenever the values of any $n - 2$ monotones in $\mathcal{S}$ are fixed, the remaining two monotones exhibit trade-off. Note that our definition for two monotones to exhibit trade-off requires $A$ and $B$ to be nontrivial on the set \cref{eq:monotone_fibre}.
\end{itemize}
When each element in $\mathcal{S}$ belongs to a different resource theory, such a synergy or trade-off relation is a dependence relation among the respective resource theories.

	\begin{enumerate}[resume*]
		\item Extract order-theoretic conclusions about the dependence relations between resources.
	\end{enumerate}
	
	If each $\mathcal{M}_i$ is a complete set of monotones, they contain full information about resource orderings for all resource theories under consideration.
	In this case, the constraints from step (2) constitute dependence relations among the complete nonfreeness properties of the resources \emph{at the order-theoretic level}. 
	That is, one can attempt to find aspects of the dependence relations that can be described independently of any particular choice of the complete set of monotones.

	For example, when the set of equality constraints on monotones in $\mathcal{M}$ is such that all monotones in $\mathcal{M}$ can be expressed as functions of the monotones in $\mathcal{M}_i$, we learn that fixing a resource's location within the partial order $\succeq_1$ leaves no freedom for its value within the remaining resource orderings.
	That is, fixing $F_1$-nonfreeness completely specifies all $F_i$-nonfreeness properties for any $F_i\in\mathcal{F}$.

	As another example, consider a pair of generalized monotones, $\mathcal{M}_1$ and $\mathcal{M}_2$, that are complete sets of monotones. If, for any ${\tt r,s}\in R_0$, we have
	\begin{equation}
		\label{eq_gmsy}
		\forall \, \alpha \in \mathcal{I}_1 \; : \; {M}_{1,\alpha}({\tt r})\geq {M}_{1,\alpha}({\tt s}) \quad \iff \quad \forall \, \beta \in \mathcal{I}_2 \; : \;  {M}_{2,\beta}({\tt r})\geq {M}_{2,\beta}({\tt s}),
	\end{equation}
	and that at least one monotone in $\mathcal{M}_1$ and another monotone in $\mathcal{M}_2$ are nontrivial functions on $R_0$, then we say that there is a \emph{complete synergy relation} within $R_0$ between the two resource theories. 
	That is, when restricted to $R_0$, an upward movement (this subsumes the cases of strictly upward movement and no movement, but \emph{not} movements between incomparable elements) within the partial order in one resource theory ensures an upward movement in the partial order of the other. 
	
	We can also define a \emph{complete trade-off relation} between the two resource theories for $R_0$ in a similar way: an upward movement within the partial order in one resource theory ensures a {\em downward} movement (i.e., a strictly downward movement or no movement) in the partial order of the other. 
	Note that here we do not require the two resource theories to be fully dependent on each other. 
	As such, the definition of a synergy relation between resource theories introduced in \cref{sec_prelim}{\,---\,}which applies to fully dependent resource theories{\,---\,}is a special case of the complete trade-off relation defined here.

	However, there are also aspects of dependence relations between resources that we can learn in the absence of complete sets of monotones.
	For example, if the maximal values of $M_{1,\alpha}$ and $M_{2,\alpha}$ are \emph{not} jointly realizable for any ${\tt r}\in R$, then a resource cannot be simultaneously at the top of both the orderings $\succeq_1$ and $\succeq_2$. 
	Remarkably, we can arrive at this conclusion regardless of how little information the monotones $\mathcal{M}_i$ carry about the resource orderings. 
	
	\begin{enumerate}[resume*]
		\item Identify aspects of the dependence relations that have operational significance.
	\end{enumerate}

If one has operational interpretations of each of a set of monotones, then their dependence relations also acquire operational significance. 
	For instance, some monotones quantify the optimal performance (or success probability) in a specific task for which the resources are used.
	Dependences among such monotones reveal synergies and trade-offs between the performance in the associated tasks.

	In the rest of \cref{sec_summary}, we illustrate the recipe from \cref{sec_recipe} by a simple example that can be completely solved. 
	The derivation and more detailed justification of the reported results is, however, postponed to \cref{sec_aboutface,sec_relations}. 
	These later sections also provide an intuitive account of our results by using geometric representations within the Bloch ball.

\subsection{Summary of results in our simple example: resources of about-face asymmetry along different axes}
\label{sec_aboutface0}

Let $\mathbb{Z}_2(\hat{n})$ denote the representation of the discrete group $\mathbb{Z}_2$ given by the identity and a $\pi$ rotation about the axis $\hat{n}$ of the Bloch ball representation of the qubit. 
$\mathbb{Z}_2(\hat{n})$ is referred to as the \enquote{about-face} symmetry group as its only nontrivial element is the 180$^\circ$-rotation, which corresponds to performing an about-face rotation around $\hat{n}$. 
Each axis thus gives a reasource theory of $\mathbb{Z}_2(\hat{n})$-asymmetry, whose free operations are the $\mathbb{Z}_2(\hat{n})$-covariant ones.
The associated nonfreeness properties are then $\mathbb{Z}_2(\hat{n})$-asymmetry properties.

For our concrete example, we consider dependence relations among three about-face asymmetry properties.
Specifically, we consider these with respect to three mutually orthogonal axes, providing the $\mathbb{Z}_2(\hat{x})$-, the $\mathbb{Z}_2(\hat{y})$-, and the $\mathbb{Z}_2(\hat{z})$-asymmetry properties respectively.

We further restrict our attention to the resourcefulness of states of a single qubit. 
It therefore suffices to spell out which completely-positive and trace-preserving (CPTP) maps on a qubit are $\mathbb{Z}_2(\hat{n})$-covariant. 
To this end, let $\vec{\sigma}= (\sigma_x, \sigma_y,\sigma_z)$ denote the vector of Pauli operators. 
The three about-face symmetry groups of interest to us, $\mathbb{Z}_2(\hat{x})$, $\mathbb{Z}_2(\hat{y})$, and $\mathbb{Z}_2(\hat{z})$, can thus be represented by $\{\id, \sigma_z \}$, $\{\id, \sigma_x\}$ and $\{\id, \sigma_y\}$ respectively. 
In fact, for any axis $\hat{n}$, the corresponding about-face symmetry group acts on qubits as $\{\id, \sigma_n \}$, where $\sigma_n\coloneqq \vec{\sigma}\cdot\hat{n}$. 
It then follows that the set of $\mathbb{Z}_2(\hat{n})$-covariant operations consists of those CPTP maps $\mathcal{T}$ that satisfy
\begin{equation}
	\sigma_n\mathcal{T}(\rho)\,\sigma_n =\mathcal{T}(\sigma_n \,\rho\, \sigma_n), \, \forall \rho.
\end{equation}

In the following, we summarize the results concerning resource dependence relations in this example. 
The proofs and more details can be found in \cref{sec_aboutface,sec_relations} and the appendices. 

\subsubsection{A complete set of monotones and their relations for a given axis}
\label{sec_constraint1RT}

Consider step (1a) of the recipe, specialized to this example.  It calls for the characterization of the nonfreeness properties of qubit states relative to each of the three resource theories under consideration.

The three resource theories associated to about-face symmetries for the axes $\hat{x}$, $\hat{y}$, and $\hat{z}$ are related to one another by a symmetry transformation, and so it suffices to characterize one of these.  We choose the one associated to the $x$-axis. 

In \cref{app:qubitConversion}, we find a complete set of monotones for the resource theory of $\mathbb{Z}_2(\hat{x})$-asymmetry with two elements, which we refer to as the $A$-monotone and the $B$-monotone. 
The $\mathbb{Z}_2(\hat{x})$-asymmetry properties characterized by these are termed the $A_x$-asymmetry property and the $B_x$-asymmetry property respectively. The values of these two monotones for a state $\rho$ represented by the Bloch vector $\vec{r}\coloneqq\{r_x,r_y,r_z\}$ are:
\begin{equation}
	\label{eq_ABdef}
		\begin{split}
	   		A_x(\rho) &\coloneqq \sqrt{r_y^2+r_z^2}; \\
	   		B_x(\rho) &\coloneqq 
	        		\begin{cases}
					\sqrt{\frac{r_y^2+r_z^2}{1-r_x^2}}  &\text{ if } r_x^2<1, \\
					0 &\text{ if } r_x^2=1.
				\end{cases}
		\end{split}
\end{equation}

We now turn to step (1b) of the recipe, finding the dependence relations that hold among the monotones in the complete set. 
To do so, we need to determine the scope of possible pairs of real values of $\bigl(A_x(\rho), B_x(\rho) \bigr)$ for an arbitrary qubit state $\rho$. 
This set of jointly realizable pairs is specified by the following two constraints:
\begin{equation}
\label{eq_BAranges}
	A_x(\rho),B_x(\rho) \in [0,1], 
\end{equation}
and 
\begin{equation}
\label{eq_BgeA0}
	B_x(\rho) \ge A_x(\rho). 
\end{equation}
The first of these simply specifies the bounds on the possible values of $A_x$ and $B_x$ individually, while the second describes a dependence relation between them.

\begin{figure}[h!]
	\centering
	\includegraphics[width=0.3\textwidth]{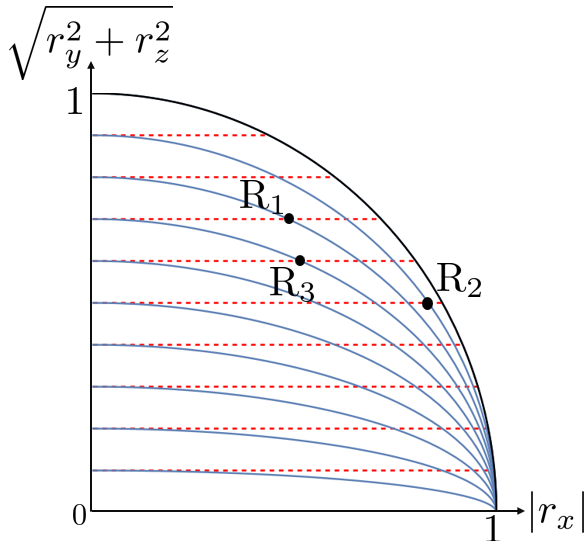}
	\hspace{1cm}
	\includegraphics[width=0.3\textwidth]{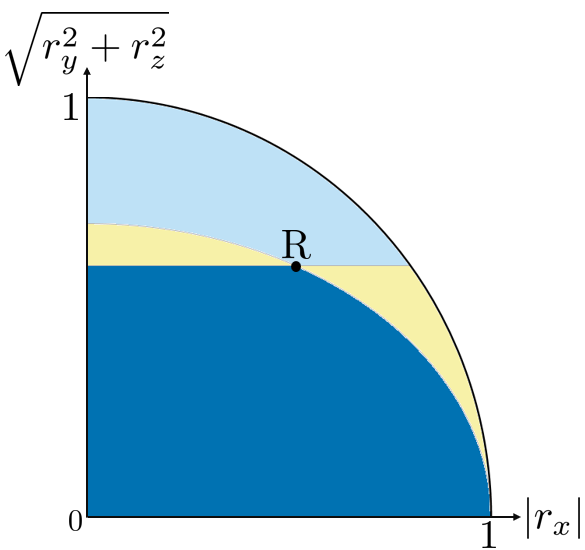}
	\caption{Left: The characterization of three equivalence classes of resources ${\rm R}_1$, ${\rm R}_2$ and ${\rm R}_3$ in terms of $A_x$ and $B_x$, the complete set of monotones for $\mathbb{Z}_2(\hat{x})$-asymmetry properties.  Level sets of $A_x$ are denoted by the red dashed lines.  Level sets of $B_x$ are denoted by the blue curves. Equivalence classes of states are parametrized by $\sqrt{\smash[b]{r_y^2+r_z^2}}$ and $|r_x|$ (since $A_x$ and $B_x$ can be expressed in terms of these) and so we choose these for our axes.  The value of $A_x$ characterizing a given level set is given by the corresponding curve's height on the $\sqrt{\smash[b]{r_y^2+r_z^2}}$ axis.  Similarly for $B_x$. The resources ${\rm R}_1$ and ${\rm R}_2$ are incomparable, as are ${\rm R}_3$ and ${\rm R}_2$, while ${\rm R}_1$ and ${\rm R}_3$ are strictly ordered. Right: For a given equivalence class of resources ${\rm R}$, the dark-blue region is the set of equivalence classes that are below ${\rm R}$ in the partial order, while the light-blue region is the set of equivalence classes that are above ${\rm R}$ in the partial order, and the yellow region is the set of equivalence classes that are incomparable to ${\rm R}$ in the partial order.
}
	\label{fig_coordinate}
\end{figure}

Since there are no equality constraints, $A_x$ and $B_x$ are irredundant. Because there are two monotones in this complete and irredundant set, the partial order of the set of equivalence classes of resources is not a total order. \cref{fig_coordinate} shows how the states in the partial order of $\mathbb{Z}_2(\hat{x})$-asymmetry are parametrized by $A_x$ and $B_x$, and examples of pairs of resources that are incomparable and pairs that are strictly ordered.

In the left plot of \cref{fig_coordinate}, we draw some of the level sets of $A_x$ (the red lines) and of $B_x$  (the blue lines). 
Inequality \eqref{eq_BgeA0} expresses that a given level set of $A_x$ does not intersect all the level sets of $B_x$ and vice-versa. 
Furthermore, since every pair of values of $A_x$ and $B_x$ satisfying \cref{eq_BAranges,eq_BgeA0} is realizable by some resource, the partial order induced by $\mathbb{Z}_2(\hat{x})$ has infinite width and is not weak.

At this point, let us comment on how our conclusions are related to the choice of a complete set of monotones.
If one used a monotone given by $A'_x \coloneqq f \circ A_x$ instead of $A_x$, for an invertible monotone function $f$, then the specific set jointly realizable values would change.
However, the types of dependence relations would not change.
The scope of jointly realizable values would be isomorphic (as partially ordered sets).

\subsubsection{For three orthogonal axes: dependence relations among monotones}
\label{sec_eqineq0}

We now consider step 2\,---\,finding the constraints that hold among the monotones in the union of the complete sets for all resource theories under consideration, namely the resource theories of $\mathbb{Z}_2(\hat{x})$-, $\mathbb{Z}_2(\hat{y})$-, and $\mathbb{Z}_2(\hat{z})$-asymmetry. 
The union of the complete sets, which we denote by $\mathcal{M}$ as in \cref{sec_recipe}, consists of $A_x$, $B_x$, $A_y$, $B_y$, $A_z$, and $B_z$.

\paragraph{Equality constraints}
\label{sec_eq0}

As will be shown in \cref{sec_equalities}, there are three equality constraints among the six monotones in $\mathcal{M}$:
\begin{equation}
	\label{eq_equality0}
	\begin{split}
		2 \left[ B_x^2(\rho) - A^2_x(\rho) \right] - B_x^2(\rho) \left[-A_x^2(\rho) +  A_y^2(\rho) + A_z^2(\rho)\right] &= 0, \\
		2 \left[ B_y^2(\rho) - A^2_y(\rho) \right] - B_y^2(\rho) \left[ \phantom{-} A_x^2(\rho) -  A_y^2(\rho) + A_z^2(\rho)\right]&= 0, \\
		2 \left[ B_z^2(\rho) - A^2_z(\rho) \right] - B_z^2(\rho)\left[ \phantom{-} A_x^2(\rho) +  A_y^2(\rho) - A_z^2(\rho)\right] &= 0.
	\end{split}
\end{equation}
The fact that there are no fewer than three such constraints can be understood through parameter counting.
The state space of a qubit has three degrees of freedom: 
For instance, it can be parametrized by the three Bloch vector components $r_x$, $r_y$, and $r_z$. 
Since the $6$ functions in $\mathcal{M}$ are smooth, their jointly realizable values form an (at most) $3$-dimensional manifold in the $6$-dimensional vector space $\mathbb{R}^{\mathcal{I}}$. 
The remaining (at least) $3$ degrees of freedom are removed by the equality constraints.

\paragraph{Inequality constraints} 
\label{sec_ineq}

From the equality constraints and the fact that $B$-monotones are nonnegative, we know that fixing the values of all the $A$-monotones uniquely specifies the values of all the $B$-monotones.
Thus, it is sufficient to use the set of all inequality constraints among $A_x$, $A_y$, and $A_z$ as a generating set. 
As shown in \cref{sec_ineq_A}, these inequalities are:
\begin{equation}\label{eq_inequalityA0}
	\begin{split}
		- A_x^2(\rho) +  A_y^2(\rho) +  A_z^2(\rho) \ge 0, \\
		A_x^2(\rho) -  A_y^2(\rho) +  A_z^2(\rho) \ge0, \\
		A_x^2(\rho) +  A_y^2(\rho) -  A_z^2(\rho) \ge 0, \\
		A_x^2(\rho) + A_y^2(\rho) + A_z^2(\rho)  \le 2.
	\end{split}
\end{equation}
It follows that the full set of dependence relations among the six monotones 
	in $\mathcal{M}$ is given by the equalities in \cref{eq_equality0} and the inequalities in \cref{eq_inequalityA0}.

In particular, from the inequalities in \eqref{eq_inequalityA0}, we can use the equalities in \eqref{eq_equality0} to obtain the inequalities for any other triple of monotones. 
For example, in \cref{sec_ineq_B}, the inequalities holding among $\{B_x,B_y,B_z\}$ are shown to be those of \eqref{eq_inequalityB} and in \cref{sec_AxBxAy}, the inequalities holding among $\{A_x,B_x,A_y\}$ are shown to be those of \eqref{eq_ineqAxBxAy}.

	Only those inequalities that {\em cannot} be inferred from the constraints obtained in step 1, i.e., from \eqref{eq_BAranges} and \eqref{eq_BgeA0}, describe nontrivial dependence relations between about-face asymmetry properties relative to {\em different axes}.  
	For instance, consider the one inequality that follows from individual constraints is $A_x^2(\rho) + A_y^2(\rho) + A_z^2(\rho) \leq 3$.
	It is a direct consequence of the fact that $A_x$, $A_y$ and $A_z$ all have a maximum value of 1. 
	By contrast, each of the inequalities in \eqref{eq_inequalityA0} is not derivable from the constraints derived in step 1 and consequently describes a nontrivial dependence relation among $\{A_x, A_y, A_z\}$.

\subsubsection{Conceptual understandings of these dependence relations}

	In step 3, we establish conceptual understandings of the dependence relations given by the equality and inequality constraints.
	
	\paragraph{Characterize the relations among monotones}
	\label{sec_interpret3way}
	Consider step (3a), which is to characterize the dependence relations among monotones.
	
	$\mathtt{JointRealize}(\mathcal{M})$, namely, the jointly realizable values of $A_x$, $B_x$, $A_y$, $B_y$, $A_z$ and $B_z$, are those that satisfy the equalities in \eqref{eq_equality0} and the inequalities in \eqref{eq_inequalityA0}, together with the constraint that they are all nonnegative (which follows from the definitions of the $A$- and $B$-monotones).  
	As these are polynomial constraints, $\mathtt{JointRealize}(\mathcal{M})$ describes a semi-algebraic set in six dimensions.  
	In order to try to visualize some of its aspects, it is useful to consider various 3-dimensional projections of it. 
	
\begin{figure}[h]
	    \centering
	    \includegraphics[width=0.95\textwidth]{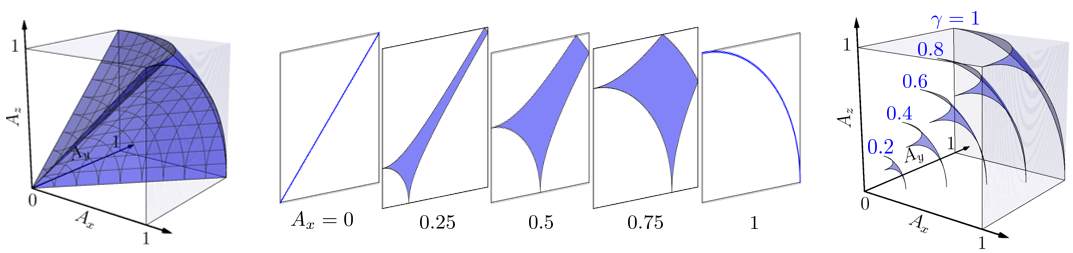}
	    \caption{Left: The region of jointly realizable values of $A_x$, $A_y$ and $A_z$; the mesh is added to indicate the curvature. 
	    	Middle: $A_y$--$A_z$ cross-sections of this region for five different values of $A_x$. 
	    	Right: The regions of jointly realizable values of $A_x$, $A_y$ and $A_z$ given a fixed purity, represented by the Bloch vector radius $r$. }
	    \label{fig_qt3dA}
	\end{figure}

	For example, the projection of this semi-algebraic set onto the triple of axes corresponding to  $A_x$, $A_y$, and $A_z$ yields the region depicted in  \cref{fig_qt3dA}, which we denote as $\mathtt{JointRealize}(A_x,A_y,A_z)$.

	We now seek to interpret the dependence relations among $A_x$, $A_y$, and $A_z$ by leveraging the visualization in \cref{fig_qt3dA}.

	Now consider projecting $\mathtt{JointRealize}(A_x,A_y,A_z)$  %the region of jointly realizable values of $A_x$, $A_y$ and $A_z$ 
	(the blue region depicted in  the left plot of \cref{fig_qt3dA}) onto the $A_x$--$A_y$ plane, the $A_y$--$A_z$ plane and the $A_z$--$A_x$ plane. 
	This helps in understanding the dependence relations that might hold among pairs of the $A$-monotones. 
	It is not difficult to see from \cref{fig_qt3dA} that the projection into the $A_x$--$A_y$ plane is the square region $[0,1]\times [0,1]$, implying no nontrivial dependence relation between $A_x$ and $A_y$.  
	By symmetry, the same conclusion holds for $A_y$ and $A_z$ and for $A_z$ and $A_x$. 

	We now consider different {\em cross-sections} of the $\mathtt{JointRealize}(A_x,A_y,A_z)$. These are depicted in the middle of \cref{fig_qt3dA}. 
	The shapes of these reveal the dependence relations between two of the three $A$-monotones, $A_y$ and $A_z$, as one varies the value of the third, $A_x$.  As the value of $A_x$ increases, we observe a transition from a synergy relation between $A_y$ and $A_z$ to a trade-off relation.
	Specifically, whenever $A_x=0$ holds, there is a synergy relation between $A_y$ and $A_z$ (as defined in \eqref{eq_mmsy}).
	On the other hand, whenever we have $A_x=1$, there is a trade-off relation between $A_y$ and $A_z$ (as defined in \eqref{eq_mmtr}).
	For intermediate values, we observe a dependence relation that interpolates between a synergy and a trade-off relation. 
	Similar conclusions hold for every other pair of $A$-monotones conditioned on a value of the third.

	If we consider fixing the {\em purity} of the state instead, then there is a trade-off among the three $A$-monotones. 
	This is depicted in the rightmost plot of \cref{fig_qt3dA} for various values of the purity.\footnote{The purity of a qubit is often expressed in terms of its Bloch vector radius $r$ as $1/2(1+r)$. In this paper, we use $r$ to directly represent the purity for simplicity. } 
	In particular, we can observe that additionally fixing the value of any one of the three $A$-monotones necessarily leads to a simple trade-off among the remaining two, as defined in \eqref{eq_mmtr}.

	Thus, the complete dependence relations among $A_x$, $A_y$, and $A_z$ are not simply characterized as either synergy or trade-off relations.
	Rather, they are nuanced and contingent on the specific additional constraints imposed. 

\begin{figure}[h!]
	\centering
	\includegraphics[width=\textwidth]{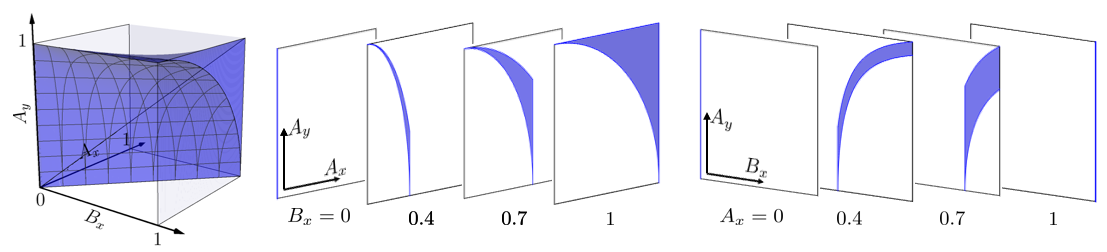}
	\caption{Left: The region of jointly realizable values of $A_x$, $B_x$ and $A_y$; the mesh is added to indicate the curvature. 
	Middle: $A_x$--$A_y$ cross-sections of this region for four different values of $B_x$. 
	Right: $B_x$--$A_y$ cross-sections of this region for four different values of $A_x$.  }
	\label{fig_AxBxAy}
\end{figure}

	One can also consider other 3-dimensional projections of the semi-algebraic set $\mathtt{JointRealize}(\mathcal{M})$.
	For instance, the projection onto the $B_x$--$B_y$--$B_z$ subspace is visualized and analyzed in \cref{sec_ineq_B}. It is also possible to consider triple of axes corresponding to monotones of different types, namely, a mixture of $A$-monotones and $B$-monotones. 
	For example, the projection onto the $A_x$--$B_x$--$A_y$ subspace is depicted in the leftmost plot of  \cref{fig_AxBxAy}. 

	As before, consider further projecting this region, namely, $\mathtt{JointRealize}(A_x,B_x,A_y)$	onto the three planes respectively. 
	The 2-dimensional projections onto the $A_x$--$A_y$ and $B_x$--$A_y$ planes are both the full $[0,1]\times [0,1]$ squares.
	There is thus no nontrivial (unconditional) dependence relation among these pairs of monotones. 
	However, the projection onto the $A_x$--$B_x$ plane is a triangle, whose vertices are the $(0,0)$, $(1,1)$, and $(0,1)$, as a direct consequence of Inequality \eqref{eq_BgeA0}. 
	This is a feature of our chosen $\mathbb{Z}_2(\hat{x}$)-asymmetry monotones rather than a resource dependence relation. 

	The middle plot of \cref{fig_AxBxAy} illustrates that when $B_x\in(0,1)$ and is fixed, then as $A_x$ increases, both the upper and lower bounds of possible values of $A_y$ decrease. 
	Consequently, if the increase in $A_x$ is significant enough to make the upper bound of $A_y$ after the increase smaller than its lower bound before, then $A_y$ must decrease. 
	Therefore, there is a kind of trade-off relation between $A_x$ and $A_y$.

	Conversely, the rightmost plot shows that when $A_x \in(0,1)$ is fixed, as $B_x$ increases, both the upper and lower bounds of possible values of $A_y$ increase. 
	Therefore, if the increase in $B_x$ is significant enough to make the lower bound of $A_y$ after the increase larger than the upper bound of $A_y$ before the increase, then the increase in $B_x$ by this amount necessarily leads to an increase in $A_y$. 
	In other words, there is a kind of synergy (rather than trade-off) relation between $B_x$ and $A_y$.

\paragraph{Order-theoretic characterization}
\label{sec_order}

	Now consider step (3b), wherein one seeks an order-theoretic characterization of the dependence relations.
	
	Recall that there are three resource theories under consideration, namely those of $\mathbb{Z}_2(\hat{x})$-, $\mathbb{Z}_2(\hat{y})$-, and $\mathbb{Z}_2(\hat{z})$-asymmetry. For a given axis $\hat{n}$, the corresponding $B$-monotone $B_n$ can be written in terms of the respective $A$-monotone $A_n$ and the Bloch vector radius $r$ as 
	\begin{equation}
		B_n(\rho) = \frac{A_n}{\sqrt{A_n^2 + 1 - r^2}}
	\end{equation}
	whenever the respective $r_n$ satisfies $r_n^2 \neq 1$.
	For a given $r$ this is a monotonic function.
	Therefore, states of a fixed purity form a \emph{totally} preordered set in terms of its $\mathbb{Z}_2(\hat{n})$-asymmetry properties.
	Moreover, this preorder can be characterized by the $A_n$-monotone alone.
	%\footnotetext{A preorder is total if there are not incomparable elements.
	%That is, for all $\mathtt{r}$ and $\mathtt{s}$ we have either $\mathtt{r} \succeq \mathtt{s}$ or $\mathtt{s} \succeq \mathtt{r}$ (or both).}
	Importantly, since fixing the values of $A_n$ and $B_n$ fixes the purity, it follows that states with a given value of $A_n$ and a given value of $B_n$---i.e., those in the same equivalence class within the partial order for $\mathbb{Z}_2(\hat{n})$-asymmetry---must also form a \emph{totally} preordered set in terms of their $\mathbb{Z}_2(\hat{n}')$-asymmetry properties for any axis $\hat{n}'$ (and such a total preorder can be completely characterized by $A_{n'}$ alone).  
	
	Furthermore, the equality constraints in \eqref{eq_equality0} tell us that when the state's locations (i.e., the equivalence classes) in the partial orders for two of the three resource theories are fixed, its location in the partial order of the third resource theory is also fixed. 
	This is because fixing a state's location in two of these three partial orders corresponds to fixing the values of two pairs of the $A$- and $B$-monotones, while the three equality constraints then allow one to solve for the remaining pair.

	The equality constraints in \cref{eq_equality0} also enable us to extract order-theoretic characterization of dependence relations from any of the 3-way dependence relations analyzed in \cref{sec_interpret3way}.
	For example, consider \cref{fig_AxBxAy}. 
	There, we can use the values of $A_x$ and $B_x$ to determine a state's location in the partial order for $\mathbb{Z}_2(\hat{x})$-asymmetry. 
	Given the values of $A_x$ and $B_x$, the states form a total {preorder} for $\mathbb{Z}_2(\hat{y})$-asymmetry and $A_y$ alone is sufficient to determine its location.
	Thus, the middle and the rightmost plots of \cref{fig_AxBxAy} show that the type of dependence relations observed is determined by {\em how} one moves upward in the $\mathbb{Z}_2(\hat{x})$-asymmetry ordering. 
	Specifically, increasing $A_x$ while keeping $B_x$ constant exhibits a kind of trade-off relation between $A_y$ and $A_x$. 
	On the other hand, if $A_x$ is kept constant and $B_x$ is varied, we obtain a kind of synergy between $A_y$ and $B_x$.
	
	As another example of extracting order-theoretic dependence relations from the 3-way relations analyzed in \cref{sec_interpret3way}, consider \cref{fig_qt3dA}. Since we have $1 \geq B_n \geq A_n$, as indicated by inequalities \eqref{eq_BAranges} and \eqref{eq_BgeA0}, a state $\rho$ is at the top of the order induced by $\mathbb{Z}_2(\hat{n})$ if and only if $A_n=1$. 
	Thus, the figure depicts that a state can simultaneously be at the top of any two of the three partial orders induced by $\mathbb{Z}_2(\hat{x})$, $\mathbb{Z}_2(\hat{y})$, and $\mathbb{Z}_2(\hat{z})$. However, the inability of $A_x$, $A_y$, and $A_z$ to simultaneously reach their maximum values indicates that a state cannot be at the top of all three partial orders simultaneously. 
	Let us further focus on the rightmost plot of  \cref{fig_qt3dA}.
	Since the states of fixed purity form a \emph{totally} preordered set in the resource theory of $\mathbb{Z}_2(\hat{n})$-asymmetry 
		and can be completely characterized by $A_n$,
		the trade-off relation among $\{A_x, A_y, A_z\}$ observed in this plot can also be interpreted as a trade-off relation among $\mathbb{Z}_2(\hat{x})$-, $\mathbb{Z}_2(\hat{x})$- and $\mathbb{Z}_2(\hat{z})$-asymmetry properties for a given purity.

\paragraph{Operational significance of the dependence relations}
\label{sec_opsg}

Now consider step (3c), seeking dependence relations with operational significance.

In our example of the resource theory of about-face asymmetry for a given axis $\hat{n}$, we show in \cref{sec_Amonotone} that the $A$-monotone has a simple operational interpretation.  It quantifies the optimal probability of success for a $\mathbb{Z}_2(\hat{n})$-phase estimation task, that is, the task of distinguishing between a 0-degree rotation and a $\pi$-degree rotation about axis $\hat{n}$. Consequently, the dependence relations that hold between $A_x$, $A_y$, and $A_z$ imply a dependence relation among the degree of success that can be achieved in the phase estimation tasks for $\mathbb{Z}_2(\hat{x})$, $\mathbb{Z}_2(\hat{y})$, and $\mathbb{Z}_2(\hat{z})$.  For example, the inequality constraint $A_x^2(\rho) + A_y^2(\rho) + A_z^2(\rho)  \le 2$ in \eqref{eq_inequalityA0} indicates a form of trade-off among the success rates for these three tasks. That is, if one must prepare a state $\rho$ without prior knowledge of which of these three tasks one will face, the average probability of success will be lower compared to a scenario where the state $\rho$ can be tailored to the task.

In \cref{sec_bound}, we also describe an operational interpretation of the dependence relation from inequality \eqref{eq_BgeA0} (which asserts that the value of the $A$-monotone of a state is a lower bound on the value of the $B$-monotone for this state).  Specifically, it describes a gap between the cost and the yield of a state relative to a particular ``gold standard'' chain\footnote{A chain is a subset of a partial order and must be totally ordered.} of resources.

\section{The partial order for the resource theory of an about-face symmetry} 
\label{sec_aboutface}

	\subsection{A complete characterization of the partial order}\label{sec_partialorder}
	
	Recall from \cref{sec_aboutface0} that the resource theory of $\mathbb{Z}_2(\hat{n})$-asymmetry is defined by taking the free operations to be the  $\mathbb{Z}_2(\hat{n})$-covariant operations. As noted in \cref{sec_constraint1RT}, we take the case of the $\hat{x}$ axis as our illustrative example.  
	In \cref{app:qubitConversion}, we provide a parametrization of all $\mathbb{Z}_2(\hat{x})$-covariant operations on a qubit, based on the results of \cite{beth_ruskai_analysis_2002}. 
	We use it to describe the set of all states that can be converted from a given $\rho$ by $\mathbb{Z}_2(\hat{x})$-covariant operations. This in turn gives us necessary and sufficient conditions for the existence of a free conversion among qubit states in this resource theory. Specifically, in \cref{app:qubitConversion} we prove the following result.
	\begin{theorem}[Resource order for $\mathbb{Z}_2(\hat{x})$-asymmetry]\label{thm:complete_characterization}
				A qubit state $\rho$ can be converted to another qubit state $\sigma$ under $\mathbb{Z}_2(\hat{n})$-covariant operations if and only if 
				\begin{align}
				\label{eq_complete_characterization}
					A_x(\rho) \ge A_x(\sigma)  \;\; \text{ and } \;\;  B_x(\rho ) \ge B_x(\sigma).
				\end{align}
			\end{theorem}
	\cref{thm:complete_characterization} indicates that both $A_x$ and $B_x$ are resource monotones in the resource theory of $\mathbb{Z}_2(\hat{x})$-asymmetry, and that together they form a \emph{complete set} of monotones in this resource theory. $A_x$ and $B_x$ completely characterize the preorder in the resource theory of $\mathbb{Z}_2(\hat{x})$-asymmetry, and consequently also the partial order. 
	
	We repeat the definitions of $A_x$ and $B_x$ from \cref{eq_ABdef}:
	\begin{equation}
	\label{eq_ABx1}
		\begin{split}
	   		A_x(\rho) &\coloneqq \sqrt{r_y^2+r_z^2}, \\
	   		B_x(\rho) &\coloneqq 
	        		\begin{cases}
					\sqrt{\frac{r_y^2+r_z^2}{1-r_x^2}}  &\text{ if } r_x^2<1, \\
					0 &\text{ if } r_x^2=1.
				\end{cases}
		\end{split}
	\end{equation}

	Geometrically, $A_x(\rho)$ is the radius of the cylinder with $\hat{x}$ as its central axis and whose surface contains the Bloch vector representing $\rho$, denoted as $\vec{r}\coloneqq\{r_x,r_y,r_z\}$. 
	All states on the surface of such a cylinder have the same value of $A_x$. 
	The value $A_x(\rho)$ is also the length of the projection of $\vec{r}$ onto the $\hat{y}--\hat{z}$ plane, i.e., the plane orthogonal to $\hat{x}$. 
	
	$B_x(\rho)$, on the other hand, is the length of the minor axis of a prolate spheroid\footnote{A prolate spheroid is the surface of revolution obtained by rotating an ellipse about its major axis~\cite[p.10]{hilbertGeometry2020}. It is also an ellipsoid whose two minor radii are equal.} whose major axis is $\hat{x}$ and whose surface contains $\vec{r}$. 
	All states on the surface of the corresponding prolate spheroid except for $\pm\hat{x}$ have the same value of $B_x$. 
	When $B_x(\rho)=0$, the prolate spheroid becomes a line whose endpoints are $\pm\hat{x}$, these are precisely the about-face symmetric states as one would expect.

	As we noted previously, \cref{eq_ABx1} implies 
	\begin{equation}\label{eq_BgeA}
		1 \ge B_x(\rho) \ge A_x(\rho)\ge 0, 
	\end{equation}
	as well as the following implications between assignments of extremal values of $A_x$ and $B_x$:
	\begin{align}
		\label{eq_AxBx0} &A_x(\rho)=0 \iff B_x(\rho)=0, \\
		\label{eq_AxBx1} &A_x(\rho)=1 \overset{\centernot\impliedby}{\implies} B_x(\rho)=1.
	\end{align}

	\cref{thm:complete_characterization} implies the following corollary.
	\begin{corollary}[Equivalence classes of states in the resource theory of $\mathbb{Z}_2(\hat{x})$-asymmetry]	
	\label{thm:equivalence_characterization}
		Two qubit states, $\rho$ and $\sigma$, are equivalent in the resource theory of $\mathbb{Z}_2(\hat{x})$-asymmetry  if and only if
		\begin{align}\label{eq_equivalence_characterization}
		    A_x(\rho) = A_x(\sigma)  \;\;\text{ and } \;\; B_x(\rho) = B_x(\sigma).
		\end{align}
	\end{corollary}

\begin{figure}[h]
		\centering
		\includegraphics[width=0.3\textwidth]{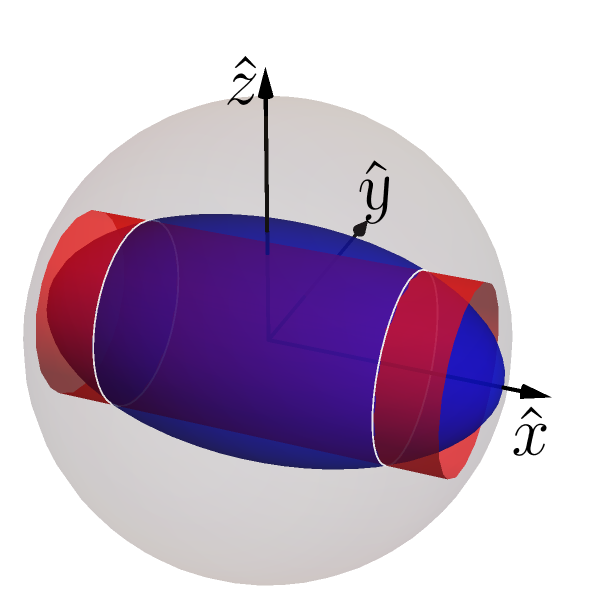}
		\caption{The intersection of all states with $A_x=0.4$ (represented by the cylinder) and all states with $B_x=0.5$ (represented by the prolate spheroid) defines an equivalence class in the resource theory of $\mathbb{Z}_2(\hat{x})$-asymmetry. }
		\label{fig_AxBx}
	\end{figure}

	Geometrically, an equivalence class of states corresponds to the intersection of the set of states with a given value of $A_x$ and the set of states with a given value of $B_x$.  \cref{fig_AxBx} shows an example of such an intersection, where $A_x=0.4$ and $B_x= 0.5$. 
	
\begin{figure}[h]
		\centering
		\includegraphics[width=0.25\textwidth]{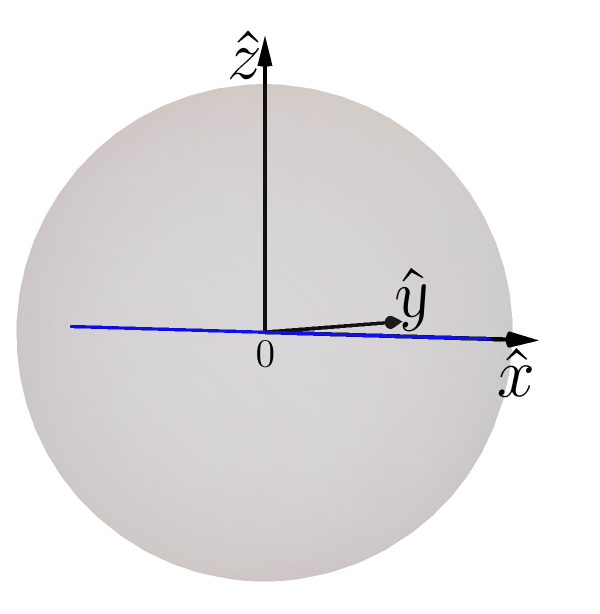}
		\hspace{2mm}
		\includegraphics[width=0.25\textwidth]{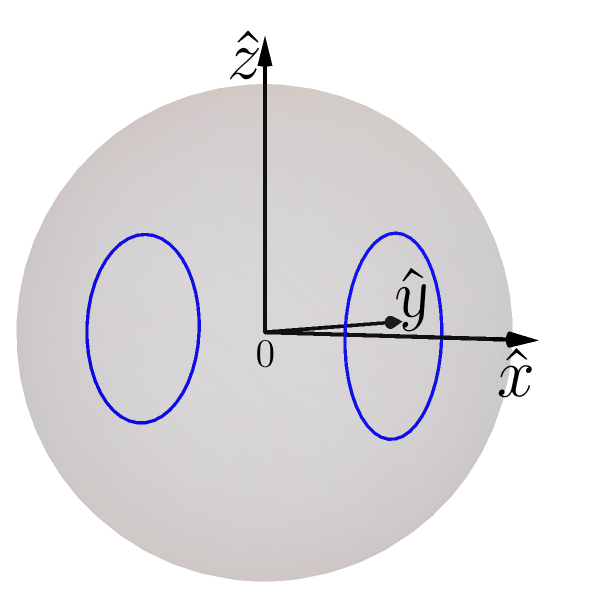}
		\hspace{2mm}
		\includegraphics[width=0.25\textwidth]{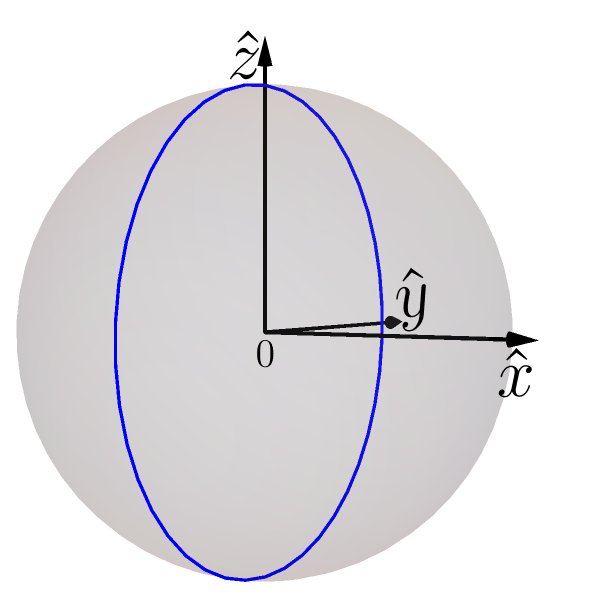}
		\caption{Left:  the equivalence class of states with $A_x=B_x= 0$, which is also the set of free states in the resource theory of $\mathbb{Z}_2(\hat{x})$-asymmetry. Mid: the equivalence class of states with $A_x=0.4$, $B_x=0.5$. Right: the equivalence class of states with $A_x=B_x = 1$, which is also the unique maximal element in the partial order in the resource theory of $\mathbb{Z}_2(\hat{x})$-asymmetry. }
		\label{fig_eqclX}
	\end{figure}

	Given the expressions for $A_x$ and $B_x$ in terms of the state's Bloch vector (\cref{eq_ABx1}), we can derive that the Bloch vectors associated with the states in the equivalence class of \cref{fig_AxBx} are those satisfying $r_y^2+r_z^2=0.4$ and $r_x^2= 0.6$, which describes a pair of circles in the Bloch ball. We depict this equivalence class of states on its own in the middle plot of \cref{fig_eqclX}.     
		
	The leftmost plot of \cref{fig_eqclX} depicts the equivalence class of states at the bottom of the partial order of $\mathbb{Z}_2(\hat{x})$-asymmetry, which is the one that achieves the minimum values for the complete set of monotones, i.e., the one with $A_x=B_x=0$.\footnote{The states at the bottom of a resource order can also be characterized as those that are preparable using the free operations, which in this case means the states that are invariant under the action of the $\mathbb{Z}_2(\hat{x})$-symmetry transformations.}  
	These are the states whose Bloch vectors satisfy $r_y^2+r_z^2=0$, and thus, they form the $\hat{x}$ axis of the Bloch ball. 
	
	The rightmost plot of \cref{fig_eqclX} depicts the equivalence class at the top of the $\mathbb{Z}_2(\hat{x})$-asymmetry order, which in this case simultaneously achieves the maximum $A_x$ and $B_x$ values, i.e., the one with $A_x=B_x=1$.  
	These are the states whose Bloch vectors satisfy $r_y^2+r_z^2=1$ (and $r_x = 0$) and thus they form the circle of pure states in the $\hat{y}--\hat{z}$ equatorial plane of the Bloch ball. 

	We have seen that the equivalence classes of states in the resource theory of $\mathbb{Z}_2(\hat{x})$-asymmetry can be parameterized by the values of $r_y^2+r_z^2$ and $r_x^2$, or equivalently, in terms of their square roots, namely, $\sqrt{\smash[b]{r_y^2+r_z^2}}$ and $|r_x|$.  
	In \cref{fig_coordinate}, we use this parametrization to depict the partial order over equivalence classes in the resource theory of $\mathbb{Z}_2(\hat{x})$-asymmetry.

	\subsection{Properties of the partial order}
	\label{sec_properties}

		\cref{fig_coordinate} makes it clear that we do not have a total order, i.e., there exist incomparable elements in the resource theory of $\mathbb{Z}_2(\hat{x})$-asymmetry. An example is the states in the equivalence class of ${\rm R}_1$ and ${\rm R}_2$ in the figure. The states in ${\rm R}_1$ cannot be converted to the ones in ${\rm R}_2$ because ${\rm R}_2$ has a larger value of $B_x$. On the other hand, the states in ${\rm R}_2$ cannot be converted to the ones in ${\rm R}_1$ because ${\rm R}_1$ has a larger value of $A_x$. 
		
		The partial order in the resource theory of $\mathbb{Z}_2(\hat{x})$-asymmetry exhibits many common traits of a partial order. For example, the partial order is \emph{locally infinite} and is not \emph{weak}; both the \emph{height} and the \emph{width} of the partial order are infinite. We will not spell out these properties here. We refer the interested readers to~\cite[Sec. 4.1]{Wolfe_2020} for details.

		A noteworthy property is that the partial order over the equivalence classes of resources has a \emph{unique} maximal element, which is the equivalence class consisting of pure states with $r_x = 0$ shown in the rightmost graph in \cref{fig_eqclX}. 
		This follows from Implication \eqref{eq_AxBx1}, which asserts that whenever $A_x$ achieves its maximal value of $1$, so does $B_x$.
		Thus, states in the equivalence class constituting the unique top of the order are optimal for \emph{both} sorts of operational tasks described in \cref{sec_connection}. 
		
		Having a unique top-of-the-order equivalence class of states is a feature of many quantum resource theories, such as the resource theory of bipartite entangled states under LOCC and the resource theory of athermality, but there are some, such as the resource theory of bipartite entangled states under LOSR  or of quantum common-cause boxes under LOSR, where there are many incomparable equivalence classes that are all maximal. 

		\cref{fig_coordinate} also illustrates the implications described in \eqref{eq_AxBx0} and \eqref{eq_AxBx1}. 
		Specifically, the shape of the level sets for $A_x$ and $B_x$ makes evident that if a state has the minimal value of $A_x$ or the minimal value of $B_x$, then it has the minimal value for both, and is therefore bottom-of-the-order. 
		Similarly, they make evident that if a state has the maximal value of $A_x$, then, as mentioned above, it has the maximal value of $B_x$ and consequently is top-of-the-order, while if it has the maximal value of $B_x$, it need not have the maximal value of $A_x$ and so is not necessarily top-of-the-order.
	
		From \cref{thm:complete_characterization}, we also learn that by fixing the value of either $A_x$ or $B_x$, we obtain a totally ordered set of qubit states in the resource theory of $\mathbb{Z}_2(\hat{x})$-asymmetry, whose ordering is characterized by the other monotone. 
		This fact is also evident from an examination of \cref{fig_coordinate}.

		In addition, once we fix the purity of the qubit states, i.e., consider states where $r \coloneqq \sqrt{\smash[b]{r_y^2+r_x^2+r_z^2}}$ takes a particular value, we also obtain a set of states that form a {\em total order} in the resource theory of $\mathbb{Z}_2(\hat{x})$-asymmetry. Specifically, when the Bloch vector radius $r$ is fixed to 1, i.e., if only considering pure states, the expressions of $A_x$ and $B_x$ are reduced to
		\begin{equation}\label{eq_ABxpure}
			\begin{split}
				A_x(\rho) &= \sqrt{1-r_x^2}, \\
				B_x(\rho) &= 
				\begin{cases}
				1, &\text{ if } r_x^2<1\\
				0, &\text{ if } r_x^2=1
				\end{cases}.
			\end{split}
		\end{equation}
		Since $B_x$ assigns the value 1 for all nonfree pure states and the value 0 for all free pure states while $A_x$ assigns a different value for each state with different $r_x^2$, $A_x$ alone 
		completely characterizes the total order for pure qubit states while $B_x$ alone does not. 
		
		On the other hand, when $r<1$, i.e., if only considering impure states of the same purity, $r_x^2$ must also be smaller than 1, and consequently, the two monotones are simplified to
		\begin{equation}
			\begin{split}
				A_x(\rho) &= \sqrt{r^2-r_x^2}, \\
				B_x(\rho) &= \sqrt{\frac{r^2-r_x^2}{1-r_x^2}}
			\end{split}
		\end{equation}
		In this case, either $A_x$ or $B_x$ can completely characterize the total order. 

		Note that there are other ways to select a set of qubit states forming a total order in the resource theory of $\mathbb{Z}_2(\hat{x})$-asymmetry, such as the set of states with the same value of $|r_x|$. 

	\subsection{Connection to known monotones}
	\label{sec_connection}
		\subsubsection{The $A$-monotone}
		\label{sec_Amonotone}

			The monotone defined by the cylindrical radius of the state relative to $\hat{x}$, denoted $A_x(\rho)$ above, is an asymmetry measure based on trace distance introduced in Ref. \cite{marvianExtending2014}, specialized to the case of the about-face symmetry relative to $\hat{x}$. 
			
			The measure of $\mathbb{Z}_2(\hat{x})$-asymmetry of a state $\rho$ based on trace distance is defined as
			\begin{equation}\label{eq_dist_monotone}
				\begin{split}
					\frac12 \norm{ \rho - \sigma_x\rho \sigma_x }_1
				\end{split}
			\end{equation}
			where $\norm{A}_1 \coloneqq \tr \left( \sqrt{A^{\dag}A} \right)$.

			The trace distance between $\rho$ and $\rho'$ is defined as $D(\rho,\rho') \coloneqq \frac12 \norm{ \rho -\rho' }_1$. 
			It is well-known (see, e.g., Ref. \cite{nielsenQuantum2010}) that if the Bloch vectors of $\rho$ and $\rho'$ are $\vec{r}$ and $\vec{s}$ respectively, then the trace distance between the two states is given by $\frac12 \abs{\vec{r} -\vec{s}}${\,\textemdash\,}half of the Euclidean distance between the Bloch ball vectors. 
			Using this fact, we can express \eqref{eq_dist_monotone} as $\sqrt{r^2-r_x^2}$, which is just $A_x(\rho)$.

		Note that trace distance is a measure of distinguishability between $\rho$ and $\rho'${\,\textemdash\,}it satisfies the data-processing inequality.
		Furthermore, the two states $\rho$ and $\sigma_x \rho \sigma_x$ make up the orbit of $\rho$ under $\mathbb{Z}_2(\hat{x})$. As explained in \cite[Section 3.4.2]{marvian2012symmetry}, every measure of distinguishability gives an asymmetry monotone when applied to the elements of the group orbit.
		Thus, our monotone $A_x$ can be seen as an instance of this general procedure, explored also in \cite[Section 4.3]{gonda2021resource}.
		
		More concretely, trace distance corresponds to the optimal probability of distinguishing two states by some quantum measurement \cite{nielsenQuantum2010}.
		It follows that $A_x(\rho)$ is the optimal probability of guessing whether a rotation of $0$ or of $\pi$ degrees about the $\hat{x}$ axis was applied to the state $\rho$. 
		This provides an operational interpretation of the $A$-monotone. 

		\subsubsection{The $B$-monotone}
		\label{sec_Bmonotone}

			The monotone $B_x$ can be understood as a special case of a general class of monotones studied in \cite[Section 3.2]{gondaMonotones2019}, namely, a monotone obtained from a resource cost construction.%\footnote{As a side note, when $B_x$ is viewed as a monotone in the resource theory of $U(1)$-asymmetry about the $\hat{x}$ axis (rather than as a monotone in the resource theory of $\mathbb{Z}_2(\hat{x})$-asymmetry), it is a function of another known monotone called the purity of coherence defined in Ref. \cite{marvianCoherence2020}, which is related to the probability of coherence distillation. See  \cref{app:puritycoherence} for more details.} 

			The data required for the resource cost construction consists of a set $W$ of reference resources, which one could call a ``gold standard'', and a real-valued function $f_W$ with domain $W$. 
			The resulting cost monotone, called the {\em $f_W$-cost} of a state $\rho$, is defined as the smallest value of $f_W$ among the resources in $W$ that can be used to obtain $\rho$ by free operations.
			
			Here, we let $W$ be the set of equivalence classes of states corresponding to the vertical axis in \cref{fig_coordinate}, i.e., the set of states with $r_x = 0$.
			Since $W$ is a chain{\,\textemdash\,}a totally ordered set{\,\textemdash\,}there is an essentially unique way to evaluate these states, namely, by their value of $\smallsqrt{r_y^2+r_z^2}$. 
			Thus, we take $f_W$ to be $\smallsqrt{r_y^2+r_z^2}$.
			
			Note that $W$ contains a resource at the top of the resource partial order, which is the one for which $\smallsqrt{r_y^2+r_z^2} = 1$, and contains a resource at the bottom{\,\textemdash\,}a free state{\,\textemdash\,}which is the one for which $\smallsqrt{r_y^2+r_z^2} = 0$.
			
			Since the maximal resource in $W$ can perfectly encode the $\mathbb{Z}_2(\hat{x})$ symmetry group, it is a perfect reference frame for it~\cite{GouR_2008}.
			We thus refer to the maximal resource in $W$ as the ``$\hat{x}$-refbit'' and call $W$ the ``noisy $\hat{x}$-refbit'' chain. 

			From \cref{fig_coordinate} it is clear that the lowest resource in the noisy $\hat{x}$-refbit chain that can be used to generate a state $\rho$, which we will denote by $\sigma$, is the one that intersects the $B_x$-level set of $\rho$. 
			Thus, $B_x(\sigma) = B_x(\rho)$. 
			Since $\sigma$ is in the noisy $\hat{x}$-refbit chain, its Bloch vector can be written as $\vec{r}'=(0, r'_y, r'_z)$, and thus, $B_x(\sigma)=\smallsqrt{{r'}_y^2+{r'}_z^2}$, which coincides with the $f_W$ value of $\sigma$. 
			It follows that $B_x(\rho)$ is the smallest value of $f_W$ among states in $W$ from which $\rho$ can be produced by free operations. Hence,  $B_x(\rho)$ is the $f_W$-cost of $\rho$.  
			We refer to it as the ``noisy $\hat{x}$-refbit cost'' of a state.

			Note that if we consider the subset of {\em pure} states that are not free, then from \cref{eq_ABxpure}, we see that the noisy $\hat{x}$-refbit cost for these is always maximal.

\subsubsection{Cost-yield gap}\label{sec_bound}

	We noted in \cref{sec_constraint1RT} that there is a nontrivial dependence relation between $A_x$ and $B_x$ expressed by inequality \eqref{eq_BgeA0} which says $A_x(\rho) \le B_x(\rho)$.
	In this section we provide an operational interpretation thereof. 
		
	To begin, let us point out a second operational interpretation of the $A$-monotone, distinct from the one described in \cref{sec_Amonotone}. 
	It is similar to the one we provided for the $B$-monotone, but in terms of a resource {\em yield} construction rather than a cost construction. 
	Given the same data of a reference set $W$ and a function $f_W$, the \emph{$f_W$-yield} of a state $\rho$ is defined as the largest value of $f_W$ among the resources in $W$ that can be obtained from $\rho$ by the free operations (see \cite[section 3.2]{gondaMonotones2019}).  
	$A_x$ can be shown to be the $f_W$-yield where $W$ is the noisy $\hat{x}$-refbit chain and the function $f_W$ is $\smallsqrt{r_y^2+r_z^2}$ just as above.
	The proof is analogous to the proof that $B_x$ is the noisy $\hat{x}$-refbit cost, provided in \cref{sec_Bmonotone}.
	Therefore, $A_x(\rho)$ can be called the ``noisy $\hat{x}$-refbit yield'' of $\rho$. 
	
	The dependence relation between $A_x$ and $B_x$ thus says that (noisy $\hat{x}$-refbit) yield cannot exceed the (noisy $\hat{x}$-refbit) cost.
	Whenever the inequality is strict, there is a nonzero \emph{cost-yield gap}. 
	All qubit states except for the free states and the states in the noisy $\hat{x}$-refbit chain have such a gap.

	Cost-yield gaps are relevant in many resource theories.
	For example, consider the case where the yield is zero while the cost is nonzero (though this does not occur in our case).
	In the context of entanglement theory, this phenomenon is analogous to the notion of bound entanglement~\cite{Horodecki1998}, i.e., the distillable entanglement of a state is zero but the entanglement cost~\cite{Hayden_2001} is nonzero.  
	One difference is that the distillable entanglement and the entanglement cost are defined in terms of asymptotic rates of conversion rather than in a single-shot setting. 
	A notion that is a bit closer to our cost-yield gap is that of a 1-bound entangled state:
	A state is 1-bound if it is neither locally preparable nor 1-distillable, where 1-distillability is a single-copy notion of distillability~\cite{Dur2000,Bartlett2006}. 
	The notion of bound states for both asymptotic and single-shot cases can be also studied in other resource theories~\cite[Section V.B]{Chitambar2019}.
	
	Cost and yield monotones are useful for operational scenarios where it is expedient to convert all resource states into the ``gold standard'' form. 
	The fact that all nonfree states outside the noisy $\hat{x}$-refbit chain have cost-yield gaps means that in such operational scenarios, converting a gold standard resource into a generic resource $\rho$ inevitably leads to a loss of value, in the sense that one cannot retrieve a gold standard resource as valuable as the original one from $\rho$. 

\section{Quantum asymmetry dependence relations}
\label{sec_relations}

	Our characterization of the resource ordering implies that if we are interested in the dependence relations among properties relative to the three asymmetries $\mathbb{Z}_2(\hat{x})$, $\mathbb{Z}_2(\hat{y})$, and $\mathbb{Z}_2(\hat{z})$, it suffices to focus on the complete set of monotones for each type of $\mathbb{Z}_2$-asymmetry, that is, the complete set of monotones for $\mathbb{Z}_2(\hat{x})$,
	\begin{subequations}\label{eq_ABx}
		\begin{align}
	    A_x(\rho) & \coloneqq \sqrt{r_y^2+r_z^2}, \label{eq_Ax}\\
		%\sqrt{r^2-r_x^2}, \label{eq_Ax}\\
	    B_x(\rho) &\coloneqq 
	    \begin{cases}
	    \sqrt{\frac{r_y^2+r_z^2}{1-r_x^2}} , &\text{ if } r_x^2<1\\
	    0, &\text{ if } r_x^2=1
	    \end{cases}
	    ,\label{eq_Bx}
		\end{align}
	\end{subequations}
	the complete set for $\mathbb{Z}_2(\hat{y})$,
	\begin{subequations}
		\label{eq_ABy}
	\begin{align}
	    A_y(\rho) &\coloneqq  \sqrt{r_x^2+r_z^2}, \label{eq_Ay}\\
		%\sqrt{r^2-r_y^2}, \label{eq_Ay}\\
	    B_y(\rho) &\coloneqq     
	    \begin{cases}
	        \sqrt{\frac{r_x^2+r_z^2}{1-r_y^2}} , &\text{ if } r_y^2<1\\
	        0, &\text{ if } r_y^2=1
	        \end{cases}, \label{eq_By}
	\end{align}
	\end{subequations}
	and the complete set for $\mathbb{Z}_2(\hat{z})$,
	\begin{subequations}
		\label{eq_ABz}
	\begin{align}
	    A_z(\rho) &\coloneqq  \sqrt{r_x^2+r_y^2}, \label{eq_Az}\\
		%\sqrt{r^2-r_z^2}, \label{eq_Az}\\
	    B_z(\rho) &\coloneqq     
	    \begin{cases}
	        \sqrt{\frac{r_x^2+r_y^2}{1-r_z^2}} , &\text{ if } r_z^2<1\\
	        0, &\text{ if } r_z^2=1
	        \end{cases}. \label{eq_Bz}
	\end{align}
	\end{subequations}

$A_x$, $B_x$, $A_y$, $B_y$, $A_z$ and $B_z$ form the set $\mathcal{M}$ as defined in \cref{eq_defM}.

\subsection{Special case: pure states}
\label{sec_pure}

Since the discontinuity in $B_n$ concerns the case when $r_n^2=1$, which necessarily concerns pure states, we first consider the dependence relations among $\mathbb{Z}_2(\hat{x})$-, $\mathbb{Z}_2(\hat{y})$-, and $\mathbb{Z}_2(\hat{z})$-asymmetry properties of pure states. 

\begin{figure}[h!]
	\centering
	\includegraphics[width=0.3\textwidth]{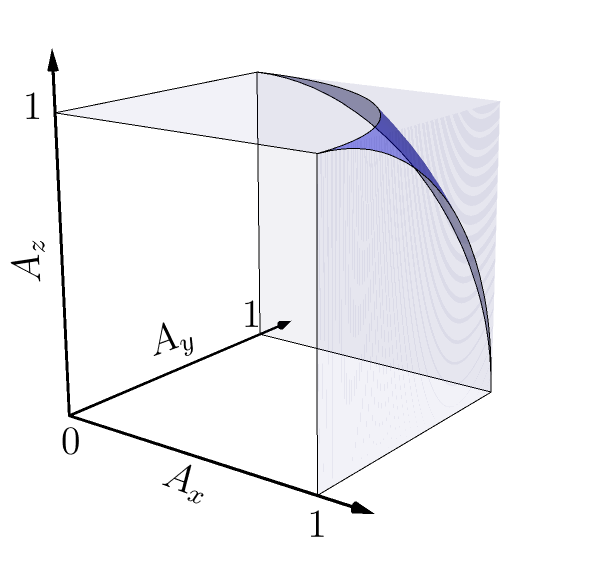}
	\hspace{1cm}
	\includegraphics[width=0.3\textwidth]{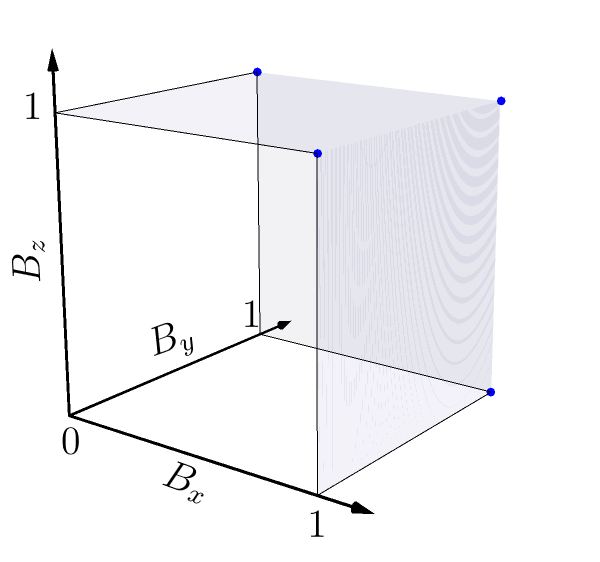}
	\caption{The set of jointly realizable values of $\{A_x, A_y, A_z\}$ (the left plot) and of $\{B_x, B_y, B_z\}$ (the right plot) for a pure state.}
	\label{fig_pure}
\end{figure}

For a pure state, we have $r_x^2+r_y^2+r_z^2=1$ and thus,
\begin{subequations}
	\label{eq_xpure}
	\begin{align}
		A_x(\rho) &= \sqrt{1-r_x^2}, \label{eq_Axpure}\\
		\label{eq_Bxpure}
		B_x(\rho) &= 
		\begin{cases}
		1, &\text{ if } r_x^2<1\\
		0, &\text{ if } r_x^2=1
		\end{cases},
	\end{align}
\end{subequations}
\begin{subequations}
	\label{eq_ypure}
	\begin{align}
		A_y(\rho) &=  \sqrt{1-r_y^2},\label{eq_Aypure}\\
		\label{eq_Bypure}
		B_y(\rho) &=     
		\begin{cases}
			1 , &\text{ if } r_y^2<1\\
			0, &\text{ if } r_y^2=1
			\end{cases},
	\end{align}
\end{subequations}
\begin{subequations}
	\label{eq_zpure}
	\begin{align}
		A_z(\rho) &=  \sqrt{1-r_z^2},\label{eq_Azpure}\\
		\label{eq_Bzpure}
		B_z(\rho) &=     
		\begin{cases}
			1, &\text{ if } r_z^2<1\\
			0, &\text{ if } r_z^2=1
			\end{cases}. 
	\end{align}
\end{subequations}
It is clear from \cref{eq_Bxpure,eq_Bypure,eq_Bzpure} that there are only four triples of values of $B_x$, $B_y$ and $B_z$ that are jointly realizable by a pure state. 
They are 
\begin{align}
	\label{eq_Bpure}
		\{B_x(\rho),B_y(\rho),B_z(\rho)\} =\begin{cases}
			\{0,1,1\}, &\text{ if } r_x^2=1\\
			\{1,0,1\}, &\text{ if } r_y^2=1\\
			\{1,1,0\}, &\text{ if } r_z^2=1\\
			\{1,1,1\}, &\text{ otherwise. }
			\end{cases}
\end{align}
That is, when the Bloch vector is along one of the coordinate axes $\hat{n}$, the corresponding $B_n$ takes the value 0 while the other two take the value 1.
If the Bloch vector is not aligned with any of the coordinate axes, $B_x$, $B_y$, and $B_z$ are all maximal. 
These four possibilities are depicted as the four blue points in the right plot of \cref{fig_pure}. 

Thus, the minimal values of at least two $B$-monotones drawn from $B_x$, $B_y$, and $B_z$ cannot be jointly realized by a pure state, indicating that it is impossible for a pure state to simultaneously be bottom of any two of the resource orders for $\mathbb{Z}_2(\hat{x})$-asymmetry, $\mathbb{Z}_2(\hat{y})$-, and $\mathbb{Z}_2(\hat{z})$-asymmetry. This also means that for a pure states, its $\mathbb{Z}_2(\hat{x})$-, $\mathbb{Z}_2(\hat{y})$- and $\mathbb{Z}_2(\hat{z})$-asymmetry properties are not independent of each other.

The maximal values of $B_x$, $B_y$ and $B_z$, unlike the minimal values, can be jointly realized, as long as the pure state is not aligned with any of the axes $\hat{x}$, $\hat{y}$ or $\hat{z}$. However, this does not mean that a pure state can be top-of-the-order simultaneously in the resource theories of $\mathbb{Z}_2(\hat{x})$-asymmetry, of $\mathbb{Z}_2(\hat{y})$-asymmetry, and of $\mathbb{Z}_2(\hat{z})$-asymmetry, because having each of the $B$-monotones reach its maximum is not a sufficient condition for the corresponding $A$-monotone to do so, as indicated by \cref{eq_AxBx1}.

Recalling that for pure states, $A_n$ alone completely characterizes the resource order of $\mathbb{Z}_2(\hat{n})$-asymmetry, it follows that in order to 
to fully understand the dependence relation among $\mathbb{Z}_2(\hat{x})$, $\mathbb{Z}_2(\hat{y})$, and $\mathbb{Z}_2(\hat{z})$-asymmetry properties of a pure state, we need to derive the constraints on $A_x$, $A_y$ and $A_z$. Since the Bloch vector radius $r_x^2+r_y^2+r_z^2$ equals $1$ for pure states, \cref{eq_Axpure,eq_Aypure,eq_Azpure} tell us that
\begin{equation}\label{eq_trade-offpureA}
	A_x^2(\rho)+A_y^2(\rho)+A_z^2(\rho) = 2.
\end{equation}
Thus, the maximal values of $A_x$, $A_y$ and $A_z$ cannot be jointly realized, otherwise the right-hand side could reach 3 instead of 2. 
It is not possible for a pure state to simultaneously be top-of-the-order for $\mathbb{Z}_2(\hat{x})$-asymmetry, $\mathbb{Z}_2(\hat{y})$-asymmetry, and $\mathbb{Z}_2(\hat{z})$-asymmetry.
Furthermore, \cref{eq_trade-offpureA} shows that for pure states, $A_x$, $A_y$, and $A_z$ satisfy our sufficient condition for them to exhibit trade-off, as defined in \cref{sec_step3}. 
This is depicted in the left plot of \cref{fig_pure}. 

We can also see that knowing the values of any two of $A_x$, $A_y$ and $A_z$ is sufficient to determine the third. 
Thus, for a pure state, any two of the three $A$-monotones completely specify the $\mathbb{Z}_2$-asymmetry properties for all three axes $\hat{x}$, $\hat{y}$ and $\hat{z}$.

For a general state, we will see that its $\mathbb{Z}_2(\hat{x})$-, $\mathbb{Z}_2(\hat{y})$- and $\mathbb{Z}_2(\hat{z})$-asymmetry properties do not trade off against each other in the same sense as for pure states. Nevertheless, they still constrain each other.
\subsection{Equality constraints}
\label{sec_equalities}

In this subsection, we derive all of the equality constraints among the six monotones of our three resource theories. 
To do so, we start by expressing the Bloch vector components as functions of $A_x$, $A_y$ and $A_z$ and as functions of $B_x$, $B_y$ and $B_z$.

Specifically, \cref{eq_Ax,eq_Ay,eq_Az} imply that the expressions for  $r^2_x$, $r_y^2$, and $r^2_z$ in terms of  $A_x$, $A_y$ and $A_z$ are:
\begin{equation}\label{eq_rintermsofA}
	\begin{split}
		r_x^2 &= \frac12 \left( - A_x^2(\rho) +  A_y^2(\rho) + A_z^2(\rho) \right), \\
		r_y^2 &= \frac12 \left( A_x^2(\rho) -  A_y^2(\rho) + A_z^2(\rho) \right), \\
		r_z^2 &= \frac12 \left( A_x^2(\rho) +  A_y^2(\rho) - A_z^2(\rho) \right). 
	\end{split}
\end{equation}
Similarly, \cref{eq_Bx,eq_By,eq_Bz} imply the expressions for $r^2_x$, $r_y^2$, and $r^2_z$ in terms of  $B_x$, $B_y$ and $B_z$ when the Bloch vector radius $r$ is smaller than 1 (otherwise, we refer back to \cref{eq_Bpure}):
\begin{equation}
\label{eq_rintermsofB}
\begin{split}
   r^2_x &= \frac{1}{1-B_x^2} \left( \frac{B_x^2 + B_y^2 + B_z^2 - 2 (B_x^2 B_y^2 + B_x^2 B_z^2 + B_y^2 B_z^2 )+ 
   3 B_x^2 B_y^2 B_z^2}{2 - B_x^2 - B_y^2 - B_z^2 + B_x^2 B_y^2 B_z^2} -B_x^2 \right),\nonumber\\
   r^2_y &= \frac{1}{1-B_y^2} \left( \frac{B_x^2 + B_y^2 + B_z^2 - 2 (B_x^2 B_y^2 + B_x^2 B_z^2 + B_y^2 B_z^2 )+ 
   3 B_x^2 B_y^2 B_z^2}{2 - B_x^2 - B_y^2 - B_z^2 + B_x^2 B_y^2 B_z^2} -B_y^2 \right),\nonumber\\
   r^2_z &= \frac{1}{1-B_z^2} \left(\frac{B_x^2 + B_y^2 + B_z^2 - 2 (B_x^2 B_y^2 + B_x^2 B_z^2 + B_y^2 B_z^2 )+ 
   3 B_x^2 B_y^2 B_z^2}{2 - B_x^2 - B_y^2 - B_z^2 + B_x^2 B_y^2 B_z^2} -B_z^2 \right),
\end{split}
\end{equation}
where we omit the argument $\rho$ for brevity and note that for impure states, we have $B_x,B_y,B_z\in[0,1)$ and consequently both $1-B_n^2$ and $2 - B_x^2 - B_y^2 - B_z^2 + B_x^2 B_y^2 B_z^2$ are strictly positive.

Equating the expressions in \cref{eq_rintermsofA,eq_rintermsofB} for $r^2_x, r_y^2, r^2_z$, we arrive at the three equalities of \cref{eq_equality0}. We repeat these equalities here:
	 \begin{subequations}
		\label{eq_equality}
		\begin{align}
			\label{eq_1eq}
			2 \left[ B_x^2(\rho) - A^2_x(\rho) \right] - B_x^2(\rho) \left[-A_x^2(\rho) +  A_y^2(\rho) + A_z^2(\rho)\right] &= 0, \\ 
			\label{eq_2eq}
			2 \left[ B_y^2(\rho) - A^2_y(\rho) \right] - B_y^2(\rho) \left[A_x^2(\rho) -  A_y^2(\rho) + A_z^2(\rho)\right]&= 0, \\
			\label{eq_3eq}
			2 \left[ B_z^2(\rho) - A^2_z(\rho) \right] - B_z^2(\rho)\left[A_x^2(\rho) +  A_y^2(\rho) - A_z^2(\rho)\right] &= 0.
		\end{align}
	\end{subequations}
\cref{eq_equality} also applies to pure states, even though it is derived by assuming $r<1$, i.e., by assuming an impure state. To see this, note that
% $A_x^2(\rho) +  A_y^2(\rho) + A_z^2(\rho)=2$ (i.e., \cref{eq_trade-offpureA}), a necessary and sufficient condition for pure states, together with \cref{eq_equality} implies that $\{B_x, B_y, B_z\}$ can only take one of the four possibilities specified in \cref{eq_Bpure}, which also are the only possibilities for $\{B_x, B_y, B_z\}$ for pure states as explained in \cref{sec_pure}. Conversely, 
when $\{B_x, B_y, B_z\}$ takes one of the four possibilities specified in \cref{eq_Bpure} for pure states, \cref{eq_equality} simply reduces to \cref{eq_trade-offpureA} (and that $A_n=0$ if $B_n=0$, agreeing with \cref{eq_BAranges}). For example, when $\{B_x,B_y,B_z\}=\{0,1,1\}$, \cref{eq_1eq} reduces to $A_x=0$ while \cref{eq_2eq,eq_3eq} both reduce to \cref{eq_trade-offpureA}. 

The connection between the number of equality constraints and the dimension of the state space of a qubit has already been discussed in \cref{sec_eq0}. Here we comment on some other aspects.

Given that $A_x, B_x, A_y, B_y, A_z, B_z\geq 0$  (as noted in \cref{eq_BAranges}), when the values of any three of these six monotones are known, the set of equalities in \cref{eq_equality} always has a unique solution for the remaining monotones, except when the known values include two of the $B$-monotones simultaneously being 1 (and thus must be a pure state). This exception is due to the discreteness of the $B$-monotones for pure states as shown in \cref{sec_pure}. Hence, fixing the values of at least three of the six monotones, with at least two of the fixed ones being $A$-monotones, always uniquely determines the values of the remaining monotones and thus all the $\mathbb{Z}_2(\hat{x})$-, $\mathbb{Z}_2(\hat{y})$-, and $\mathbb{Z}_2(\hat{z})$-asymmetry properties.
%the values of $A_x$, $A_y$, and $A_z$ are known, or whenever the values of any triple consisting of two $A$-monotones drawn from $\{A_x,A_y,A_z\}$ and one $B$-monotone drawn from $\{B_x,B_y,B_z\}$ are known. %That is, fixing the values of the triple either consisting of $A_x$, $A_y$, and $A_z$ or consisting of two of the $A$-monotones and one of the $B$-monotone, such as $A_x$, $B_x$ and $A_y$, always uniquely fixes the value of the other three. 
%Fixing any four or more of the six monotones is always sufficient to uniquely determine the values of the rest. Furthermore, for impure states, i.e, when $A_x, B_x, A_y, B_y, A_z, B_z\in [0,1)$, fixing the values of any \emph{three} or more of the six monotones always uniquely determines the values of the other three. Whenever the values of all six monotones are known, 

\begin{figure}[h]
	\centering
	\includegraphics[width=0.3\textwidth]{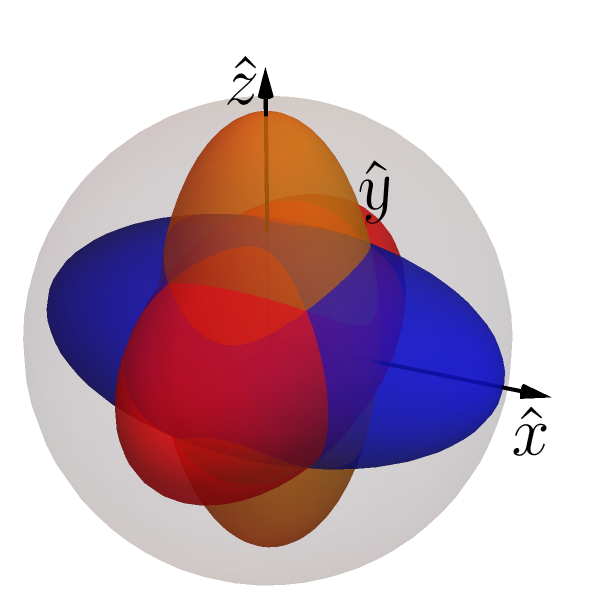}
	\hspace{0.2cm}
	\includegraphics[width=0.3\textwidth]{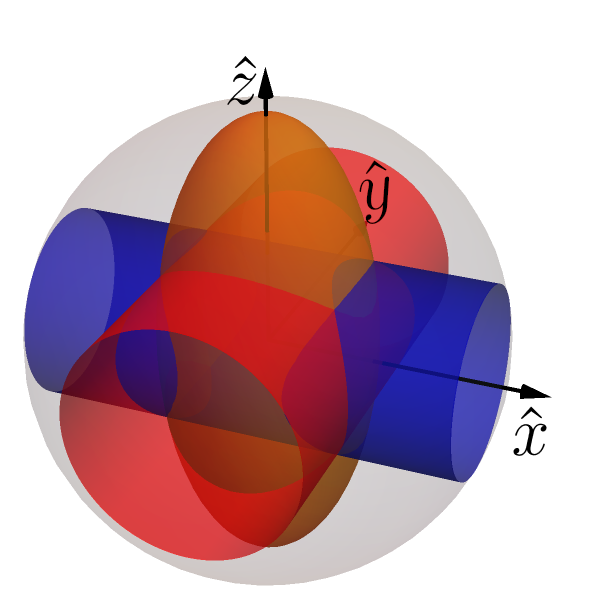}
	\hspace{0.2cm}
	\includegraphics[width=0.3\textwidth]{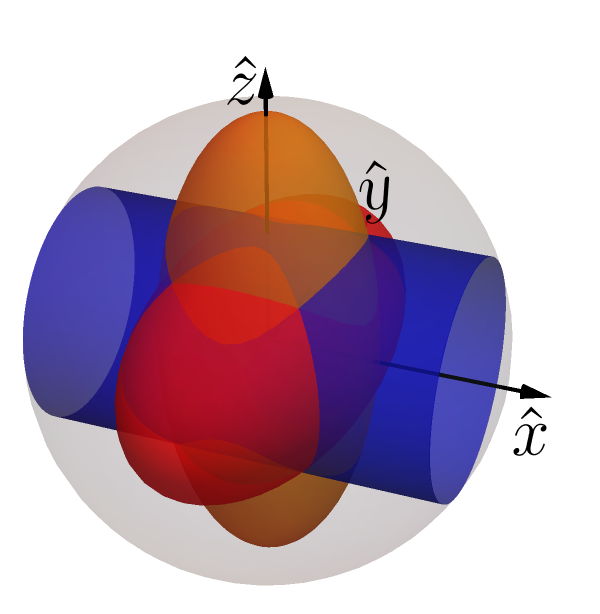}
	\caption{Left: Intersection of three prolate spheroids representing the states with a given $B_x$, $B_y$, and $B_z$, respectively. 
		Middle: Intersection of two cylinders and a prolate spheroid representing the states with a given $A_x$, $A_y$, and $B_z$, respectively.  
		Right: Intersection of a cylinder and two prolate spheroids representing the states with a given $A_x$, $B_y$, and $B_z$, respectively.  }
	\label{fig_8points}
\end{figure}

\cref{fig_8points} provides three examples for visualizing the set of states that share the same $\mathbb{Z}_2(\hat{x})$, $\mathbb{Z}_2(\hat{y})$, and $\mathbb{Z}_2(\hat{z})$-asymmetry properties, where each plot uses a different triple of monotones to establish this set via an intersection of the corresponding level sets. In general, the intersection is a set of eight states, but it may degenerate to four, two, or one state(s). 
This is because in the expressions for all six monotones in $\mathcal{M}$, only the squares of Bloch vector components appear. When the values of $r_x^2$, $r_y^2$, and $r_z^2$ are fixed, the ambiguity in the sign of $r_x$, $r_y$, and $r_z$ gives rise to up to $8$ states.

\subsection{Dependence relations among $A_x$, $A_y$, $A_z$: a generating set of inequality constraints}
\label{sec_ineq_A}

Now, we turn to inequality constraints.
These can be derived from the nonnegativity requirements on the Bloch vector components
\begin{equation}\label{eq_rspositive}
	r^2_x \ge 0,  \qquad  r_y^2 \ge 0,  \qquad r^2_z \ge 0,
\end{equation}
and the fact that the maximum Bloch vector radius of a state is 1,
\begin{equation}\label{eq_r2range}
	r^2_x + r_y^2 + r^2_z \le 1.
\end{equation}
%To apply these two requirement, we now express $r^2_x, r_y^2, r^2_z$ as functions of the resource monotones.

% \subsubsection{For $A_x$, $A_y$ and $A_z$}
% \label{sec_ineq_A}

Combining the inequalities \eqref{eq_rspositive} with the expressions for $r^2_x, r_y^2, r^2_z$ as functions of $A_x$, $A_y$ and $A_z$ (Eq.~\eqref{eq_rintermsofA}), we obtain
\begin{equation}\label{eq_inequalityA}
	\begin{split}
		- A_x^2(\rho) +  A_y^2(\rho) +  A_z^2(\rho) \ge 0, \\
		A_x^2(\rho) -  A_y^2(\rho) +  A_z^2(\rho) \ge0, \\
		A_x^2(\rho) +  A_y^2(\rho) -  A_z^2(\rho) \ge 0,
	\end{split}
\end{equation}
and the inequality \eqref{eq_r2range} leads to another inequality constraint\footnote{This inequality constraint was also derived in Ref.\ \cite{cheng_complementarity_2015}.}
\begin{equation}\label{eq_r2Aineq}
	A_x^2(\rho) + A_y^2(\rho) + A_z^2(\rho)  \le 2.
\end{equation}
From the equality constraints of \eqref{eq_equality}, we know that fixing the values of $A_x$, $A_y$ and $A_z$ uniquely fixes the values of $B_x$, $B_y$ and $B_z$. 
Thus, the inequalities in \eqref{eq_inequalityA} and \eqref{eq_r2Aineq} together form a generating set of all inequality constraints (as defined in \cref{sec_recipe}) on $A_x$, $A_y$, $A_z$, $B_x$, $B_y$ and $B_z$, i.e., on $\mathcal{M}$. 

The set of equalities in \eqref{eq_equality} do not impose constraints on $A_x$, $A_y$, $A_z$ when $B_x$, $B_y$, $B_z$ are not constrained. Thus, the inequalities in
\eqref{eq_inequalityA} and \eqref{eq_r2Aineq}, together with the constraint that $A_x, A_y, A_z \geq 0$ (which follow from the definitions of the $A$-monotones) completely characterize $\mathtt{JointRealize}(A_x,A_y,A_z)$, the set of jointly realizable values of $A_x$, $A_y$ and $A_z$. The region of such values is depicted in \cref{fig_qt3dA}. The key features and implications of \cref{fig_qt3dA} have already been analyzed in \cref{sec_interpret3way}. Here, we provide geometric intuitions with Bloch ball pictures and algebraic descriptions for some of the features and implications. 

\begin{figure}[h]
	\centering
	\includegraphics[width=0.3\textwidth]{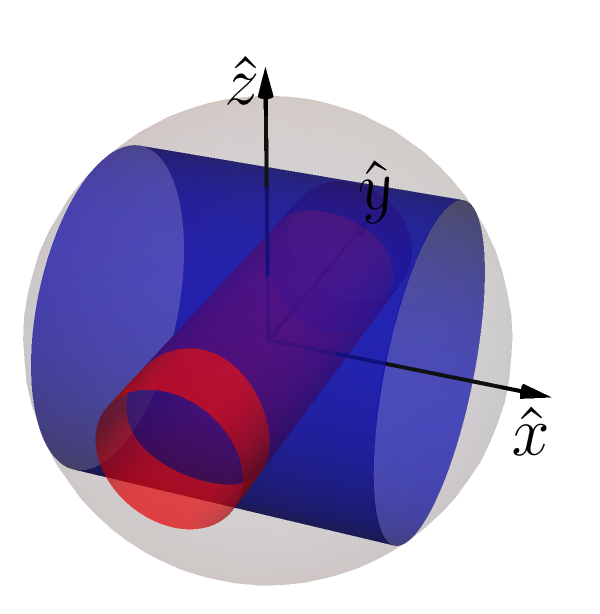}
	\hspace{0.2cm}
	\includegraphics[width=0.3\textwidth]{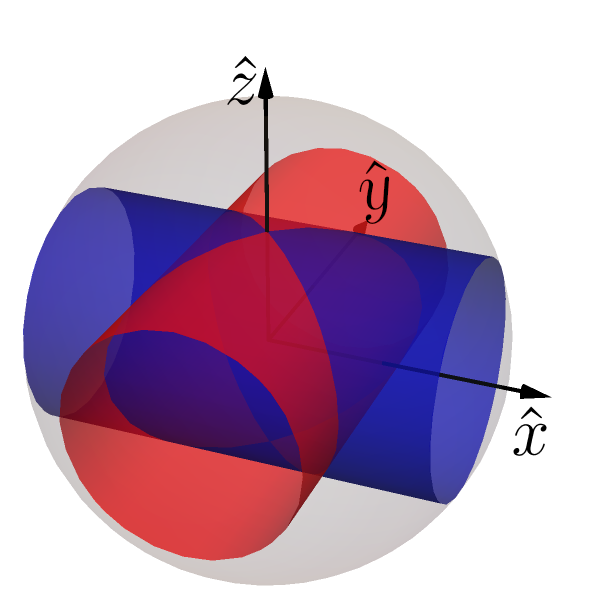}
	\hspace{0.2cm}
	\includegraphics[width=0.3\textwidth]{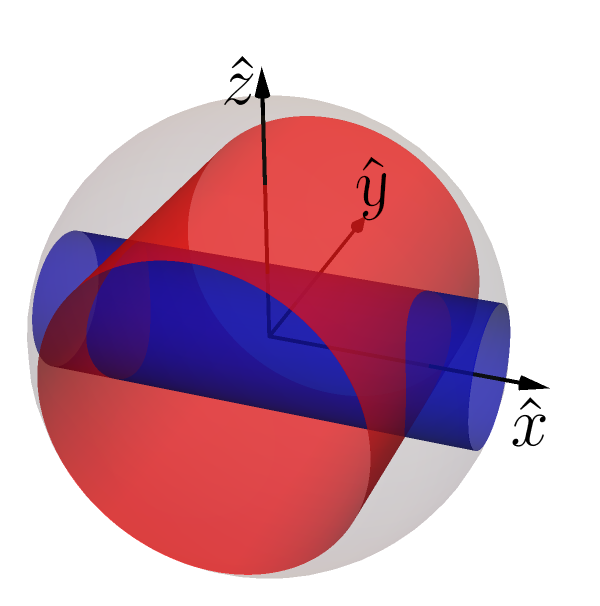}
	\caption{Examples of the intersections between the blue cylinder (representing the states with a particular value of $A_x$) and the red cylinder (representing the states with a particular value of $A_y$). Left: $A_x>A_y$. Middle: $A_x=A_y$. Right: $A_x<A_y$.}
	\label{fig_AxAy}
\end{figure}

Recall from \cref{sec_partialorder} that states with a given value of $A_n$ form the surface of a cylinder centered around the $\hat{n}$ axis with radius $A_n$. 
To determine if a pair of values for two $A$-monotones is viable, we only need to check if the two respective cylinders intersect within the Bloch ball. \cref{fig_AxAy} displays three examples of such intersections.

Since any two cylinders with radii in $[0,1]$ and centred around orthogonal axes do intersect in the Bloch ball, a given pair of $A$-monotones can take an arbitrary pair of values in $[0,1]$. This corresponds to the fact that the projection of $\mathtt{JointRealize}(A_x,A_y,A_z)$ onto the plane spanned by any two of these (such as the projection onto the  $A_x{-}A_y$ plane for instance)  yields the full $[0,1]\times[0,1]$ square, as can be inferred from the left plot of \cref{fig_qt3dA}.

However, if we fix the value for one of  $A_x$, $A_y$ and $A_z$, the range of jointly realizable values for the other two is limited. This is clear from inspecting the $A_y$--$A_z$ cross-sections of $\mathtt{JointRealize}(A_x,A_y,A_z)$, a few of which are depicted in the middle part of \cref{fig_qt3dA}. 
Geometrically, for a given value of $A_x$, a pair of values of $A_y$ and $A_z$ is jointly realizable with it when the intersection region of the cylinders representing the valuations of $A_y$ and $A_z$ also has an intersection with the cylinder representing the valuation of $A_x$. 

\begin{figure}[h]
	\centering
	\includegraphics[width=0.3\textwidth]{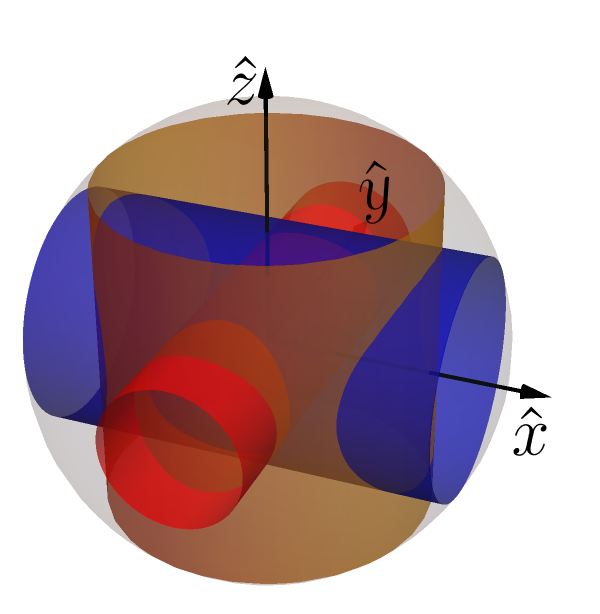}
	\hspace{0.2cm}
	\includegraphics[width=0.3\textwidth]{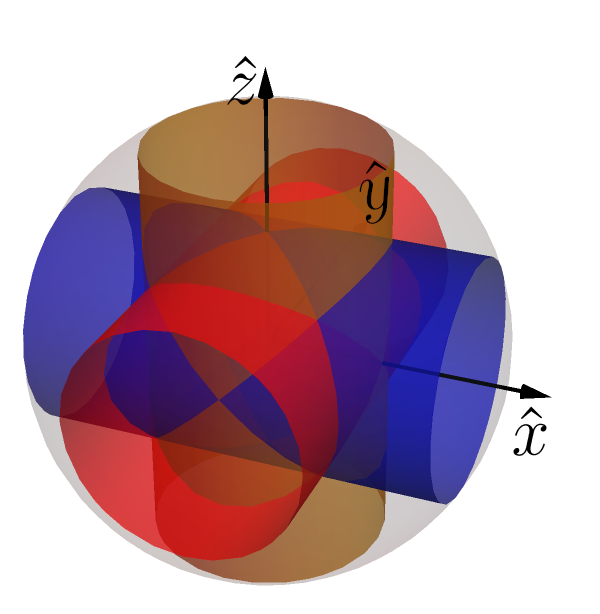}
	\hspace{0.2cm}
	\includegraphics[width=0.3\textwidth]{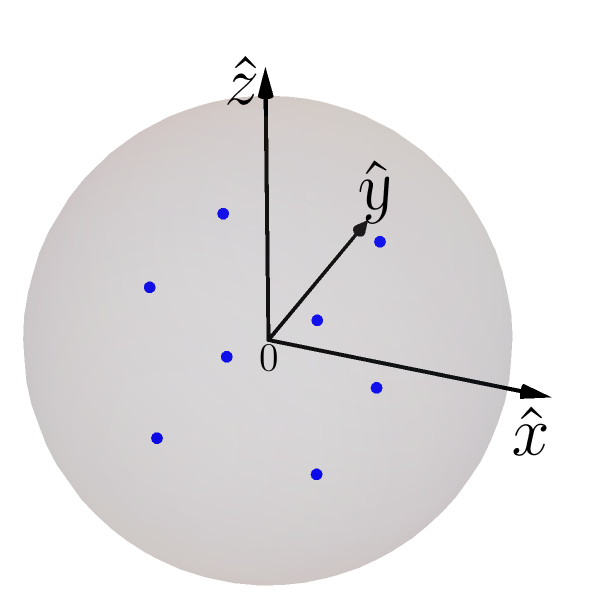}
	\caption{Left: an example where the three cylinders do not share common points. 
		Middle: an example where the three cylinders do not share common points. The two blue cylinders are identical. 
		Right: the eight points picked out by the three cylinders in the middle figure. Since $A_x=A_y=A_z$ in the middle figure, the eight points are the nodes of a cube.}
	\label{fig_AxAyAz}
\end{figure}

For example, in the leftmost plot of \cref{fig_AxAyAz}, the three cylinders do not have a three-way intersection, while in the second plot they do. 
That is, for a fixed $A_x$, some pairs of $A_y$ and $A_z$ cannot arise from any valid quantum state (cf.\ the first plot in \cref{fig_AxAyAz}) and some can (cf.\ the second plot in \cref{fig_AxAyAz}). 

The cross-sections in the middle plot of \cref{fig_qt3dA} show a transition from $A_y$ and $A_z$ having a synergy relation to a trade-off relation as $A_x$ increases from 0 to 1. Let $\alpha$ denote the fixed value of $A_x$. In the nonextremal cases, i.e., when $\alpha\in(0,1)$, the region of jointly realizable values of $A_y$ and $A_z$ in the cross-sections has four boundaries. 
They are given by
\begin{subequations}\label{eq_AyAzBounds}
	\begin{alignat}{2}
		\text{bottom-left: }& &  - \alpha^2 +  A_y^2(\rho) +  A_z^2(\rho) &\ge 0, \label{eq_bl}\\
		\text{bottom-right: }& & \alpha^2 -  A_y^2(\rho) +  A_z^2(\rho) &\ge 0, \label{eq_br} \\
		\text{top-left: }& & \alpha^2 +  A_y^2(\rho) -  A_z^2(\rho) &\ge 0, \label{eq_tl} \\
		\text{top-right: }& &  \alpha^2 +  A_y^2(\rho) +  A_z^2(\rho) &\le 2. \label{eq_tr}
	\end{alignat}
\end{subequations}
These expressions are derived directly from inequalities \eqref{eq_inequalityA} and \eqref{eq_r2Aineq}, respectively. 
When $\alpha$ either equals 0 or 1, the four boundaries boil down to a single curve.

Rearranging inequalities \eqref{eq_AyAzBounds}, we find
\begin{align}
	\label{eq_sync} \bigl\lvert A_y^2(\rho) &-  A_z^2(\rho) \bigr\rvert \le \alpha^2,  \\
	\label{eq_sum} \alpha^2 \le A_y^2(\rho) &+  A_z^2(\rho) \le 2-\alpha^2. 
\end{align}
The first pair of inequalities, i.e., \eqref{eq_sync}, indicates that the difference between $A_y$ and $A_z$ decreases when $\alpha$ decreases. The difference disappears when $\alpha=0$, which explains the synergy relation between the two when the qubit is perfectly symmetric with respect to $\mathbb{Z}_2(\hat{x})$.

The second pair of inequalities, i.e., \eqref{eq_sum}, indicates that as $\alpha$ increases, the minimum value that can be achieved by $A_y^2+A_z^2$ increases, while the maximal value that can be achieved by $A_y^2+A_z^2$ decreases. When $\alpha=1$, we have $\alpha^2=2-\alpha^2$, i.e., the two extremal values coincide, resulting in the trade-off relation between $A_y$ and $A_z$ as shown in the corresponding cross-section in \cref{fig_qt3dA}.

\cref{eq_sync,eq_sum} together (or, equivalently, the boundaries in \cref{eq_AyAzBounds}) indicate that in general, the dependence relation between $A_y$ and $A_z$ given $\alpha$ is neither a synergy nor a trade-off relation. Specifically, if the dependence relation between $A_y$ and $A_z$ given $\alpha$ had been given by the equality condition in \eqref{eq_br} or \eqref{eq_tl}, namely, $A_y^2(\rho)-A_z^2(\rho)=\alpha^2$ or  $A_z^2(\rho)-A_y^2(\rho)=\alpha^2$, it would have been a synergy relation. Similarly, if it had been given by the equality condition in \eqref{eq_bl} or \eqref{eq_tr}, it would have been a trade-off relation. Instead, the dependence relation between $A_y$ and $A_z$ given $\alpha$ describes a jointly realizable region that is {\em bounded} by these trade-off and synergy relations. 

The rightmost plot of \cref{fig_qt3dA} shows that the semi-algebraic set for the possible values of the tuple $\{A_x, A_y, A_z\}$ can be decomposed into slices that each correspond to a fixed purity. 
For $r=1$, we get the one discussed in \cref{sec_pure} for pure states. 
The other fixed-purity slices have the same shape as the one for pure states, namely, a subregion of a sphere, but smaller radii as $r$ decreases. 
This is because, for a fixed Bloch vector radius $r\in[0,1]$, we have
	\begin{equation}
		\label{eq_ApurityTrade}
		A_x^2(\rho) +  A_y^2(\rho) +  A_z^2(\rho) =2r^2.
	\end{equation}
See \cref{sec_fixedPurity} for the derivation. This equation also reveals that, when the purity is fixed, there is a trade-off relation among the $A$-monotones for axes $\hat{x}$, $\hat{y}$ and $\hat{z}$. This is because when the value for one of the three $A$-monotones $A_x$, $A_y$ and $A_z$ is fixed, the other two monotones satisfy \cref{eq_mmtr} and thus $A_x$, $A_y$ and $A_z$ satisfy our sufficient condition for them to exhibit trade-off, as defined in \cref{sec_step3}.

Furthermore, since states with a given purity form a total order in the resource theory of $\mathbb{Z}_2(\hat{n})$-asymmetry that is captured by $A_n$, the trade-off among $A_x$, $A_y$ and $A_z$ also implies a trade-off among the $\mathbb{Z}_2(\hat{x})$-, $\mathbb{Z}_2(\hat{y})$-, and $\mathbb{Z}_2(\hat{z})$-asymmetry properties.

\begin{figure}[h]
	\centering
	\includegraphics[width=0.3\textwidth]{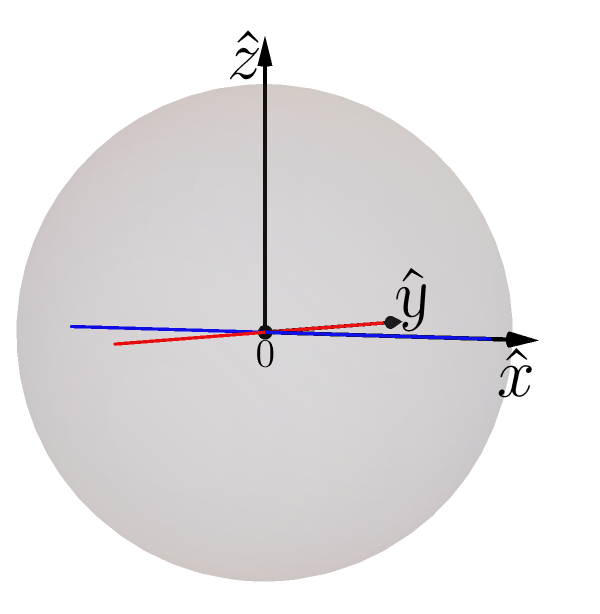}
	\hspace{0.2cm}
	\includegraphics[width=0.3\textwidth]{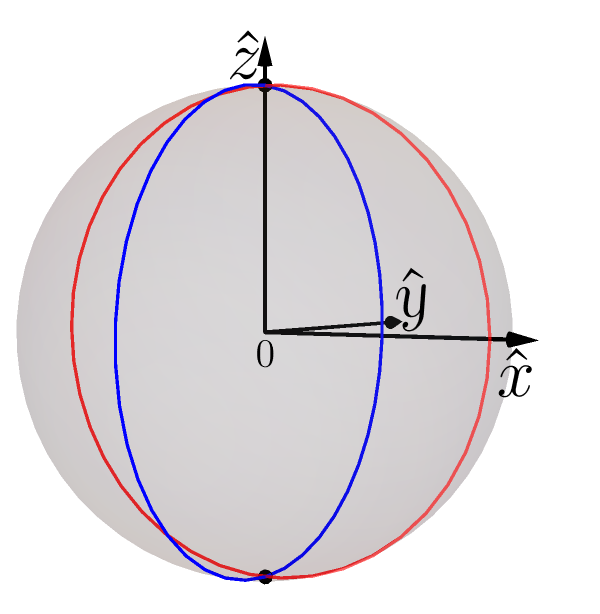}
	\caption{Left: A state can be bottom-of-the-order for $\mathbb{Z}_2$-asymmetry relative to both axes $\hat{x}$ and $\hat{y}$ simultaneously since the equivalence class at the bottom of the partial order in the resource theory of $\mathbb{Z}_2(\hat{x})$-asymmetry (represented by the blue line) intersects with that for $\mathbb{Z}_2(\hat{y})$-asymmetry (represented by the red line). Right: A state can be top-of-the-order for $\mathbb{Z}_2$-asymmetry relative to both axes $\hat{x}$ and $\hat{y}$  simultaneously since the equivalence class at the top of the partial order in the resource theory of $\mathbb{Z}_2(\hat{x})$-asymmetry (represented by the blue circle) intersects with that for $\mathbb{Z}_2(\hat{y})$-asymmetry (represented by the red circle).}
	\label{fig_XYextreme}
\end{figure}

Finally, recall from \cref{sec_properties} that when $A_n=0$, a state must be bottom-of-the-order for $\mathbb{Z}_2(\hat{n})$-asymmetry and when $A_n=1$, it must be top-of-the-order instead.  Also recall that projecting $\mathtt{JointRealize}(A_x,A_y,A_z)$ onto the $A_x$--$A_y$, $A_y$--$A_z$, or $A_z$--$A_x$ plane always includes the point ${0,0}$. This suggests that if we only consider the $\mathbb{Z}_2(\hat{n})$-asymmetry properties for two orthogonal axes, a state can simultaneously be bottom-of-the-order for both axes or top-of-the-order for both axes.  The geometric account of this fact is provided in \cref{fig_XYextreme}.

However, when we consider all three axes $\hat{x}$, $\hat{y}$ and $\hat{z}$, the situation changes. While a state can be bottom-of-the-order for $\mathbb{Z}_2(\hat{n})$-asymmetry relative to all three axes simultaneously (since $A_x=A_y=A_z=0$ is realizable), it is impossible for a state to be top-of-the-order for all three axes at the same time (since $A_x=A_y=A_z=1$ is not realizable).

\begin{figure}[h]
	\centering
	\includegraphics[width=0.3\textwidth]{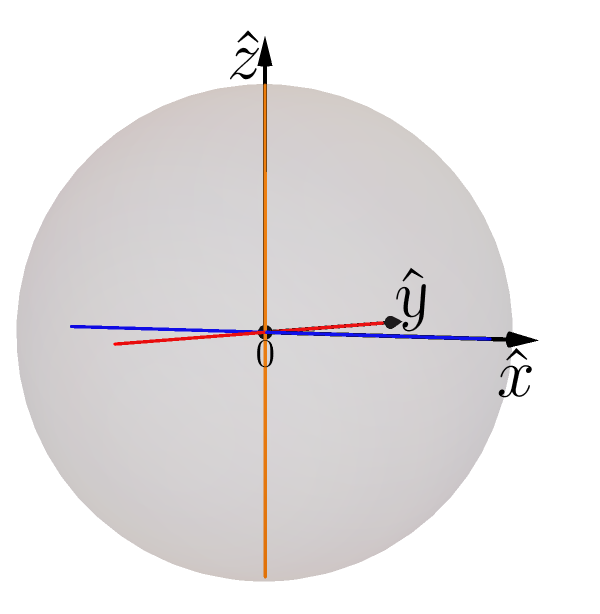}
	\hspace{0.2cm}
	\includegraphics[width=0.3\textwidth]{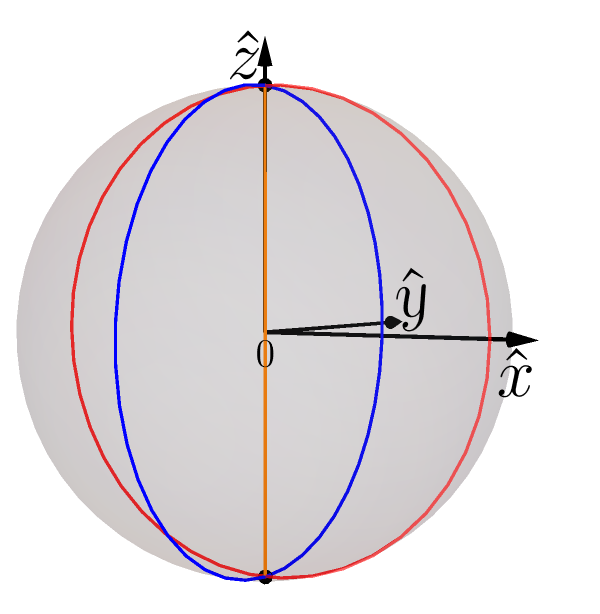}
	\caption{Left: 
	The intersection of the set of free states in the resource theory of $\mathbb{Z}_2(\hat{x})$-asymmetry (represented by the blue line) and that for $\mathbb{Z}_2(\hat{y})$-asymmetry (represented by the red line) is the maximally mixed state. The only equivalence class for $\mathbb{Z}_2(\hat{z})$-asymmetry that intersects with the maximally mixed state is the set of free states for $\mathbb{Z}_2(\hat{z})$-asymmetry (represented by the orange line). 
	Right: the intersection of the equivalence class at the top of the order for $\mathbb{Z}_2(\hat{x})$-asymmetry (represented by the blue circle) and that for $\mathbb{Z}_2(\hat{y})$-asymmetry (represented by the red circle) is the pair of points $\{\ket{+\hat{z}},\ket{-\hat{z}}\}$. The only equivalence class for $\mathbb{Z}_2(\hat{z})$-asymmetry that intersects this pair of points is the set of free states for $\mathbb{Z}_2(\hat{z})$-asymmetry (represented by the orange line). 
	}
	\label{fig_XYZextreme}
\end{figure}

Geometrically, as shown in \cref{fig_XYZextreme}, no matter if the state is simultaneously bottom-of-the-order or simultaneously top-of-the-order for $\hat{x}$ and $\hat{y}$, it always must be at the bottom of the resource order for $\mathbb{Z}_2(\hat{z})$-asymmetry.

\subsection{Dependence relations among $B_x$, $B_y$ and $B_z$}
	\label{sec_ineq_B}

	Since the equalities of \cref{eq_equality} do not impose any constraints on $B_x$, $B_y$, $B_z$ when $A_x$, $A_y$ and $A_z$ are unconstrained, the dependence relations among $B_x$, $B_y$ and $B_z$ are fully characterized by inequality constraints. 
	
	There are two ways to derive the inequality constraints on $B_x$, $B_y$ and $B_z$. The first one mirrors the approach used for $A_x$, $A_y$ and $A_z$ in \cref{sec_ineq_A}. That is, to use the constraints on the Bloch vector components, i.e., \cref{eq_rspositive,eq_r2range}, and their expressions in terms of $B_x$, $B_y$, $B_z$, i.e., \cref{eq_rintermsofB} for impure states and \cref{eq_Bpure} for pure states. The second way is to use \cref{eq_equality}, the equalities on $A_x$, $B_x$, $A_y$, $B_y$, $A_z$ and $B_z$, i.e., on $\mathcal{M}$, to express $A_x$, $A_y$, $A_z$ in terms of $B_x$, $B_y$, $B_z$, and then use these expressions to convert the generating set of inequalities on $A_x$, $A_y$ and $A_z$ (described in \cref{eq_inequalityA0}) into inequalities on $B_x$, $B_y$ and $B_z$. Here we present the former method. (We will present the second method when deriving inequalities on a different triple of monotones later in \cref{sec_AxBxAy}.) 

	In \cref{eq_Bpure}, we already listed all tuples of values of $B_x$, $B_y$ and $B_z$ that are jointly realizable when the Bloch vector radius $r$ is 1. 
	For now, let us derive the inequalities on $B_x$, $B_y$ and $B_z$ for impure states, i.e., assuming $r < 1$.

	For impure states, by the definition of the $B$-monotone, we have
	\begin{equation}\label{eq_Brange}
		B_x(\rho), B_y(\rho), B_z(\rho) \in [0,1).
	\end{equation}

	The nonnegativity conditions on the Bloch vector components in \cref{eq_rspositive} and their expressions in terms of $B_x$, $B_y$ and $B_z$ in \cref{eq_rintermsofB} gives
	 \begin{equation}
		\label{eq_inequalityB}
		\begin{split}
			-B_x^2(\rho) + B_y^2(\rho) + B_z^2(\rho) - 2 B_y^2(\rho) B_z^2(\rho) + B_x^2(\rho) B_y^2(\rho) B_z^2(\rho) \ge 0,  \\
			B_x^2(\rho) - B_y^2(\rho) + B_z^2(\rho) - 2 B_z^2(\rho) B_x^2(\rho) + B_x^2(\rho) B_y^2(\rho) B_z^2(\rho) \ge 0, \\
			B_x^2(\rho) + B_y^2(\rho) - B_z^2(\rho) - 2 B_x^2(\rho) B_y^2(\rho) + B_x^2(\rho) B_y^2(\rho) B_z^2(\rho) \ge 0.
		\end{split}
	\end{equation}
	These inequalities are analogous to the three inequalities for $A_x$, $A_y$ and $A_z$ in \cref{eq_inequalityA}.

	In \cref{sec_ineq_A}, the condition on the Bloch vector components, namely, $r^2_x+r^2_y+r^2_z\leq 1$ (\cref{eq_r2range}), was shown to give rise to a fourth inequality constraint on $A_x$, $A_y$ and $A_z$, namely, \cref{eq_r2Aineq}. %, namely $A_x^2(\rho) + A_y^2(\rho) + A_z^2(\rho)  \le 2$. 
	However, in the case of $B_x$, $B_y$ and $B_z$, this condition gives no additional inequality constraint on $B_x$, $B_y$ and $B_z$. 
	This is because the inequalities in Eqs. \eqref{eq_inequalityB} and the range of $B_x$, $B_y$ and $B_z$ specified in \cref{eq_Brange} already ensure that \cref{eq_r2range} is satisfied.

\begin{figure}[h]
	    \centering
	    \includegraphics[width=0.95\textwidth]{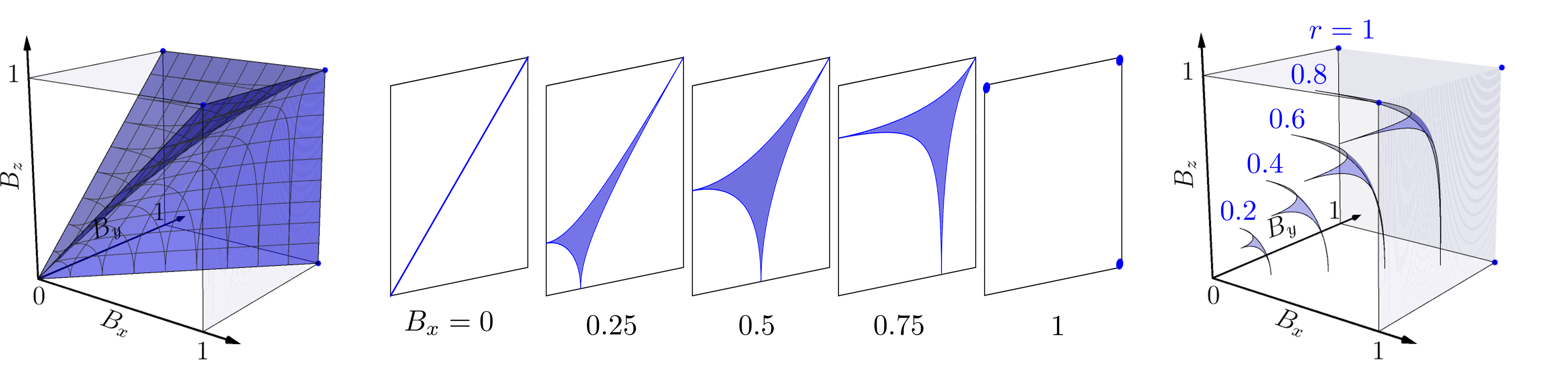}
	    \caption{Left: The region of jointly realizable values of $B_x$, $B_y$ and $B_z$; the mesh is added to indicate the curvature. 
	    	Middle: $B_y$--$B_z$ cross-sections of this region for five different values of $B_x$. 
	    	Right: The regions of jointly realizable values of  $B_x$, $B_y$ and $B_z$ given a fixed value of purity, represented by the Bloch vector radius $r$. }
	    \label{fig_qt3dB}
	\end{figure}

	Given that $B_x,B_y,B_z\geq0$ (as required by \cref{eq_Brange}), inequalities \eqref{eq_inequalityB} fully determine the set of jointly realizable values of $B_x$, $B_y$ and $B_z$ for an impure state.	This set, together with the set of jointly realizable values of $B_x$, $B_y$ and $B_z$ for a pure state, 
	 is depicted in the leftmost plot of \cref{fig_qt3dB}.

	The projection of $\mathtt{JointRealize}(B_x,B_y,B_z)$ (the blue region depicted in the left plot of \cref{fig_qt3dB}) into any of the three planes (the $B_x$--$B_y$ plane, the $B_y$--$B_z$ plane, or the $B_z$--$B_x$ plane) results in the union of $[0,1) \times [0,1)$ and the three points $(1,1)$ $(0,1)$ and $(1,0)$. Take the projection into the $B_y$--$B_z$ plane for example. We see that when no constraint is placed on $B_x$, any pair of values of $B_y$ and $B_z$ within the range given by \cref{eq_Brange} (which is for impure states) is jointly realizable, and if one of $B_y$ and $B_z$ is 1, the other must be either 0 or 1, due to the discreteness in $B$-monotones for pure states as mentioned earlier in \cref{sec_pure}.
	If, on the other hand, $B_x$ is fixed, then the set of jointly realizable values of $B_y$ and $B_z$ is more limited. 
	This is indicated by the $B_y$--$B_z$ cross-sections in the middle plot of \cref{fig_qt3dA}.
	By symmetry, the same conclusions hold for $B_x$ and $B_y$ and for $B_y$ or $B_z$.
	
	One can understand these conclusions using geometric analyses similar to what was done in \cref{sec_ineq_A}. The difference is that rather than considering intersections of the level sets for particular values of the $A$-monotones, which are cylinders (see \cref{fig_AxAy,fig_AxAyAz}), one must consider intersections of the level sets for particular values of the $B$-monotones, which are prolate spheroids. 

	We now analyze in more detail the $B_y$--$B_z$ cross-sections with respect to different values of $B_x$ as depicted in the middle plot of \cref{fig_qt3dB}.
	When $B_x= 0$, we have $B_y = B_z$, implying a synergy between $B_y$ and $B_z$. 
	This is analogous to what is observed for $A_y$ and $A_z$ when $A_x=0$, as noted in \cref{sec_ineq_A}.
	On the other hand, the cross-section for $B_x= 1$ is rather different from the one for $A_x= 1$ in \cref{fig_qt3dB}.
	This peculiar feature of $B_x$, $B_y$ and $B_z$ has been discussed in \cref{sec_pure}.

	In the nonextremal cases, i.e., whenever $\beta$, defined as $\beta\coloneqq B_x$, is an element of $(0,1)$, the cross-section has three boundaries. They are
	\begin{subequations}\label{eq_ByBzBounds}
		\begin{alignat}{4}
			\label{eq_ByBzBounds_a} 
			\text{bottom-left: } 		&& \frac{B_y^2(\rho) + B_z^2(\rho) - 2 B_y^2(\rho) B_z^2(\rho)}{1-B_y^2(\rho) B_z^2(\rho) } \ge \beta^2,  \\
			\label{eq_ByBzBounds_b} 
			\text{bottom-right: } 	&&  \frac{B_y^2(\rho) - B_z^2(\rho)}{1-2B_z^2(\rho)+B_y^2(\rho) B_z^2(\rho)  } \le \beta^2,  \\
			\label{eq_ByBzBounds_c} 
			\text{top-left: } 		&&  \frac{	-B_y^2(\rho) + B_z^2(\rho)	}{1-2B_y^2(\rho)+ B_y^2(\rho) B_z^2(\rho)}\le \beta^2,
		\end{alignat}
	\end{subequations}
	When $\beta=0$, the three boundaries boil down to a single line. 
	
	The shapes of the $B_y$--$B_z$ cross-sections in the middle three subfigures of \cref{fig_qt3dB} indicate that the dependence relation between $B_y$ and $B_z$ when $\beta\in(0,1)$ is neither strictly trade-off nor strictly synergy. If this dependence relation had been given by the equality constraint in \eqref{eq_ByBzBounds_a} (the bottom-left boundary of the blue region in the three middle subfigures of \cref{fig_qt3dB}) then it would have been a trade-off relation.  If it had been given by the equality constraint in \eqref{eq_ByBzBounds_b} or \eqref{eq_ByBzBounds_c} (the top-left and bottom-right boundaries of the blue region in the three middle subfigures of \cref{fig_qt3dB} respectively), then it would have been a synergy relation.  Though neither a trade-off nor a synergy relation, the dependence relation between $B_y$ and $B_z$ given $\beta$ describes a region of jointly realizable values of $B_y$ and $B_z$ that is bounded by these trade-off and synergy relations. 

	The right plot of \cref{fig_qt3dB} shows the decomposition of $\mathtt{JointRealize}(B_x,B_y,B_z)$ into slices that each correspond to a fixed purity. Slices with $r < 1$ share the same shape, whose size decreases as $r$ decreases. 
	This is because 
	\begin{equation}\label{eq_fixedPurity_B}
		\frac{1}{1-B_x^2} + \frac{1}{1-B_y^2} + \frac{1}{1-B_z^2} = \frac{3-r^2}{1-r^2},
	\end{equation}
	for a fixed Bloch vector radius $r \in [0,1)$ and one can check that the right-hand side decreases as $r$ decreases. See \cref{sec_fixedPurity} for the derivation. The pure state case has to be treated separately and yields the four solutions mentioned in \cref{eq_Bpure}.
	Although not immediate from the algebraic expression, \cref{eq_fixedPurity_B} represents a trade-off relation because for a fixed purity, if one of $B_x$, $B_y$ and $B_z$ increases, at least one of the other two has to decrease without any increase in the rest. This trade-off relation is also indicated by the shape of the fixed-purity slices.
	
	Furthermore, since in the resource theory of $\mathbb{Z}_2(\hat{n})$-asymmetry, $B_n$ fully characterizes the total order over impure states with a given Bloch vector radius $r<1$, the trade-off relation among $B_x$, $B_y$ and $B_z$ also implies the trade-off relations among the three corresponding kinds of $B$-asymmetry properties of an impure state.

\subsection{Dependence relations among $A_x$, $B_x$, and $A_y$}

\label{sec_AxBxAy}

Similar to the dependence relations among $A_x$, $A_y$, and $A_z$ (analyzed in \cref{sec_ineq_A}) or among $B_x$, $B_y$, $B_z$ (analyzed in \cref{sec_ineq_B}), the dependence relations among  $A_x$, $B_x$, and $A_y$ are fully characterized by inequality constraints, since the equalities of \cref{eq_equality} do not impose any constraints on $A_x$, $B_x$, $A_y$ when $B_y$, $A_z$ and $B_z$ are unconstrained.

We use the generating set of inequalities of \cref{eq_inequalityA,eq_r2Aineq} to derive inequalities on $A_x$, $B_x$ and $A_y$. To do so, we need to express $A_z$ in terms of $A_x$, $B_x$ and $A_y$ using the equalities of \cref{eq_equality}.

When $B_x>0$, from the equality constraint \cref{eq_1eq}, we have
\begin{equation}
\label{eq_axay0}
	A_z^2(\rho)=2-2\frac{A_x^2(\rho)}{B_x^2(\rho)}+A_x^2(\rho)-A_y^2(\rho).
\end{equation}
Substituting $A_z^2(\rho)$ with the above expression in the inequalities in \cref{eq_inequalityA,eq_r2Aineq}, we get a pair of inequalities that are joint-realizability constraints on $A_x$, $B_x$ and $A_y$ in addition to \cref{eq_BgeA}.
Recall that \cref{eq_BgeA} expresses the constraints on $A$- and $B$-monotones {\em within} a single resource theory of $\mathbb{Z}_2(\hat{n})$-asymmetry. 
This pair of inequalities is 
\begin{subequations}
	\label{eq_ineqAxBxAy}
\begin{align}
	%B_x^2(\rho)-A_x^2(\rho) & \ge 0, \\
	\label{eq_2AxBxAy}
	B_x^2(\rho)-A_x^2(\rho)+A_x^2(\rho)B_x^2(\rho)-A_y^2(\rho)B_x^2(\rho) & \ge0, \\
	\label{eq_3AxBxAy}
	-  B_x^2(\rho)+A_x^2(\rho)+A_y^2(\rho)B_x^2(\rho)& \ge 0.
	%\\B_x^2(\rho) &\le 1. 
\end{align}
\end{subequations}

When $B_x=0$, from \cref{eq_AxBx0}, we know that $A_x$=0, which means that $r_y=r_z=0$. Then, by the definition of $A_y$ and $A_z$ (\cref{eq_Ay,eq_Az}, respectively), we have 
\begin{align} 
\label{eq_azay}
	A_z(\rho)=A_y(\rho).
\end{align}
Substituting $A_z(\rho)$ with $A_y(\rho)$ and setting $A_x(\rho)=0$ in \cref{eq_inequalityA,eq_r2Aineq}, only gives $A_y^2\in [0,1]$, which is already implied by the definition of $A_y$, as indicated by \cref{eq_BgeA}. 

Thus, \cref{eq_ineqAxBxAy}, together with the constraints $B_x \geq A_x$ and ${A_x,B_x,A_y\in [0,1]}$  (see \cref{eq_BgeA}), completely characterize $\mathtt{JointRealize}(A_x,B_x,A_y)$, the region of jointly realizable values of $A_x$, $B_x$, and $A_y$. 
The depiction of $\mathtt{JointRealize}(A_x,B_x,A_y)$, has been provided in \cref{fig_AxBxAy}, whose key features and implications have been discussed in \cref{sec_interpret3way}.

Recall from \cref{sec_order} that \cref{fig_AxBxAy} shows {\em how} one moves upward in the partial order for $\mathbb{Z}_2(\hat{x})$-asymmetry properties, specifically, whether one does so by increasing the value of $A_x$ while fixing $B_x$ or by increasing the value of $B_x$ while fixing $A_x$, determines the kind of dependence relations one sees between $\mathbb{Z}_2(\hat{x})$-asymmetry properties and the  $\mathbb{Z}_2(\hat{y})$-asymmetry properties. Here, we provide a geometrical account in terms of the Bloch representations for the relation between the location of a state in the partial order of the resource theory of $\mathbb{Z}_2(\hat{x})$-asymmetry and that for $\mathbb{Z}_2(\hat{y})$-asymmetry. 

\begin{figure}[h]
		\centering
		\includegraphics[width=0.3\textwidth]{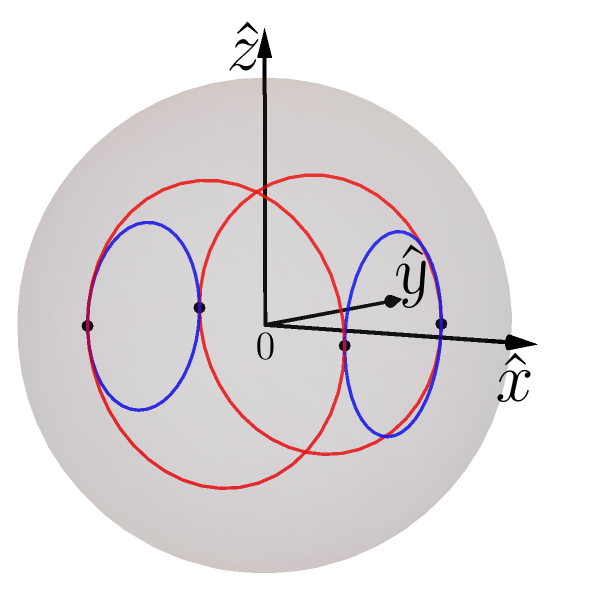}
		\hspace{0.2cm}
		\includegraphics[width=0.3\textwidth]{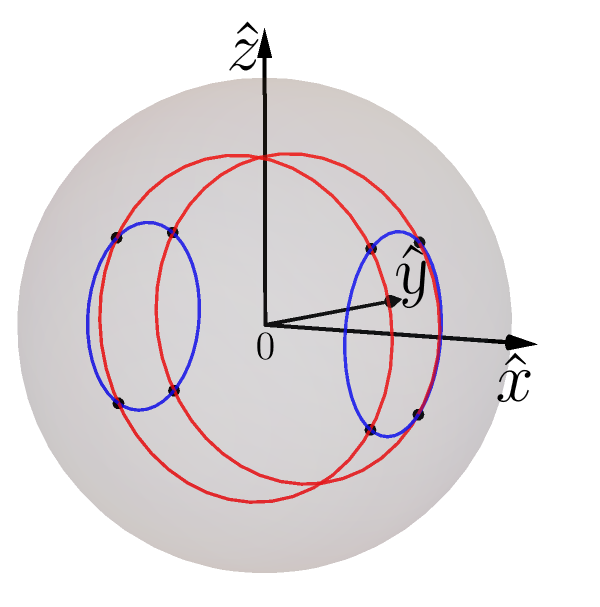}
		\hspace{0.2cm}
		\includegraphics[width=0.3\textwidth]{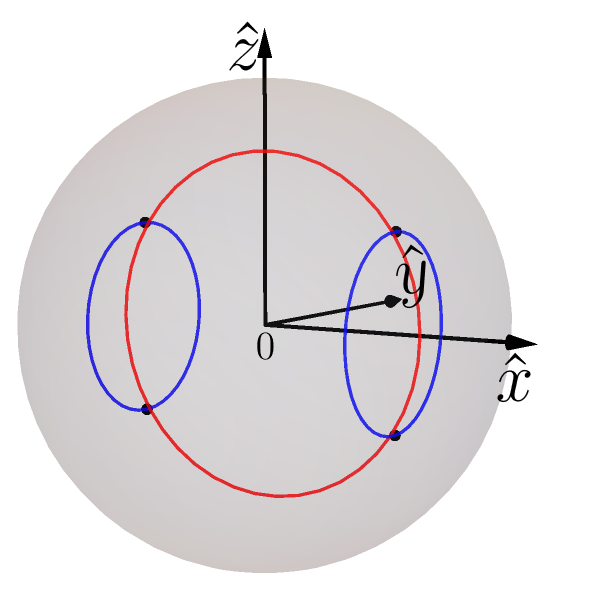}
		\caption{Three examples of equivalence classes (represented by pairs of red circles or a single red circle) in the resource theory of $\mathbb{Z}_2(\hat{y})$-asymmetry intersecting with the equivalence class determined by $A_x=0.4$ and $B_x=0.5$ (represented by a pair of blue circles) in the resource theory of $\mathbb{Z}_2(\hat{x})$-asymmetry. Left: $A_y=0.6$ and $B_y=\sqrt{\frac{3}{7}}\approx 0.65$. Middle: $A_y=\frac{2\sqrt{3}}{5}\approx 0.69$ and $B_y=\frac{1}{\sqrt{2}}\approx 0.71$. Right: $A_y=B_y=\frac{\sqrt{13}}{5}\approx 0.72$.}
		\label{fig_XYequi}
	\end{figure}

Fixing the location of a state in the partial order of the resource theory of $\mathbb{Z}_2(\hat{x})$-asymmetry means to specify its equivalence class in the resource theory of $\mathbb{Z}_2(\hat{x})$-asymmetry. As an illustrative example, we consider the $\mathbb{Z}_2(\hat{x})$-asymmetry properties associated to the equivalence class of states depicted in the middle plot in \cref{fig_eqclX}, i.e., when $A_x=0.4$ and $B_x=0.5$, and ask what $\mathbb{Z}_2(\hat{y})$-asymmetry properties are consistent with it. In \cref{fig_XYequi}, we depict our chosen example of an equivalence class relative to $\mathbb{Z}_2(\hat{x})$-asymmetry as a pair of blue circles, and we plot three examples of equivalence classes relative to $\mathbb{Z}_2(\hat{y})$-asymmetry, one for each of the three subfigures.  The latter are depicted as a pair of red circles (or as a single red circle). 

Geometrically, the constraint ensuring that some given $\mathbb{Z}_2(\hat{x})$-asymmetry properties are consistent with some given $\mathbb{Z}_2(\hat{y})$-asymmetry properties is that the associated pair of equivalence classes of states have a nontrivial intersection. The leftmost and rightmost examples in \cref{fig_XYequi} describe the equivalence classes of $\mathbb{Z}_2(\hat{y})$-asymmetry that are respectively lowest and highest in the partial order of $\mathbb{Z}_2(\hat{y})$-asymmetry while still being consistent with the given $\mathbb{Z}_2(\hat{x})$-asymmetry properties. These are the lowest and highest such equivalence classes because the corresponding red circles have, respectively, the smallest and largest radii while still intersecting the blue circles.  Recall that the cylindrical radius relative to the $\hat{y}$ axis is simply the value of $A_y$, which, as noted above, is sufficient for fixing the $\mathbb{Z}_2(\hat{y})$-asymmetry properties when the $\mathbb{Z}_2(\hat{x})$-asymmetry properties are fixed. The middle example depicts an equivalence class of $\mathbb{Z}_2(\hat{y})$-asymmetry intermediate between these. 

\begin{figure}[h]
	\centering
	\includegraphics[width=0.3\textwidth]{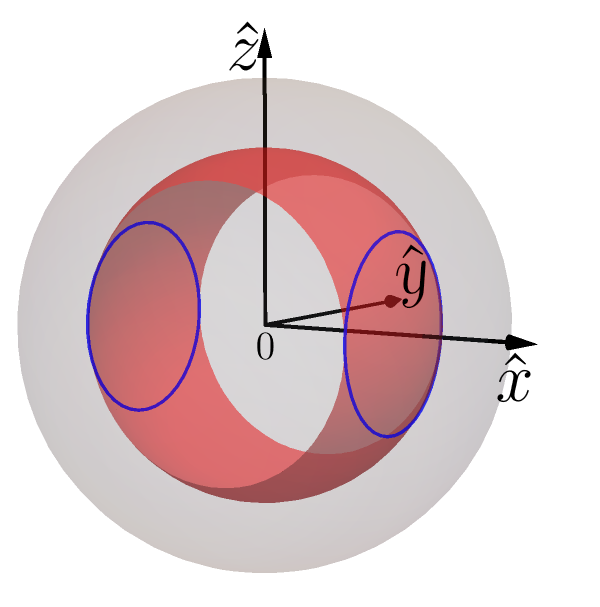}
	\hspace{0.2cm}
	\includegraphics[width=0.3\textwidth]{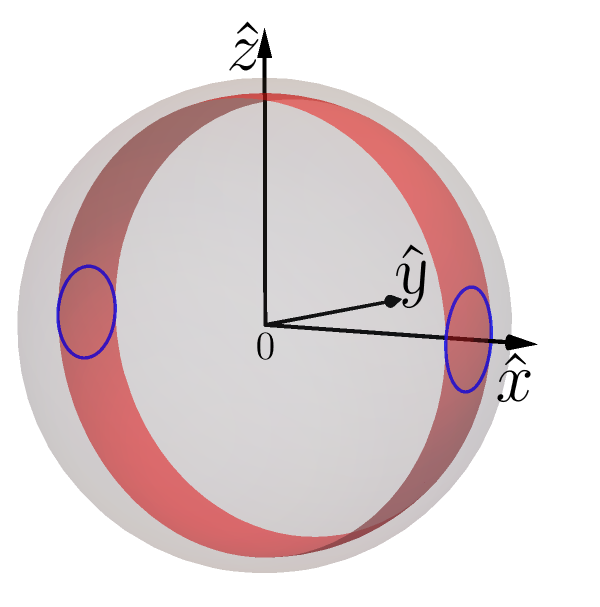}
	\hspace{0.2cm}
	\includegraphics[width=0.3\textwidth]{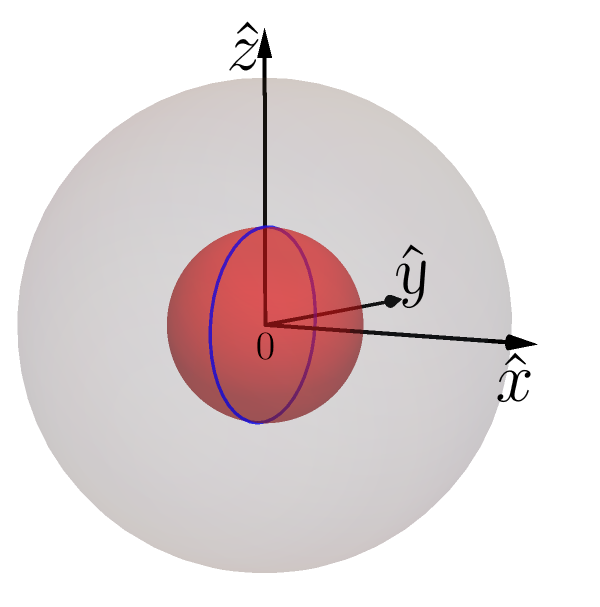}
	\caption{The range (represented by the red surface) of the $\mathbb{Z}_2(\hat{y})$-asymmetry properties that are consistent with different $\mathbb{Z}_2(\hat{x})$-asymmetry properties (represented by a blue circle or a pair of blue circles). Middle: Left: $A_x=0.4$ and $B_x=0.5$. Middle: $A_x=0.2$ and $B_x=0.5$. Right: $A_x=0.4$ and $B_x=0.4$.}
	\label{fig_XYrange}
\end{figure}

As the geometry makes clear, the range of $\mathbb{Z}_2(\hat{y})$-asymmetry properties that are possible for a given choice of $\mathbb{Z}_2(\hat{x})$-asymmetry properties is limited. The entire range of the equivalence classes in the $\mathbb{Z}_2(\hat{y})$-asymmetry resource theory that are consistent with our chosen example of $\mathbb{Z}_2(\hat{x})$-asymmetry equivalence class, i.e., when  $A_x=0.4$ and $B_x=0.5$, is depicted in the leftmost plot of \cref{fig_XYrange}, which is a sphere with radius $\sqrt{1-\frac{A_x^2}{B_x^2}+A_x^2}$ that is truncated at $y=\pm A_x$. The range when $B_x$ is kept the same but $A_x$ changes is depicted in the middle plot of \cref{fig_XYrange}, and the range when $A_x$ is kept the same but $B_x$ changes is depicted in the rightmost plot of \cref{fig_XYrange}. 

Recall from \cref{sec_interpret3way} the kind of trade-off relation between $A_x$ and $A_y$ for a given $B_x$ observed in the middle plot of \cref{fig_AxBxAy} (excluding the extremal cases of $B_x=0$ and $B_x=1$). In fact, when $B_x\in(0,1)$ and is fixed, we get a kind of trade-off relation between $\mathbb{Z}_2(\hat{x})$-asymmetry properties and $\mathbb{Z}_2(\hat{y})$-asymmetry properties, since a sufficiently big increase in $A_x$ necessarily results in the state moving downward in the partial order for $\mathbb{Z}_2(\hat{y})$-asymmetry properties.  The comparison between the middle plot with the leftmost plot of \cref{fig_XYrange} provides a concrete example of such a kind of trade-off relation. Specifically, when $B_x=0.5$,  the increase from $A_x=0.2$ to $A_x=0.4$ is big enough for the ranges of possible $\mathbb{Z}_2(\hat{y})$-asymmetry properties before and after to not overlap and in particular, the range after is strictly lower in the partial order for $\mathbb{Z}_2(\hat{y})$-asymmetry properties than the range before. Similarly, the comparison between the rightmost plot of \cref{fig_XYrange} with the leftmost plot provides a concrete example for the kind of synergy relation between $\mathbb{Z}_2(\hat{x})$-asymmetry and $\mathbb{Z}_2(\hat{y})$-asymmetry properties observed when $A_x$ is fixed instead.
Specifically, when $A_x=0.4$, the increase from $B_x=0.4$ to $B_x=0.5$ is big enough for the ranges of possible $\mathbb{Z}_2(\hat{y})$-asymmetry properties before and after to not overlap and in particular, the range after is strictly higher in the partial order for $\mathbb{Z}_2(\hat{y})$-asymmetry properties than the range before.

\subsection{Dependence relations among $A_x$, $B_x$, $A_y$ and $A_z$}

Previously, when we only considered dependence relations among three monotones, drawn from $\mathcal{M}$, i.e., from $A_x$, $B_x$, $A_y$, $B_y$, $A_z$ and $B_z$, the dependence relations were always fully characterized by inequalities. However, if we consider more than three monotones from $\mathcal{M}$, equality constraints will appear even if we are not fixing the values of the rest. This is because, as explained in \cref{sec_eq0}, there are only three degrees of freedom in the state space of a qubit.
Here, we will investigate the dependence relations among four monotones, namely $A_x$, $B_x$, $A_y$ and $A_z$.

In \cref{sec_AxBxAy}, we have in fact already noted the equality constraints on $A_x$, $B_x$, $A_y$ and $A_z$,  namely, \cref{eq_1eq}, which is one of the three equalities of \eqref{eq_equality}, i.e., the set of equality constraints on $\mathcal{M}$.

In terms of inequality constraints, both the generating set of inequalities, i.e., \cref{eq_inequalityA0}, the inequalities on $A_x$, $A_y$, $A_z$, and the set of inequalities on $A_x$, $B_x$, $A_y$, i.e., \cref{eq_ineqAxBxAy}, hold. As demonstrated in \cref{sec_AxBxAy}, the former set can be converted to the latter set via the equalities of \cref{eq_axay0,eq_azay}. That is, the generating set of inequality constraints of \cref{eq_inequalityA0}, together with the equality of \cref{eq_1eq}, completely characterizes the dependence relations among $A_x$, $B_x$, $A_y$ and $A_z$.

Now we seek to gain a conceptual understanding of the dependence relations among $A_x$, $B_x$, $A_y$ and $A_z$. Recall that $A_x$ and $B_x$ completely characterize all $\mathbb{Z}_2(\hat{x})$-asymmetry properties of a state, and that as mentioned in \cref{sec_order}, when $A_x$ and $B_x$ are fixed, $A_y$ alone characterizes all of its $\mathbb{Z}_2(\hat{y})$-asymmetry properties while $A_z$ alone characterizes all of its $\mathbb{Z}_2(\hat{z})$-asymmetry properties. 

The equality constraints of \cref{eq_equality} tell us that once a state's properties for both $\mathbb{Z}_2(\hat{x})$-asymmetry and $\mathbb{Z}_2(\hat{y})$-asymmetry are fixed, there is no more freedom in its $\mathbb{Z}_2(\hat{z})$-asymmetry properties. This is because, once the values for $A_x$, $B_x$ and $A_y$ are known, we can derive a unique value for $A_z$ using \cref{eq_equality} under the condition that $A_z\geq 0$. 

\begin{figure}[h]
	\centering
	\includegraphics[width=0.3\textwidth]{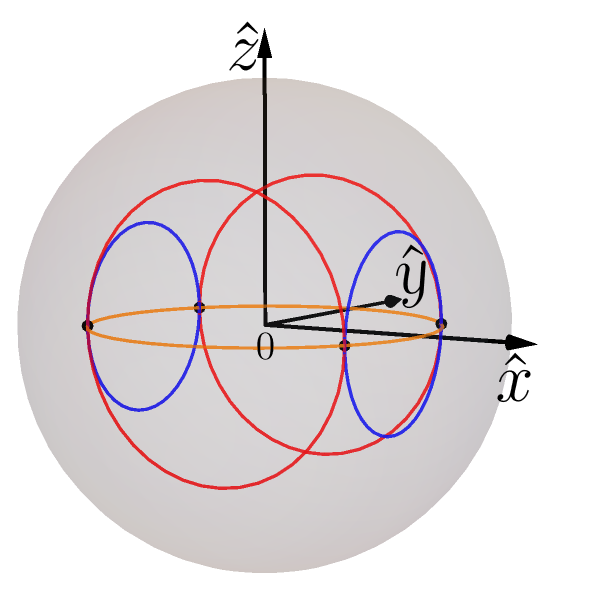}
	\hspace{0.2cm}
	\includegraphics[width=0.3\textwidth]{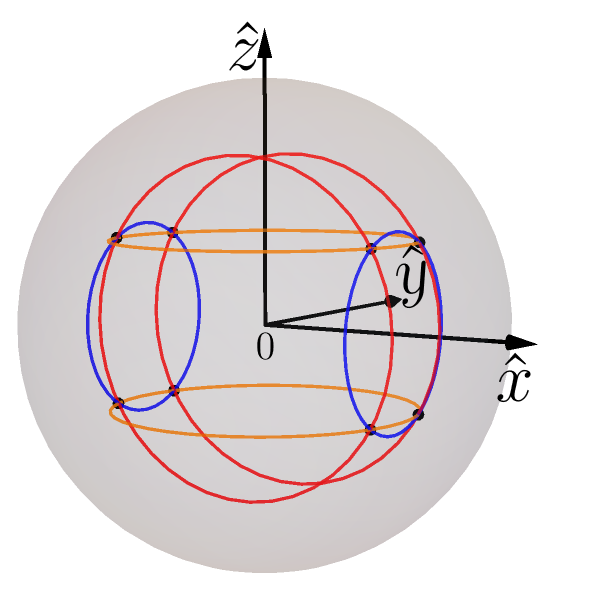}
	\hspace{0.2cm}
	\includegraphics[width=0.3\textwidth]{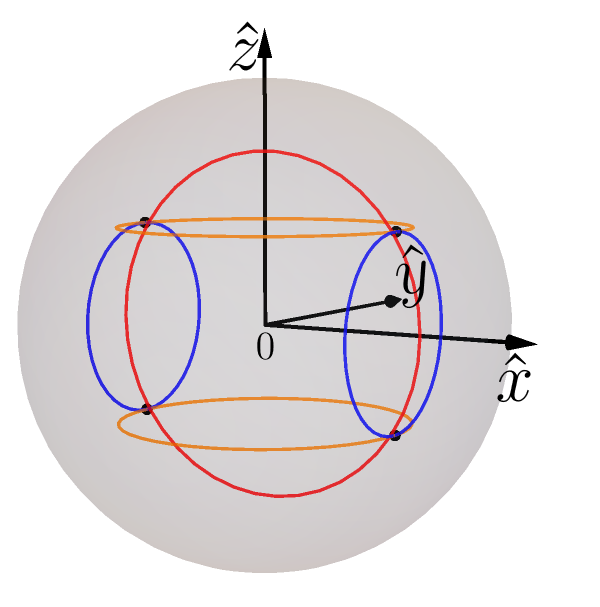}
	\caption{In each subfigure, the yellow circles represent an equivalence class in the resource theory of $\mathbb{Z}_2(\hat{z})$-asymmetry picked out by the black points in the respective plot in \cref{fig_XYequi}, which is the intersection of an equivalence class in the resource theory of $\mathbb{Z}_2(\hat{x})$-asymmetry and one in the resource theory of $\mathbb{Z}_2(\hat{y})$-asymmetry.}
	\label{fig_XYZequi}
\end{figure}

We consider again the three examples introduced in \cref{fig_XYequi}.   For each, we depict in \cref{fig_XYZequi}  the equivalence class of $\mathbb{Z}_2(\hat{z})$-asymmetry fixed by the corresponding intersection between the equivalence class of $\mathbb{Z}_2(\hat{x})$-asymmetry and that of $\mathbb{Z}_2(\hat{y})$-asymmetry.

To quantitatively understand how fixing the location of a state in the resource order for $\mathbb{Z}_2(\hat{x})$-asymmetry constraints its location in the resource order for $\mathbb{Z}_2(\hat{y})$-asymmetry and that for $\mathbb{Z}_2(\hat{z})$-asymmetry, it is sufficient to focus on how fixing $A_x=\alpha$ and $B_x=\beta$ constrains $A_y$ and $A_z$. From \cref{eq_axay0,eq_azay}, we immediately have %when $\alpha=\beta=0$,
\begin{align}
	\label{eq_ab1}
	\text{if } \alpha=\beta=0, \, &\text{then } A_y(\rho)=A_z(\rho) ;\\
	\label{eq_ab2}
	\text{if } \alpha\in(0,1] \text{ and } \beta\in(0,1], \,  &\text{then } A_z^2 + A_y^2 = 2+\alpha^2-\frac{2\alpha^2}{\beta^2}.
\end{align}
That is, when a state is at the bottom of the resource order for $\mathbb{Z}_2(\hat{x})$-asymmetry, its locations in the resource order for $\mathbb{Z}_2(\hat{y})$-asymmetry and that for $\mathbb{Z}_2(\hat{z})$-asymmetry must be the same, indicating a synergy relation between its $\mathbb{Z}_2(\hat{y})$-asymmetry properties and $\mathbb{Z}_2(\hat{z})$-asymmetry properties. In contrast, when a state is not at the bottom of the order for $\mathbb{Z}_2(\hat{x})$-asymmetry, its $\mathbb{Z}_2(\hat{y})$-asymmetry properties and $\mathbb{Z}_2(\hat{z})$-asymmetry properties strictly trade-off against each other. Such a trade-off relation can also be seen geometrically. In \cref{fig_XYZequi}, when the pair of blue circles representing $\mathbb{Z}_2(\hat{x})$-asymmetry properties is fixed, from the leftmost plot to the rightmost plot, the radius of the red circles, which is $A_y$,  increases (and therefore it moves up in the resource order for $\mathbb{Z}_2(\hat{y})$-asymmetry), while the radius of the orange circles, which is $A_y$,  decreases (and therefore it moves down in the resource order for $\mathbb{Z}_2(\hat{z})$-asymmetry).

\section{Discussion}
\label{sec_discussion}

In this work, we laid out the foundational framework and provided a step-by-step recipe for investigating resource dependence relations, namely, the relations among nonfreeness properties defined by different sets of free operations. We demonstrated the application of this framework and the recipe through a simple example involving the resource theories of about-face asymmetry for three mutually orthogonal axes, $\hat{x}$, $\hat{y}$, and $\hat{z}$. In this example, we derived a complete set of monotones for $\mathbb{Z}_2(\hat{n})$-asymmetry properties and, as a result, we succeeded in fully characterizing the dependence relations among $\mathbb{Z}_2(\hat{x})$-, $\mathbb{Z}_2(\hat{y})$-, and $\mathbb{Z}_2(\hat{z})$-asymmetry properties.

Our simple example illustrates that even in seemingly simple scenarios, resource dependence relations can be complex and nuanced. Specifically, as observed, it is uncommon to encounter synergy or trade-off relations; more often, the dependence relations involve mixtures of synergies and trade-offs, or are bounded by synergy and (or) trade-off relations.

As noted in \cref{sec_summary}, the search for resource dependence relations can be pursued even in cases where complete sets of monotones are not known, and our simple example offers indications for the kinds of answers one may obtain when addressing questions about resource dependence relations in those situations.

The study of quantum resource theories is often connected with studies of nonclassical phenomena. One natural direction for future research is to characterize what is nonclassical about resource dependence relations. This boils down to identifying aspects of the phenomenology of resource dependence relations that are not classically explainable. Our preferred definition of classical explainability of a given phenomenology is whether it can be reproduced by a generalized noncontextual ontological model~\cite{Spekkens2005contextuality,Catani2023trap}. This definition can be motivated by a version of Leibniz's principle of the identity of indiscernibles~\cite{spekkens2019ontological} and encapsulates many traditional notions of classicality such as the existence of a locally causal model~\cite{Bell_1976} and the existence of a non-negative quasiprobability representation~\cite{Spekkens2008Negativity}. 
In a future work~\cite{forthcoming}, we will present examples of aspects of resource dependence relations that witness nonclassicality in this sense.

\section*{Acknowledgements}
We thank Iman Marvian, Shuming Cheng and Rafael Wagner for useful discussions. 
YY thanks Kamil Korzekwa for hosting her at the Jagiellonian University and for useful discussions. 
YY and RWS are supported by Perimeter Institute for Theoretical Physics. 
Research at Perimeter Institute is supported in part by the Government of Canada through the Department of Innovation, Science and Economic Development and by the Province of Ontario through the Ministry of Colleges and Universities. 
YY is also supported by the Natural Sciences and Engineering Research Council of Canada (Grant No. RGPIN-2024-04419). 
TG is supported by the Austrian Science Fund (FWF) via the START Prize Y1261-N.

\appendices
\crefalias{section}{appendix}

\section{The preorder of qubit states for the resource theory of $\mathbb{Z}_2(\hat{x})$-asymmetry}
\label{app:qubitConversion}

		Here we derive necessary and sufficient conditions for deterministic conversion between states in the resource theory of $\mathbb{Z}_2(\hat{x})$-asymmetry.
		These appear as \cref{lem:monotones}.
		First, however, let us derive a convenient characterization of $\mathbb{Z}_2(\hat{x})$-covariant maps in the form of \cref{thm:Z2-cov_maps}.

	\subsection{$\mathbb{Z}_2(\hat{x})$-covariant CPTP maps}
		
		It is convenient to work in the Bloch representation. 
		This is a 4-dimensional real-valued vector representation of operators relative to the operator basis $\{\mathbb{I},\sigma_x, \sigma_y, \sigma_z\}$, where $\sigma_x$, $\sigma_y$, and $\sigma_z$ are Pauli operators, which we also denote jointly as the three-component tuple $\vec{\sigma}$. 
		
		In particular, a density operator $\rho$ is represented by the 4-dimensional real-valued vector $(1,\vec{r})$, where $\vec{r} = (r_x,r_y,r_z)$ is the vector defined uniquely via $\rho = \frac{1}{2} ( \mathbb{I}+\vec{r}\cdot\vec{\sigma})$. Linear maps of density operators are consequently represented as $4 \times 4$ real matrices. 
		Specifically, a superoperator $\mathcal{T}$ is represented as 
		\begin{equation}\label{eq_map_blochrep}
			\bm{T} \left( \vec{t},\bm{M} \right) \coloneqq 
			    \begin{bmatrix}
					1 & \vec{0} \\
					\Vec{t} & \bm{M}
				\end{bmatrix}
			    =
			    \begin{bmatrix}
					1 & 0&0&0 \\
					t_x & M_{{x}{x}} & M_{{x}{y}}&M_{{x}{z}} \\
					t_y&M_{{y}{x}}& M_{{y}{y}}&M_{{y}{z}}\\
					t_z&M_{{z}{x}}& M_{{z}{y}}&M_{{z}{z}}\\
				\end{bmatrix},
		\end{equation}
		where $\bm{M}$ is a $3 \times 3$ real matrix while $\Vec{t}$ and $\Vec{0}$ are real column and row 3-vectors respectively.
		The Bloch vector representation of the action of a  map $\mathcal{T}$ on density operator $\rho$, i.e., $\rho \mapsto \mathcal{T}(\rho)$, then takes the form
		\begin{equation}\label{eq_Blochspheremap}
			\vec{r} \; \mapsto \; \bm{M} \vec{r} + \vec{t}.
		\end{equation}
		Note that not every matrix $\bm{T}$ in the form of \cref{eq_map_blochrep} gives rise to a completely positive trace-preserving (CPTP) map in this way. 
		Necessary and sufficient conditions for a mapping in the form of \eqref{eq_Blochspheremap} to be a CPTP map are derived in \cite{beth_ruskai_analysis_2002}.

		\begin{notation}\label{def:subscript_notation}
			In order to simplify the upcoming discussion, let us introduce notation that refers to Bloch representations of specific kinds of CPTP maps.
			\begin{itemize}
				\item \emph{Rotation:}
					For any rotation $\bm{R} \in \mathrm{SO}(3)$, we write
					\begin{equation}\label{eq_T_rot}
						\bm{T}_{\text{R}}(\bm{R}) \coloneqq \bm{T}(\vec{0},\bm{R}),
					\end{equation}
					and given a unit vector $\hat{n}$ and an angle $\theta$, we denote the rotation about the $\hat{n}$ axis by $\theta$ as
					\begin{equation}\label{eq_T_rot2}
						\bm{T}_{\text{R}_{n}} (\theta) \coloneqq \bm{T}_{\text{R}} \bigl( \bm{R}_{n}(\theta) \bigr),
					\end{equation}
					where $\bm{R}_{n}(\theta)$ is the corresponding $\mathrm{SO}(3)$ representation of the rotation. 
					Every such $\bm{T}_{\text{R}_{n}}(\theta)$ is a CPTP map.
				
				\item \emph{Translation-scaling:}
					If the matrix $\bm{M}$ from \eqref{eq_map_blochrep} is a diagonal matrix denoted $\bm{\Lambda}$, we write 
					\begin{equation}\label{eq_TD}
						\bm{T}_{\rm D}(\vec{t},\bm{\Lambda}) \coloneqq \bm{T}(\vec{t},\bm{\Lambda})  
							=
					    		\begin{bmatrix}
								1 & 0&0&0 \\
								t_1 & \lambda_x & 0&0 \\
								t_2 &0& \lambda_y & 0 \\
								t_3 &0& 0 & \lambda_z \\
							\end{bmatrix}.
					\end{equation}	
					to highlight the additional property that $\bm{\Lambda}$ is diagonal.
					The map $\bm{T}_{\rm D}(\vec{t},\bm{\Lambda})$ represents the action of an arbitrary translation by $\vec{t}$ together with a particular scaling. 
					The scaling factors are $\lambda_x$, $\lambda_y$ and $\lambda_z$ and act in $\hat{x}$, $\hat{y}$ and $\hat{z}$ directions respectively. 
					Note that not every matrix $\bm{T}_{\rm D}(\vec{t},\bm{\Lambda})$ acts as a CPTP map. 
					Once again, necessary and sufficient conditions are derived in \cite{beth_ruskai_analysis_2002} and we will use them later.
				
				\item \emph{Translation:}
					For a $\bm{T}_{\rm D}(\vec{t},\bm{\Lambda})$ with trivial scaling by the identity matrix $\idmatrix$, we write
					\begin{equation}
						\bm{T}_{\rm T} \left( \vec{t} \right) \coloneqq \bm{T}_{\rm D} \left( \vec{t}, \idmatrix \right)
					\end{equation}
					for the Bloch representation of the resulting translation map.
				
				\item \emph{Generic scaling:}
					For a symmetric matrix $\bm{S}$, we write
					\begin{equation}
						\bm{T}_{\rm S}(\bm{S}) \coloneqq \bm{T}(\vec{0}, \bm{S})
					\end{equation}
					for the matrix representing a scaling operation.
					Its scaling factors are given by the eigenvalues of $\bm{S}$ and the direction of scaling is along its eigenvectors.
					For instance, $\bm{T}_{\rm S} \bigl( \bm{R}_2^{-1}\bm{\Lambda}\bm{R}_2 \bigr)$ appearing in \cref{eq_TRS} below describes a scaling by $\lambda_x$, $\lambda_y$, $\lambda_z$ in $\bm{R}_2^{-1}\hat{x}$, $\bm{R}_2^{-1}\hat{y}$, $\bm{R}_2^{-1}\hat{z}$ directions respectively.
					Whenever an eigenvalue is negative, the scaling includes a reflection about the respective axis. 
			\end{itemize}
		\end{notation}
		
		In \cref{thm:Z2-cov_maps} below, we characterize the Bloch representation of an arbitrary free operation in the resource theory of $\mathbb{Z}_2(\hat{x})$-asymmetry. 
		Before stating the result, let us examine how imposing $\mathbb{Z}_2(\hat{x})$-covariance affects the viable forms of $\bm{T}(\vec{t},\bm{M})$.
		
		The only nontrivial element of $\mathbb{Z}_2(\hat{x})$ acts as ${\rho \mapsto \sigma_x \rho \sigma_x}$.
		Its Bloch representation is 
		 \begin{equation}\label{eq_aboutfacearoundx}
		 	\bm{T}_{\text{R}_x}(\pi) = \bm{T}_{\rm D} \bigl( \vec{0},\mathrm{diag}(1,-1,-1) \bigr).
		\end{equation}
		It follows that a map $\mathcal{T}$ is $\mathbb{Z}_2(\hat{x})$-covariant if and only if its Bloch representation $\bm{T}(\vec{t},\bm{M})$ commutes with the matrix $\bm{T}_{\text{R}_x}(\pi)$. 
		As such, we must have
		\begin{equation}\label{eq_Tx}
			\bm{T}(\vec{t},\bm{M}) = 
		   		\begin{bmatrix}
					1 & 0 & 0 & 0 \\
					t_x & M_{{x}{x}} & 0 & 0 \\
					0 & 0 & M_{{y}{y}} & M_{{y}{z}}\\
					0 & 0 & M_{{z}{y}} & M_{{z}{z}}\\
				\end{bmatrix}.			
		\end{equation}
		
		For instance, a rotation as in \eqref{eq_T_rot} is a free operation if and only if it is either
		\begin{enumerate}
			\item an arbitrary rotation about the $\hat{x}$ axis:
				\begin{equation}\label{eq_rotationaboutx}
					\bm{T}_{\text{R}_x}(\theta) = 
					\begin{bmatrix}
						1 & 0 & 0 &0 \\
						0 & 1 & 0 & 0\\
						0 & 0 & \cos{\theta} & \sin{\theta}\\
						 0 & 0 & -\sin{\theta} &\cos{\theta} 
					\end{bmatrix},
				\end{equation}				
			\item a $\pi$ rotation about the $\hat{y}$ axis: 
			\begin{equation}\label{eq_rotationabouty}
				\bm{T}_{\text{R}_y}(\pi) = \bm{T}_{\rm D} \bigl( \vec{0},\mathrm{diag}(-1,1,-1) \bigr),
			\end{equation}
			\item a $\pi$ rotation about the $\hat{z}$ axis: 
			\begin{equation}\label{eq_rotationaboutz}
				\bm{T}_{\text{R}_z}(\pi) = \bm{T}_{\rm D} \bigl( \vec{0},\mathrm{diag}(-1,-1,1) \bigr),
			\end{equation}
		\end{enumerate}
		or their composition.
		
		On the other hand, a translation-scaling map $\bm{T}_{\rm D}(\vec{t},\bm{\Lambda})$ as in \eqref{eq_TD} is $\mathbb{Z}_2(\hat{x})$-covariant if and only if the translation is along the $\hat{x}$ axis. Thus, it must satisfy $\vec{t} = [t_x,0,0]^T$.
		We denote the set of all such transformations, which are also CPTP maps, by
		\begin{equation}\label{eq_Df}
			{\rm D_f} \coloneqq \Set{  \bm{T}_{\rm D}(\vec{t},\bm{\Lambda}) \given \vec{t} = [t_x,0,0]^T \text{ and } \bm{T}_{\rm D}(\vec{t},\bm{\Lambda}) \text{ is a CPTP map}}.
		\end{equation}
		
		\begin{theorem}\label{thm:Z2-cov_maps}[Characterization of $\mathbb{Z}_2(\hat{x})$-covariant maps]
			A qubit-to-qubit map is $\mathbb{Z}_2(\hat{x})$-covariant if and only if it can be written as the product
			\begin{equation}\label{eq_Z2-cov_maps}
				\bm{T}_{{\rm R}_x}(\theta_1) \, \bm{T}_{\rm D_f} \, \bm{T}_{{\rm R}_x}(\theta_2)
			\end{equation}
			of three $\mathbb{Z}_2(\hat{x})$-covariant operations, where $\bm{T}_{{\rm R}_x}(\theta_i)$ are rotations about the $\hat{x}$-axis and $\bm{T}_{\rm D_f}$ is an element of $\afcov$.
			
			Furthermore, the set $\afcov$ defined in \eqref{eq_Df} coincides with the convex hull of the set $\afcovext$ whose elements have Bloch representation
			\begin{align}\label{eq_afcovext}
%				\bm{T}_{0}(u,v)=
				\begin{bmatrix}
					1 & 0 & 0 &0 \\
					\sin{u}\sin{v} & \cos{u}\cos{v} &0&0\\
					0 & 0 & \cos{u} &0\\
					 0 & 0 & 0 &\cos{v} 
				\end{bmatrix}
			\end{align}
			for some $u\in[0,2\pi)$ and $v\in[0,\pi)$.
		\end{theorem}

		\begin{proof}%[Proof of \cref{thm:Z2-cov_maps}]
		By \cite[section 2.1]{king2001minimal}, a generic trace-preserving map $\bm{T}$ can be decomposed as a composite of two rotations with a translation-scaling map in between. Thus, it can be written as
		\begin{equation}\label{eq_decomp}
			\bm{T} = \bm{T}_{\text{R}}(\bm{R}_1) \, \bm{T}_{\text D}(\vec{t},\bm{\Lambda}) \, \bm{T}_{\text{R}}(\bm{R}_2).
		\end{equation}	
		By expanding \cref{eq_decomp} we can also write it as a scaling operation followed by a rotation and a translation:
		\begin{equation} \label{eq_TRS}
			\begin{split}
				\bm{T} &= 
					\begin{bmatrix}
						1 & \vec{0} \\
						\bm{R}_1\bm{R}_2^{-1} \vec{t} & \bm{R}_1\bm{R}_2^{-1} \bm{\Lambda}\bm{R}_2
					\end{bmatrix} \\
					&= \bm{T}_{\rm T} \bigl( \bm{R}_1\bm{R}_2^{-1} \vec{t} \, \bigr)  \, \bm{T}_{\text{R}} \bigl( \bm{R}_1 \bigr)  \, \bm{T}_{\rm S} \bigl( \bm{R}_2^{-1}\bm{\Lambda}\bm{R}_2 \bigr) .
			\end{split}
		\end{equation}
		If $\bm{T}$ is $\mathbb{Z}_2(\hat{x})$-covariant, i.e., if it is of the form as in \cref{eq_Tx}, we know that 
		\begin{enumerate}
			\item $\bm{R}_1\bm{R}_2^{-1}\vec{t}$ must be of the form $[t_x, 0, 0]^T$, and that
			\item $\bm{R}_1\bm{R}_2^{-1}\bm{\Lambda}\bm{R}_2$ must be of the form $\left[\begin{smallmatrix}
				T'_{{x}{x}} & 0&0 \\
				0& T'_{{y}{y}}&T'_{{y}{z}}\\
				0& T'_{{z}{y}}&T'_{{z}{z}}\\
			\end{smallmatrix}\right]$.
		\end{enumerate}
		Consequently, both matrices
		\begin{equation}\label{eq_TtR1R2}
			\bm{T}_{\rm T} \bigl( \bm{R}_1\bm{R}_2^{-1}\vec{t} \, \bigr) 
				=
			\begin{bmatrix}
				1 & \vec{0} \\
				\bm{R}_1\bm{R}_2^{-1}\vec{t} & \idmatrix
			\end{bmatrix}
		\end{equation} 
		and 
		\begin{equation}\label{eq_TRTS}
			\bm{T}_{\text{RS}} \coloneqq \bm{T}_{\text{R}} \bigl( \bm{R}_1 \bigr) \, \bm{T}_{\rm S} \bigl( \bm{S} \bigr)
			=\begin{bmatrix}
				1 & \vec{0} \\
				\vec{0} & \bm{R}_1 \bm{S}
			\end{bmatrix}
		\end{equation} 
		with the notation $\bm{S} \coloneqq \bm{R}_2^{-1}\bm{\Lambda}\bm{R}_2$, necessarily represent $\mathbb{Z}_2(\hat{x})$-covariant operations.
		
		Now, we prove that a necessary and sufficient condition for $\bm{T}_{\text{RS}}$ to be $\mathbb{Z}_2(\hat{x})$-covariant is that it can be expressed as
		\begin{equation}\label{eq_RS_decomp}
			\bm{T}_{\text{R}_x}(\theta_{1}) \, \bm{T}_{\rm S} (\bm{\Lambda}') \, \bm{T}_{\text{R}_x}(\theta_{2}),
		\end{equation}
		where $\bm{\Lambda}'$ is a diagonal matrix, and $\theta_{1},\theta_{2} \in [0,2\pi)$.
		The idea of the proof is that \cref{lem:invimage} gives us two \emph{necessary} conditions for $\bm{T}_{\text{RS}}$ to be $\mathbb{Z}_2(\hat{x})$-covariant, which, as we will show soon, imply the form in \eqref{eq_RS_decomp}.
		If $\bm{T}_{\text{RS}}$ can be written as in \eqref{eq_RS_decomp}, then it is evidently $\mathbb{Z}_2(\hat{x})$-covariant since each component in the product is $\mathbb{Z}_2(\hat{x})$-covariant.

		Geometrically, the map $\bm{T}_{\text{RS}}$ describes two actions on the Bloch ball. 
		The Bloch ball is first rescaled to an ellipsoid by $\bm{T}_{\rm S} \bigl( \bm{S} \bigr)$, and then rotated by $\bm{T}_{\text{R}}(\bm{R}_1)$. 
		The image of the Bloch ball under $\bm{T}_{\text{RS}}$ is thus an ellipsoid centered at the origin, i.e., a cocentric ellipsoid.
		
		We say that a nonzero vector $\vec{v}$ is an (about-face) \emph{symmetric direction} of a cocentric ellipsoid if the ellipsoid (viewed as a set of points) is invariant under a $\pi$ rotation about the axis passing through the origin along $\vec{v}$. 
		For example, if the ellipsoid is a triaxial ellipsoid (also called a scalene ellipsoid), each of its semi-axes is a symmetric direction.
		If the ellipsoid is in fact a ball, then every direction is a symmetric direction.

		A state is free in the resource theory of $\mathbb{Z}_2(\hat{x})$-asymmetry if and only if it is proportional to $\hat{x}$.
		By \cref{lem:invimage} \ref{it:invfree}, the transformation $\bm{T}_{\text{RS}}$ is $\mathbb{Z}_2(\hat{x})$-covariant only if $\bm{T}_{\text{RS}}$ maps a free state $\hat{x}$ to another free state. 
		Thus, given \cref{eq_Blochspheremap}, for $\bm{T}_{\text{RS}}$ to be $\mathbb{Z}_2(\hat{x})$-covariant, we must have,
		\begin{equation}
			\bm{R}_1 \bm{S} \hat{x}=a\hat{x},
		\end{equation}		
		where $a\in[-1,1]$. That is, the image of the vector $\hat{x}$ under the rotation-scaling transformation represented by $\bm{T}_{\text{RS}}$ is $a\hat{x}$. 

		\begin{enumerate}
			\item Case $a = 0$:
				If $a$ is $0$, then $\bm{S} \hat{x} = \bm{R}_1^{-1} a\hat{x} = 0\hat{x}$.
				That is, $\hat{x}$ is an eigenvector of $\bm{S}$ with eigenvalue 0.
			
			\item Case $a \neq 0$:
			By \cref{lem:invimage} \ref{it:invimage}, the transformation $\bm{T}_{\text{RS}}$ is $\mathbb{Z}_2(\hat{x})$-covariant only if $\hat{x}$ is an about-face symmetric direction of the cocentric ellipsoid corresponding to the image of $\bm{T}_{\text{RS}}$. Then, $\bm{S} \hat{x}$ must be a symmetric direction of the cocentric ellipsoid corresponding to the image of $\bm{T}_{\rm S} \bigl( \bm{S} \bigr) $ (instead of $\bm{T}_{\text{RS}}$), because rotations do not change the shape of the ellipsoid. Given \cref{eq_Blochspheremap}, this is equivalent to saying that $\bm{S} \hat{x}$ is a symmetric direction of the image of $\bm{S}$.
			Since $\bm{S}$ is a symmetric matrix, by \cref{lem:sym_eigen}, $\bm{S} \hat{x}$ is a nonzero eigenvector of $\bm{S}$. 
			Consequently, $\hat{x}$ is itself an eigenvector of $\bm{S}$.% with eigenvalue $a$.
		\end{enumerate}
		Thus, $\hat{x}$ is always an eigenvector of $\bm{S}$.

		Since $\bm{S}$ is a symmetric matrix, there is an orthogonal basis of eigenvectors of $\bm{S}$. % is a diagonal matrix $\Lambda'$. 
		Since $\hat{x}$ is an eigenvector of $\bm{S}$, the other two elements of such a basis necessarily lie in the \yzplane.
		Consequently, we have 
		\begin{equation}\label{eq_S_decomp}
			\bm{T}_{\rm S} \bigl( \bm{S} \bigr) = \bm{T}_{\text{R}_x}(-\theta_{2}) \,  \bm{T}_{\rm S} (\bm{\Lambda}') \, \bm{T}_{\text{R}_x}(\theta_{2})
		\end{equation}
		where $\bm{T}_{\text{R}_x}(\theta_{2})$ is the corresponding change of basis transformation and $\bm{\Lambda}'$ is a diagonal matrix whose eigenvalues coincide with those of $\bm{S}$ up to a sign.

		Since each component of the product in the right-hand side of \cref{eq_S_decomp} is $\mathbb{Z}_2(\hat{x})$-covariant, $\bm{T}_{\text{S}}(\bm{S})$ is also $\mathbb{Z}_2(\hat{x})$-covariant. Then, for $\bm{T}_{\text{RS}}$ to be $\mathbb{Z}_2(\hat{x})$-covariant, $\bm{T}_{\text{R}} \bigl( \bm{R}_1 \bigr) $ must also be $\mathbb{Z}_2(\hat{x})$-covariant. 
		As mentioned in \cref{eq_rotationaboutx,eq_rotationabouty,eq_rotationaboutz}, a rotation is $\mathbb{Z}_2(\hat{x})$-covariant if and only if it is equal to $\bm{T}_{\text{R}_x}(\theta)$, $\bm{T}_{\text{R}_y}(\pi)$, $\bm{T}_{\text{R}_z}(\pi)$ or their composition.
		Since the latter two commute with rotations about $\hat{x}$, they can be absorbed into $\bm{T}_{\rm S} (\bm{\Lambda}')$ by adjusting the signs of the eigenvalues of $\bm{\Lambda}'$.

		Therefore, if a CPTP map $\bm{T}_{\text{RS}}$ is $\mathbb{Z}_2(\hat{x})$-covariant, then it can be expressed in the form of \eqref{eq_RS_decomp}.%, where $\theta_1$ is given by $\theta - \theta_2$.
       	This is also a sufficient condition because any map in the form of \eqref{eq_RS_decomp} evidently commutes with $\bm{T}_{\text{R}_x}(\pi)$.

		\begin{figure}[h]
		    \centering
		    \includegraphics[width=0.75\textwidth]{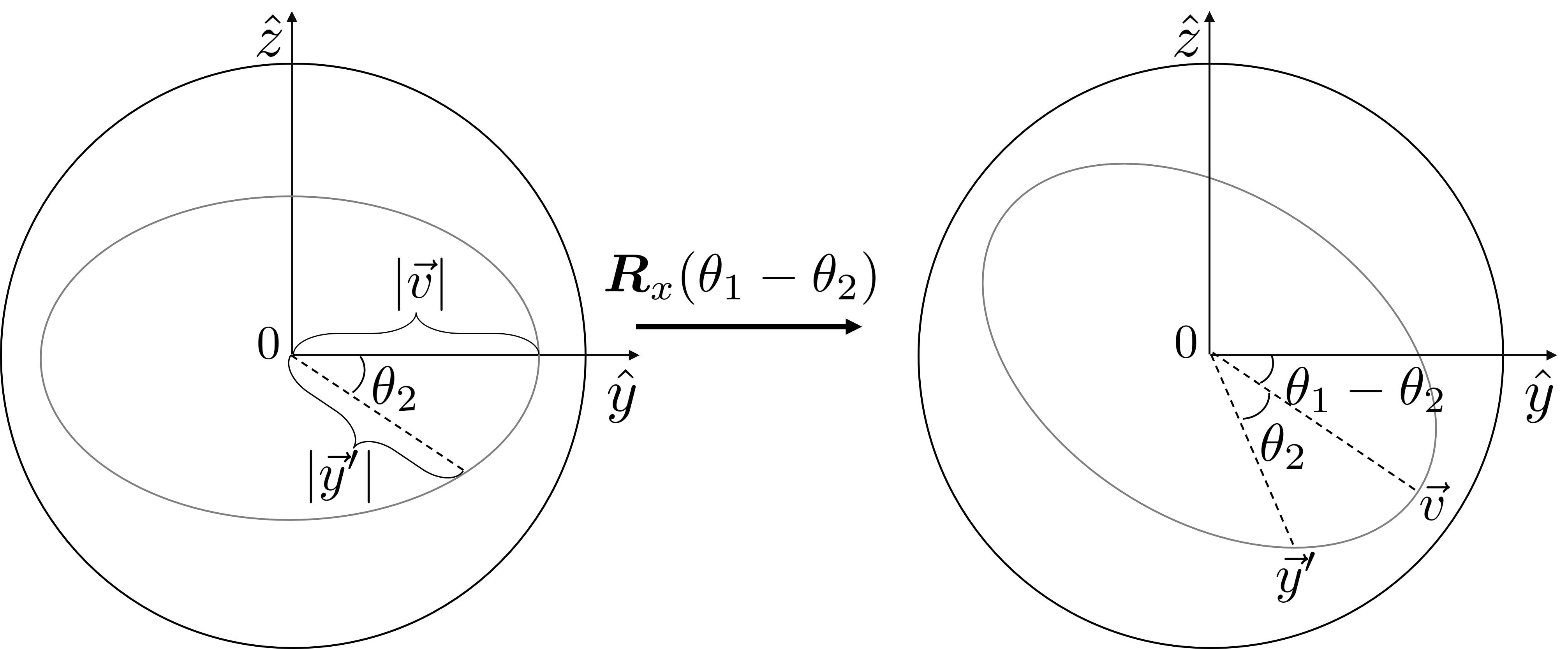}
		    \caption{
				A visual depiction of the action of $\bm{T}_{\text{RS}}$ on the \yzplane.
				Left: The \yz cross-section of the Bloch sphere and its image (the ellipsoid) under a rotation around $\hat{x}$ by $\theta_2$, followed by a rescaling. 
				Right: The image of the ellipsoid on the left under a further rotation around $\hat{x}$ by $\theta_1-\theta_2$. 
				$\vec{v}$ is a symmetric direction and $\vec{y}'$ is the image of $\hat{y}$.}
		    \label{fig_proof}
		\end{figure}
        
		Combining this with the result that $\bm{R}_1\bm{R}_2^{-1}\vec{t}$ must be proportional to $\hat{x}$ (as shown below \cref{eq_TRS}), if $\bm{T}$ is $\mathbb{Z}_2(\hat{x})$-covariant, then it must be of the form:
		\begin{equation}
			\bm{T} =  
			\bm{T}_{\text{R}_x}(\theta_{1}) \, \bm{T}_{\rm T}([t_x,0,0]) \, \bm{T}_{\rm S} (\bm{\Lambda'}) \, \bm{T}_{\text{R}_x}(\theta_{2}),
		\end{equation}
		where we used the fact that for any angle $\theta$ and any real number $t_x$, the two maps, $\bm{T}_{\text{R}_x}(\theta)$ and $\bm{T}_{\rm T} \bigl( [t_x,0,0] \bigr)$, commute.

		That is, for a trace-preserving map $\bm{T}$ to be $\mathbb{Z}_2(\hat{x})$-covariant, it has to be of the form of
		\begin{equation}\label{eq_final_decomp}
			\bm{T}_{\text{R}_x}(\theta_1) \, \bm{T}_{\rm D} \, \bm{T}_{\text{R}_x}(\theta_2),
		\end{equation}
		where $\theta_1, \theta_2 \in [0,2\pi)$.
		Conversely, every transformation as in \eqref{eq_final_decomp} is explicitly $\mathbb{Z}_2(\hat{x})$-covariant.
		
		If we want $\bm{T}$ to also be completely positive, in our notation, it has to be in the form of 
		\begin{equation}\label{eq_final_decomp_2}
			\bm{T}_{\text{R}_x}(\theta_1) \, \bm{T}_{\rm D_f} \, \bm{T}_{\text{R}_x}(\theta_2),
		\end{equation}
		since rotations do not change the positivity of the map $\bm{T}$.
		As shown in \cite[Thm.\ 4 and Eq.\ (17)]{beth_ruskai_analysis_2002}, the set $\afcovext$ of operations defined via \eqref{eq_afcovext} is the set of extreme points of $\afcov$. 
		Since $\afcov$ is both a convex and a compact set, according to the Krein-Milman Theorem, it is exactly the convex hull of its extreme points.
		\end{proof}
		
		\begin{lemma}
			\label{lem:invset}
			Consider a set $E$ of resources which is invariant under an action of a group $G$ on resources.
			A transformation $\mathcal{E}$ is then $G$-covariant in the resource theory of $G$-asymmetry only if the image of $E$ under $\mathcal{E}$ is also invariant under the $G$-action, i.e., only if we have
			\begin{equation}
				g \cdot \mathcal{E}(E) = \mathcal{E}(E)
			\end{equation}
			for all $g \in G$.
		\end{lemma}

		\begin{proof}
			In a resource theory of $G$-asymmetry, a transformation $\mathcal{E}$ is $G$-covariant if and only if it is $G$-covariant.
			That is, it should satisfy 
			\begin{equation}
				g \cdot \mathcal{E} (\ph) = \mathcal{E} ( g \cdot \ph )
			\end{equation}
			for all $g \in G$.
			It follows that if $E$ is a $G$-invariant set, i.e., if
			\begin{equation}
				g \cdot E = E
			\end{equation}
			holds for every $g$, then we have 
			\begin{equation}
				g \cdot \mathcal{E} (E) = \mathcal{E} ( g \cdot E ) = \mathcal{E} ( E ),
			\end{equation}
			which is what we wanted to show.
		\end{proof}
		\begin{corollary}\label{lem:invimage}
			As a direct consequence of \cref{lem:invset}, every $G$-covariant transformation $\mathcal{E}$ has the following two properties:
			\begin{enumerate}
				\item \label{it:invfree} The image of free resources under $\mathcal{E}$ is $G$-invariant;
				\item \label{it:invimage} The image of $\mathcal{E}$ is $G$-invariant.
			\end{enumerate}
		\end{corollary}
		
		\begin{lemma}\label{lem:sym_eigen}
			Let $S$ be a real symmetric $3 \times 3$ matrix, so that the image $E$ of the unit ball under $S$ is a (potentially degenerate) ellipsoid.
			Then $\vec{v}$ is an about-face symmetric direction of this ellipsoid if and only if $\vec{v}$ is a nonzero eigenvector of $S$.
		\end{lemma}
		\begin{proof}
			By spectral theorem, there must be an orthonormal basis $\{\hat{a}, \hat{b}, \hat{c}\}$ of eigenvectors of the symmetric matrix $S$ and consequently, $S$ is a diagonal matrix in this basis:
			\begin{equation}
				S = 
					\begin{bmatrix}
						\alpha & 0&0 \\
						0& \beta & 0 \\
						0& 0 & r \\
					\end{bmatrix}
			\end{equation}
			where $\alpha$, $\beta$, and $r$ are the corresponding eigenvalues. 
			Without loss of generality, we assume that $\abs{\alpha} \ge \abs{\beta} \ge \abs{r} \ge 0$ holds.
			
			The set of all eigenvectors of $S$ is highly dependent on which of the eigenvalues are distinct.
			For this reason, we consider three cases separately.
			\begin{enumerate}
				\item Case $\abs{\alpha} = \abs{\beta} = \abs{r}$: $E$ is the one element set $\{\vec{0}\}$ if $r$ is zero and a ball otherwise. 
				Every direction is a symmetric direction and every vector is an eigenvector. 
				
				\item Case $\abs{\alpha} > \abs{\beta} = \abs{r}$: $E$ is a line segment if $r$ is zero and a spheroid otherwise.
				The symmetric directions are those vectors proportional to $\hat{a}$ as well as the vectors lying in the $\hat{b}$--$\hat{c}$ plane.
				These coincide with the eigenvectors of $S$.

				\item Case $\abs{\alpha} = \abs{\beta} > \abs{r}$: $E$ is a disc if $r$ is zero and a spheroid otherwise.
				The symmetric directions consist of vectors proportional to $\hat{c}$ as well as vectors lying in the $\hat{a}$--$\hat{b}$ plane.
				These coincide with the eigenvectors of $S$.

				\item Case $\abs{\alpha} > \abs{\beta} > \abs{r}$: $E$ is an ellipse with distinct focal points if $r$ is zero and a triaxial ellipsoid (also called a scalene ellipsoid) otherwise.
				The symmetric directions are those vectors which are proportional to one of $\{\hat{a}, \hat{b}, \hat{c}\}$, which are also all the eigenvectors of $S$.\qedhere
			\end{enumerate}
		\end{proof}

	\subsection{Principal downward-closed sets under $\mathbb{Z}_2(\hat{x})$-covariant operations}

		In any resource theory, the set of resources that can be obtained from a given resource under the free operations is termed its \emph{principal downward-closed set} or its \emph{downward closure}. 
		%A general downward-closed subset of a preordered set is one that is closed under `descending' along the preorder.
		
		For a state $\rho$ given by the Bloch ball vector $\vec{r}$, we denote its downward closure under $\mathbb{Z}_2(\hat{x})$-covariant operations by $\dcs(\vec{r})$. 
		We proceed to characterize these sets, in the interest of studying the resource preorder.
		This is useful because there is an isomorphism between the preorder of principal downward-closed sets under set inclusion and the ordering of resources in a resource theory \cite[Lemma 23]{gondaMonotones2019}.
		That is, we have
		\begin{equation}\label{eq_order_dcs}
			\rho \succeq \sigma  \quad \iff \quad  \dcs(\vec{r}) \supseteq \dcs(\vec{s}),
		\end{equation}
		where $\succeq$ is the resource preorder and $\vec{s}$ is the Bloch ball representation of the state $\sigma$.
		
		By \cref{thm:Z2-cov_maps}, the set $\dcs(\vec{r})$ has cylindrical symmetry with respect to the $\hat{x}$ axis.
		In other words, quantum states related by a rotation around the $\hat{x}$ axis are equivalent according to the resource ordering.
		Therefore, to compute $\dcs(\vec{r})$, it suffices to consider a representative vector of the equivalence class that can be expressed as $(r_x, 0, r_z)$ with $r_z \geq 0$.
		This representative always exists, so we hereafter assume $\vec{r}$ is of this form.
		Using the characterization of $\mathbb{Z}_2(\hat{x})$-covariant operations from \cref{thm:Z2-cov_maps}, the downward closure of $\vec{r}$ can be identified by, first
		\begin{enumerate}
			\item constructing the set of states reachable from $\vec{r}$ via the operations from the set $\afcovext$, then
			\item taking the convex hull of this set, and finally
			\item considering all rotations around the $\hat{x}$ axis.
		\end{enumerate}

\begin{figure}[h]
		    \centering
		    \includegraphics[width=0.3\textwidth]{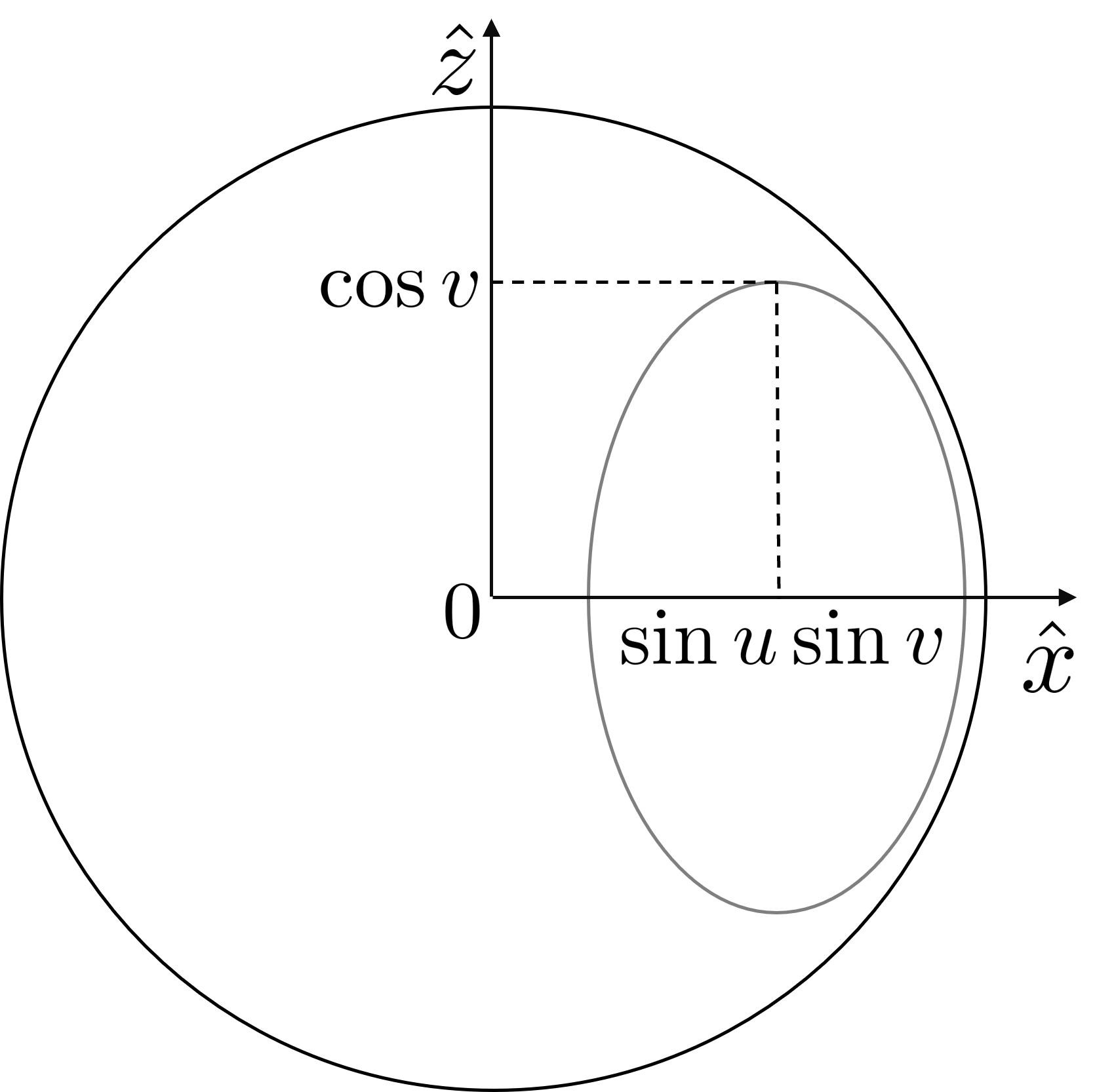}
			\hspace{1cm}
			\includegraphics[width=0.3\textwidth]{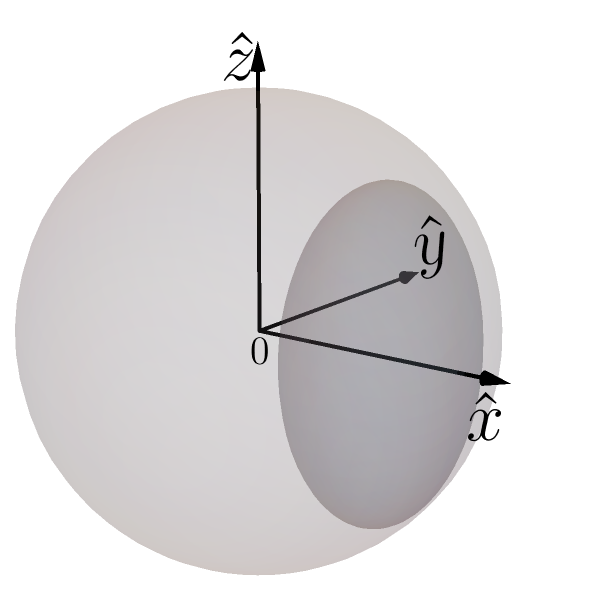}
		    \caption{Left: The \xz cross-section of the Bloch sphere and its image under a CPTP map from $\afcovext$ that satisfies $\cos{v}>0$ and $\cos{u}>0$. 
			Right: The 3D representation of the Bloch ball and its image under the same CPTP map as in the left plot.}
		    \label{fig_ellipse}
		\end{figure}

	Geometrically, the image of the Bloch ball under a generic map from $\afcovext$ is an ellipsoid centred at the point $(\sin{u}\sin{v},0,0)$ with radii $\cos{u}\cos{v}$, $\cos{u}$, and $\cos{v}$ in the $\hat{x}$, $\hat{y}$ and $\hat{z}$ directions, respectively. 
	One such ellipsoid (with $\cos{v}>0$ and $\cos{u}> 0$) is depicted in \cref{fig_ellipse}. % . 

		Let $$s_x \coloneqq \sin{u}\sin{v}+r_x \cos{u}\cos{v},$$ and let $$s_z \coloneqq r_z \cos{v}.$$ 
		We find that the region in the \xz plane to which $\vec{r}$ can be mapped by free operations is the following convex set
		\begin{equation}\label{eq_dcs}
			\begin{split} 
				\dcs(\vec{r})_{xz} = \Set*[\Big]{ \bigl( s_x ,  0 , s_z \bigr)  \given  u \in [0,2\pi), \, v \in [0,\pi] }. \\
			\end{split}
		\end{equation}

\begin{figure}[h]
		    \centering
		    \includegraphics[width=0.4\textwidth]{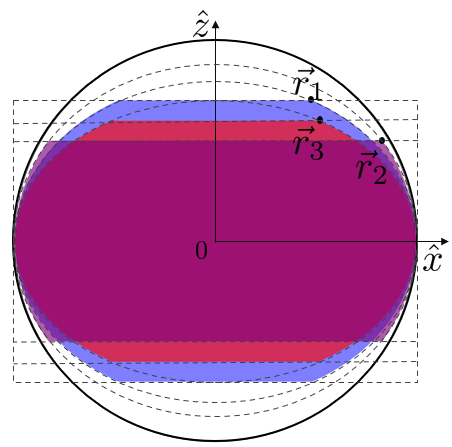}
			\hspace{1cm}
			\includegraphics[width=0.4\textwidth]{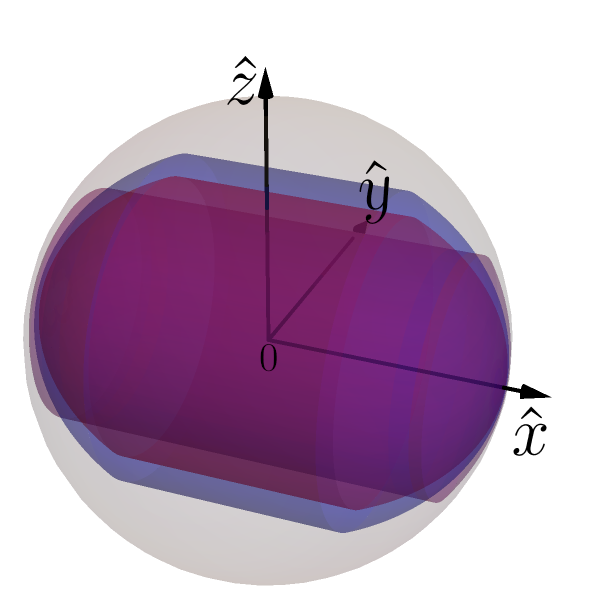}
		    \caption{Left: The \xz cross-section of the downward closures under $\mathbb{Z}_2(\hat{x})$-covariant operations of states $\vec{r}_1$, $\vec{r}_2$, and $\vec{r}_3$, represented by the blue, red and the purple region, respectively. Each downward closure
		     can be viewed as the intersection of a cylinder and a prolate spheroid. 
		    Right: The 3D representations of the downward-closed sets in the left plot.  
		    }
		    \label{fig_nototal}
		\end{figure}

		The full downward closure $\dcs(\vec{r})$ of the state $\rho$ is then this set symmetrized with respect to rotations around the $\hat{x}$ axis.
		Cross-sections of the downward-closed sets for three different initial states are plotted in \cref{fig_nototal}.
		
	\subsection{Necessary and sufficient conditions for resource conversion}\label{sec_Z2_complete_set}

		We now describe the shape of the boundary of the region $\dcs(\vec{r})_{xz}$.
		This will allow us to completely characterize the resource preorder via equivalence \eqref{eq_order_dcs}.
		
		First, notice that the maximal distance of points in this region from the $\hat{x}$ axis is $r_z$, as is evident from \cref{eq_dcs}.
		This bound is also depicted by horizontal dashed lines in the left plot of \cref{fig_nototal}.
		We can immediately read off that the function $\vec{r} \mapsto r_z = \vec{r} \cdot \hat{z}$ is a resource monotone.

		Now, consider a fixed value $s_z = r_z \cos{v}$ of the $z$ coordinate of a generic point in the set $\dcs(\vec{r})_{xz}$ defined in \cref{eq_dcs}.
		Conditioned on this constraint, let us find the maximum (and minimum) value of the $x$ coordinate $s_x$.

		When $r_z=0$, any state in the equivalence class represented by $\vec{r}$ is a free state, and its $\dcs(\vec{r})_{xz}$ is always the line segment from $-\hat{x}$ to $\hat{x}$. $r_z\neq 1$.

	When $r_z\neq0$, to simplify the expressions, we write $t_z \coloneqq \linefaktor{s_z}{r_z} \in [0,1]$ for the rescaled $z$ coordinate, so that we have $\cos{v} = t_z$ and $\sin{v} = \sqrt{1 - t_z^2}$.
		We then have
		\begin{equation}\label{eq_sx}
			s_x = \sqrt{1- t_z^2} \sin{u} + r_x t_z \cos{u}.
		\end{equation}

		To find the extremal values of $s_x$, we set its derivative with respect to $u$ to zero, i.e.,
		\begin{equation}
		   	 \frac{\rmd s_x}{\rmd u} = \sqrt{1- t_z^2} \cos{u}  - r_x t_z \sin{u} = 0,
		\end{equation}
		and we get
		\begin{equation}
			u  = \arctan{\left(\frac{1}{r_x}\sqrt{ \frac{1}{t_z^2} - 1 }\right)}.
		\end{equation}
		Substituting the above critical value of $u$ into \cref{eq_sx} and simplifying using trigonometric identities gives an equation for a boundary of $\dcs(\vec{r})_{xz}$:
		\begin{equation}
			s_x = \pm \sqrt{1 - s_z^2\frac{1-r^2_x}{r^2_z}},
		\end{equation}
		which describes an ellipse in Cartesian coordinates $(s_x,s_z)$ with its major axis $\hat{x}$ and its minor axis along $\hat{z}$ with radius $\linefaktor{r_z}{\!\sqrt{1-r^2_x}}$. 
		Such ellipses are depicted with dashed curves in the left plot of \cref{fig_nototal} for three examples of $\vec{r}$.
		
		In conclusion, $\dcs(\vec{r})_{xz}$ is the intersection of a rectangle and an ellipse.
		Instead of the parametrization in \cref{eq_dcs}, we can equivalently describe it as the set of points $\vec{s} = (s_x, 0, s_z)$ satisfying
		\begin{equation}\label{eq_dcs_boundary}
			\abs{s_z} \leq r_z  \qquad \text{and} \qquad s_x^2 + \frac{1-r^2_x}{r^2_z} s_z^2 \leq 1.
		\end{equation}
		
		To obtain the full downward closure $\dcs(\vec{r})$, we apply all rotations about the $\hat{x}$ axis to this region.
		Under such rotations, the rectangle becomes a cylinder and the ellipse becomes a prolate spheroid (i.e., an ellipsoid whose two minor radii are equal), thus the set $\dcs(\vec{r})$ is the intersection of a cylinder and a prolate spheroid (including their interiors).
		Such intersections are depicted in the right plot of \cref{fig_nototal}.
		In other words, we can replace $s_z^2$ with $s_y^2 + s_z^2$ in equations \eqref{eq_dcs_boundary} to get the two 3D boundaries. % for a state $\vec{r} = (r_x,0,r_z)$.
		That is, $\dcs(\vec{r})$ for a state $\vec{r} = (r_x, 0, r_z)$ with $r_z \geq 0$ is the set of all points $\vec{s} = (s_x, s_y, s_z)$ satisfying
		\begin{equation}\label{eq_dcs_boundary_3D}
			\sqrt{s_y^2 + s_z^2} \leq r_z  \qquad \text{and} \qquad s_x^2 + \frac{1-r^2_x}{r^2_z} (s_y^2 + s_z^2) \leq 1.
		\end{equation}
		To get it for a general state, we just need to rotate $\vec{r}$ around the $\hat{x}$ axis. This amounts to replacing $r_z^2$ with $r_y^2 + r_z^2$ in the above expressions. 
		
		When $r_y^2 + r_z^2=0$, such a state is a free state and its $\dcs(\vec{r})$ is the line segment from $-\hat{x}$ to $\hat{x}$. 

		The characterization of principal downward-closed sets can equivalently be understood as a characterization of the conditions for conversion in the resource theory \cite[Appendix A]{gondaMonotones2019}.
		We summarize the result as follows, the proof of which we just presented above.
		\begin{theorem}\label{lem:monotones}
			Let $\vec{r}$ and $\vec{s}$ denote Bloch representations of two qubit quantum states, $\rho$ and $\sigma$.
			There is a $\mathbb{Z}_2(\hat{x})$-covariant operation that maps $\rho$ to $\sigma$ if and only if $s_x^2 = 1$, or
			\begin{equation*}\label{conditionforconversion}
				\sqrt{r_y^2+r_z^2} \ge \sqrt{s_y^2+s_z^2} \qquad \text{and} \qquad \sqrt{\frac{r_y^2+r^2_z}{1-r_x^2}} \ge \sqrt{\frac{s_y^2+s^2_z}{1-s_x^2}} \qquad \text{with} \qquad r_x^2, s_x^2 \neq 1.
			\end{equation*}
		\end{theorem}
		Thus, we define the \emph{cylindrical radius} of a state $\rho$ by 
		\begin{equation}
			A_x(\rho) \coloneqq \sqrt{r_y^2+r_z^2}
		\end{equation}

		and the \emph{spheroidal minor radius} of $\rho$ by
		\begin{equation}
			B_x(\rho) \coloneqq 	        		\begin{cases}
				\sqrt{\frac{r_y^2+r_z^2}{1-r_x^2}}  &\text{ if } r_x^2<1, \\
				0 &\text{ if } r_x^2=1.
			\end{cases}
		\end{equation}

		\cref{lem:monotones} says precisely that the pair of $A_x$ and $B_x$ is a \emph{complete set of monotones} in the resource theory of $\mathbb{Z}_2(\hat{x})$-asymmetry. Furthermore, if \cref{lem:monotones} is rewritten in terms of the monotones $A_x(\rho)$ and $B_x(\rho)$, we obtain \cref{thm:complete_characterization} from the main text.  The above results, therefore, provide a proof of the latter theorem.

\section{Dependence relations for states with a fixed purity}
\label{sec_fixedPurity}

Here we consider impure states. When $\sqrt{r_x^2+r_y^2+r_z^2}=r\in[0,1)$, we have
\begin{subequations}
	\begin{align}
		A_x(\rho) &= \sqrt{r^2-r_x^2}, \label{eq_Axfixpur}\\
		B_x(\rho) &= \sqrt{\frac{r^2-r_x^2}{1-r_x^2}}, \label{eq_Bxfixpur}
		% \begin{cases}
		% 	, &\text{ if } r_z^2<1\\
		% 0, &\text{ if } r_x^2=1
		% \end{cases},	
	\end{align}
\end{subequations}
\begin{subequations}
	\begin{align}
		A_y(\rho) &=  \sqrt{r^2-r_y^2}, \label{eq_Ayfixpur}\\
		B_y(\rho) &=     \sqrt{\frac{r^2-r_y^2}{1-r_y^2}},  \label{eq_Byfixpur}
		% \begin{cases}
		% 	, &\text{ if } r_y^2<1\\
		% 	0, &\text{ if } r_y^2=1
		% 	\end{cases},
		\end{align}
	\end{subequations}
	\begin{subequations}
		\begin{align}
		A_z(\rho) &=  \sqrt{r^2-r_z^2}, \label{eq_Azfixpur}\\
		B_z(\rho) &=     \sqrt{\frac{r^2-r_z^2}{1-r_z^2}}\label{eq_Bzfixpur}
		% \begin{cases}
		% 	\sqrt{\frac{r^2-r_z^2}{1-r_z^2}}, &\text{ if } r_z^2<1\\
		% 	0, &\text{ if } r_z^2=1
		% 	\end{cases}. 
	\end{align} 
\end{subequations}
Eqs. \eqref{eq_Axfixpur}, \eqref{eq_Ayfixpur} and \eqref{eq_Azfixpur} give that, 
\begin{align}
	\label{eq_fixrA}
	A_x^2(\rho) +  A_y^2(\rho) +  A_z^2(\rho) =2r^2,
\end{align}
while Eqs. \eqref{eq_Bxfixpur}, \eqref{eq_Byfixpur} and \eqref{eq_Bzfixpur} give that,
\begin{align}
\label{eq_fixrB}
	\frac{1}{1-B_x^2} + \frac{1}{1-B_y^2} + \frac{1}{1-B_z^2} = \frac{3-r^2}{1-r^2} 
\end{align}
Note that, \cref{eq_fixrA} is valid when the purity is any value, including when $r=1$, while \cref{eq_fixrA} is only valid when $r<1$.

\section{Sufficient condition for a partial order to have infinite width and to be not weak}
\label{app_incompara}

Here, we prove the claim made in \cref{sec_recipe} that if there exist two monotones, say $M$ and $M'$, and two closed intervals $X, X' \subseteq \mathbb{R}$, such that all points in the rectangle $X \times X'$ are realizable as the values of $M$ and $M'$ for some resource, then the resource order has infinite width and it is not weak (i.e., the incomparability relation is not transitive). 

First, we review a few terminological conventions in order theory. Given a partial order, a \emph{chain} is a subset of the partial order where every pair of elements is comparable (and thus a chain is a totally order set), while an \emph{antichain} is a subset of the partial order where every pair of elements is incomparable.
The \emph{width} of the partial order is the size of its largest antichain. The partial order is said to be \emph{weak} if its incomparability relation is not transitive.

\begin{figure}[h!]
	\centering
	\includegraphics[width=0.3\textwidth]{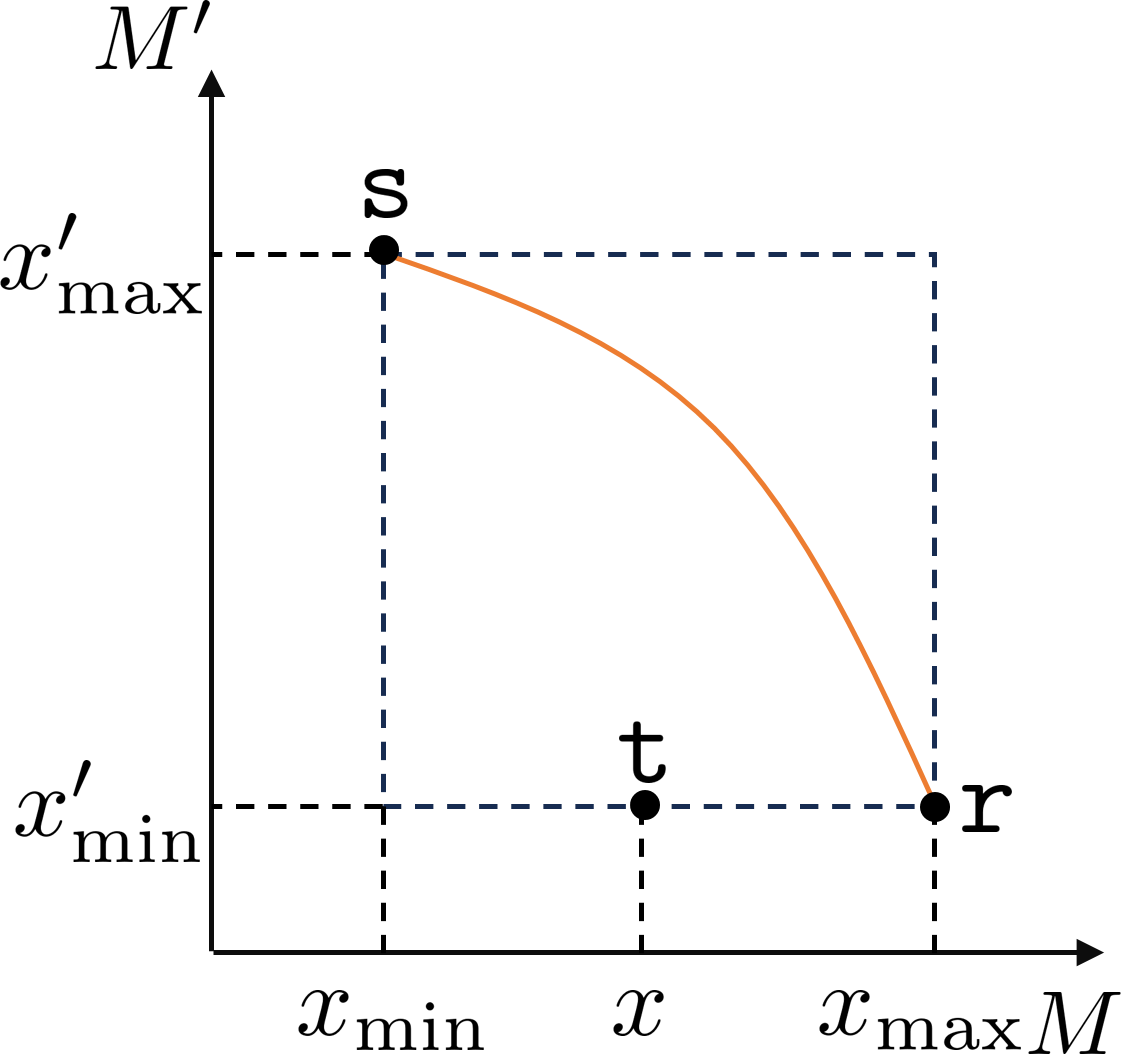}
	\caption{$M$ and $M'$ are two monotones. Any pair of values in $[x_{\rm min}, x_{\rm max}]\times[x^{\prime}_{\rm min}, x^{\prime}_{\rm max}]$ is realizable by some resource. The orange curve describes an antichain of infinite length and consequently, the width of the partial order is infinite. $\tt r$, $\tt s$, ${\tt t}$ denotes three resources. $\tt r$ and $\tt s$ are incomparable, $\tt s$ and $\tt t$ are incomparable but $\tt r$ and $\tt t$ are {\em not} incomparable. Thus, the incomparability relation is not transitive and consequently, the partial order is not weak. }
	\label{fig_weak}
\end{figure}

Suppose that $x_{\rm min}$ and $x_{\rm max}$ denote the minimum and maximum values in the interval $X$, that is, $X = [x_{\rm min}, x_{\rm max}]$.  Similarly, suppose that $X' = [x^{\prime}_{\rm min}, x^{\prime}_{\rm max}]$.  Recall that, by assumption, any pair of values $(x,x') \in X \times X'$ is realizable by some resource.   Consider a curve in  $X \times X'$ connecting the point $(x_{\rm min},x^{\prime}_{\rm max})$ to the point $(x_{\rm max},x^{\prime}_{\rm min})$ and that is monotonically increasing in $X$ and monotonically decreasing in $X'$, such as the one in \cref{fig_weak}  Such a curve describes an antichain because for any two points along the curve, $M$ is greater for the first point, while $M'$ is greater for the second point.  Because there is a continuum of points on the curve, the antichain is of infinite length. Consequently, the width of the partial order is infinite. 

Let $x$ be a value in the interval $X$ such that $x_{\rm min} <x< x_{\rm max}$.   Now consider three resources $\tt r$, $\tt s$, ${\tt t} \in R$, shown in \cref{fig_weak}, which are defined by the values of the monotones $M$ and $M'$ that they realize as follows: $\tt r$ realizes the pair of values $(x_{\rm max},x'_{\rm min})$, $\tt s$ realizes $(x_{\rm min},x'_{\rm max})$ and $\tt t$ realizes $(x,x'_{\rm min})$. $\tt r$ and $\tt s$ are incomparable since $M({\tt r})> M({\tt s})$ and $M'({\tt r})< M'({\tt s})$. $\tt s$ and $\tt t$ are incomparable since $M({\tt s})< M({\tt t})$ and $M'({\tt s})> M'({\tt t})$.  But $\tt r$ and $\tt t$ are {\em not} incomparable.  Rather, $\tt r$ is higher in the partial order than $\tt t$ because $M({\tt r})>M({\tt t})$ and $M'({\tt r})>M'({\tt t})$. 
Thus, the incomparability relation is not transitive and consequently, the partial order is not weak.

\bibliographystyle{unsrt}
\bibliography{references}
\end{document}